\documentclass[11pt]{article}
\usepackage[margin=1in]{geometry}
\usepackage{times}
\usepackage[
backend=biber,
style=alphabetic,
sorting=nyt,
maxnames=7,
maxalphanames=7,
backref=true,
url=false
]{biblatex}
\usepackage{titlesec}
\titlespacing*{\paragraph}{\parindent}{1ex plus 0.2ex minus 0.2ex}{1em}

\usepackage{makecell}  
\usepackage{booktabs}

\usepackage{tikz}
\usepackage{scalerel}
\usepackage{pict2e}
\usepackage{setspace}
\usepackage{fancyhdr}
\usepackage{tkz-euclide}
\usetikzlibrary{calc}
\usetikzlibrary{patterns,arrows.meta}
\usetikzlibrary{shadows}
\usetikzlibrary{external}
 
\usepackage{pgfplots}
\pgfplotsset{compat=newest}
\usepgfplotslibrary{statistics}
\usepgfplotslibrary{fillbetween}

\usepackage{xcolor}

\usepackage{nicefrac}

\usepackage[colorlinks=true,citecolor=blue,linkcolor=blue]{hyperref}

\hypersetup{
    colorlinks,
    linkcolor={red!50!black},
    citecolor={blue!50!black},
    urlcolor={blue!80!black}
}

\usepackage[]{amsmath,amssymb,amsfonts,latexsym,amsthm,enumerate,fullpage,xcolor,bbm}

\allowdisplaybreaks
\usepackage{wrapfig}
\usepackage{float}

\usepackage[ruled]{algorithm2e}

\usepackage[nameinlink, noabbrev, capitalize]{cleveref}
\usepackage{comment}
\usepackage{nicefrac}
\crefname{prop}{Proposition}{Propositions}
\crefname{ineq}{inequality}{inequalities}
\creflabelformat{ineq}{#2(#1)#3}
\usepackage{todonotes}
\usepackage{multicol}
\newtheorem{counter}{Counter}[section]
\newtheorem{theorem}[counter]{Theorem}
\newtheorem{thm}{Theorem}
\crefname{thm}{Theorem}{Theorems}
\Crefname{thm}{Theorem}{Theorems}
\newtheorem{lemma}[counter]{Lemma}

\newtheorem{claim}[counter]{Claim}
\newtheorem{fact}[counter]{Fact}

\newtheorem{corollary}[counter]{Corollary}

\newtheorem{definition}[counter]{Definition}

\newtheorem{remark}[counter]{Remark}
\newtheorem{example}[counter]{Example}

\usepackage{subcaption}
\usepackage{graphicx}
\usepackage{color, colortbl}
\definecolor{LightCyan}{rgb}{0.88,1,1}
\definecolor{Gray}{gray}{0.9}
\usetikzlibrary{positioning}
\usetikzlibrary{decorations.pathmorphing}
\usetikzlibrary{automata,positioning}
\usetikzlibrary{arrows}
\usetikzlibrary{arrows.meta}
\usetikzlibrary{calc}

\usepackage{enumitem}
\newlength\caselen
\newlist{casesenum}{enumerate}{2}
\setlist[casesenum,1]{label=\textbf{Case~\arabic*.}, 
  itemindent=*,leftmargin=0pt}
\setlist[casesenum,2]{label=\textbf{Case~\roman*.}, 
  itemindent=*,leftmargin=\parindent}

\newcommand{\N}{\mathbb{N}}

\newcommand{\R}{\mathbb{R}}
\newcommand{\poly}{\operatorname{poly}}
\newcommand{\polylog}{\operatorname{polylog}}
\newcommand{\pr}{{\prime}}

\newcommand{\U}{\mathbf{U}}
\newcommand{\V}{\mathbf{V}}
\newcommand{\X}{\mathbf{X}}
\newcommand{\OO}{\mathbf{O}}
\newcommand{\Xpr}{\mathbf{X}^\pr}
\newcommand{\Y}{\mathbf{Y}}

\newcommand{\A}{\mathbf{A}}
\newcommand{\B}{\mathbf{B}}

\newcommand{\W}{\mathbf{W}}

\newcommand{\ZZ}{\mathbf{Z}}
\newcommand{\mC}{\mathcal{C}}

\newcommand{\T}{\mathbf{T}}

\renewcommand{\multicitedelim}{\addsemicolon\space}

\newcommand\numberthis{\addtocounter{equation}{1}\tag{\theequation}}

\newcommand{\bits}{\{0,1\}}
\newcommand{\zo}{\{0,1\}}

\newcommand{\abs}[1]{\left\lvert #1 \right\rvert}

\DeclareMathOperator*{\Var}{Var}
\DeclareMathOperator{\Addr}{Addr}

\renewcommand{\O}{\mathbf{O}}

\usepackage{xspace}

\newcommand{\NOSF}[1][\relax]{%
\ifx\relax#1\textrm{NOSF source}\xspace%
\else\ensuremath{{\left(#1\right)}}\textrm{-NOSF source}%
\fi
}
\newcommand{\uniNOSF}[1][\relax]{%
\ifx\relax#1\textrm{uniform NOSF source}\xspace%
\else\textrm{uniform }\ensuremath{{\left(#1\right)}}\textrm{-NOSF source}%
\fi
}
\newcommand{\oNOSF}[1][\relax]{%
\ifx\relax#1\textrm{oNOSF source}\xspace%
\else\ensuremath{{\left(#1\right)}}\textrm{-oNOSF source}%
\fi
}
\newcommand{\unioNOSF}[1][\relax]{%
\ifx\relax#1\textrm{uniform oNOSF source}\xspace%
\else\textrm{uniform }\ensuremath{{\left(#1\right)}}\textrm{-oNOSF source}%
\fi
}
\newcommand{\oNOBF}[1][\relax]{%
\ifx\relax#1\textrm{oNOBF source}\xspace%
\else\ensuremath{{\left(#1\right)}}\textrm{-oNOBF source}%
\fi
}
\newcommand{\CG}[1][\relax]{%
\ifx\relax#1\textrm{aCG source}\xspace%
\else\ensuremath{{\left(#1\right)}}\textrm{-aCG source}%
\fi
}
\newcommand{\uniCG}[1][\relax]{%
\ifx\relax#1\textrm{uniform aCG source}\xspace%
\else\textrm{uniform }\ensuremath{{\left(#1\right)}}\textrm{-aCG source}%
\fi
}
\newcommand{\NOSFs}[1][\relax]{%
\ifx\relax#1\textrm{NOSF sources}\xspace%
\else\ensuremath{{\left(#1\right)}}\textrm{-NOSF sources}%
\fi
}
\newcommand{\NOBFs}[1][\relax]{%
\ifx\relax#1\textrm{NOBF sources}\xspace%
\else\ensuremath{{\left(#1\right)}}\textrm{-NOBF sources}%
\fi
}
\newcommand{\uniNOSFs}[1][\relax]{%
\ifx\relax#1\textrm{uniform NOSF sources}\xspace%
\else\textrm{uniform }\ensuremath{{\left(#1\right)}}\textrm{-NOSF sources}%
\fi
}
\newcommand{\oNOSFS}[1][\relax]{%
\ifx\relax#1\textrm{oNOSF Sources}\xspace%
\else\ensuremath{{\left(#1\right)}}\textrm{-oNOSF Sources}%
\fi
}
\newcommand{\oNOSFs}[1][\relax]{%
\ifx\relax#1\textrm{oNOSF sources}\xspace%
\else\ensuremath{{\left(#1\right)}}\textrm{-oNOSF sources}%
\fi
}
\newcommand{\oNOBFs}[1][\relax]{%
\ifx\relax#1\textrm{oNOBF sources}\xspace%
\else\ensuremath{{\left(#1\right)}}\textrm{-oNOBF sources}%
\fi
}
\newcommand{\unioNOSFs}[1][\relax]{%
\ifx\relax#1\textrm{uniform oNOSF sources}\xspace%
\else\textrm{uniform }\ensuremath{{\left(#1\right)}}\textrm{-oNOSF sources}%
\fi
}

\newcommand{\CGs}[1][\relax]{%
\ifx\relax#1\textrm{aCG sources}\xspace%
\else\ensuremath{{\left(#1\right)}}\textrm{-aCG sources}%
\fi
}
\newcommand{\uniCGs}[1][\relax]{%
\ifx\relax#1\textrm{uniform aCG sources}\xspace%
\else\textrm{uniform }\ensuremath{{\left(#1\right)}}\textrm{-aCG sources}%
\fi
}

\newcommand{\Ext}{\mathsf{Ext}}

\newcommand{\sExt}{\mathsf{sExt}}

\newcommand{\Opt}{\mathsf{Opt}}
\newcommand{\Actual}{\mathsf{Actual}}
\newcommand{\Cond}{\mathsf{Cond}}
\newcommand{\sCond}{\mathsf{sCond}}

\newcommand{\Samp}{\mathsf{Samp}}

\newcommand{\eps}{\varepsilon}
\newcommand{\epspr}{\varepsilon^\pr}

\newcommand{\Supp}{\mathsf{Supp}}

\newcommand{\Adv}{\mathsf{Adv}}

\newcommand{\cX}{\mathcal{X}}

\renewcommand{\W}{\mathbf{W}}
\newcommand{\minH}{H_\infty}
\newcommand{\sminH}{\minH^\varepsilon}
\newcommand{\avgcondminH}{\widetilde{H}_\infty}
\newcommand{\floor}[1]{\left\lfloor#1\right\rfloor}
\newcommand{\ceil}[1]{\left\lceil#1\right\rceil}

\DeclareMathOperator{\supp}{Supp}

\DeclareMathOperator{\I}{\mathbf{I}}
\DeclareMathOperator{\oI}{\mathbf{oI}}

\DeclareMathOperator{\MaxI}{\I_{\max}}
\DeclareMathOperator{\MaxoI}{\oI_{\max}}

\DeclareMathOperator*{\argmax}{\arg\!\max}

\DeclareMathOperator*{\E}{\mathbb{E}}

\DeclareMathOperator{\Reduce}{\mathsf{Reduce}}
\DeclareMathOperator{\Rect}{\mathsf{Rect}}
\DeclareMathOperator{\NOSFSamp}{\mathsf{oNOSFSamp}}
\DeclareMathOperator{\TCond}{\mathsf{2Cond}}

\usepackage{subfiles}
\newcommand{\dobib}{
    \printbibliography
}

\newcommand{\eshan}[1]{{\color{red} \footnotesize(Eshan: #1)}}
\newcommand{\mohit}[1]{{\color{orange} \footnotesize(Mohit: #1)}}
\newcommand{\nomi}[1]{{\color[RGB]{0, 100, 0} \footnotesize(Nomi: #1)}}
\newcommand{\rocco}[1]{{\color{blue} \footnotesize{Rocco: #1}}}

\newcommand{\nocomments}{
    \renewcommand{\eshan}[1]{}
    \renewcommand{\mohit}[1]{}
    \renewcommand{\nomi}[1]{}
    \renewcommand{\rocco}[1]{}
}

\nocomments 

\usepackage{tocloft}

\begin{document}
\renewcommand{\dobib}{}
\renewcommand*{\multicitedelim}{\addcomma\space}

\title{Condensing and Extracting Against Online Adversaries}

\author{  Eshan Chattopadhyay\thanks{Supported by a Sloan Research Fellowship, NSF CAREER Award 2045576, and NSF Award CCF-2514586.}\\ Cornell University\\ \texttt{eshan@cs.cornell.edu}  \and Mohit Gurumukhani\footnotemark[1] \\ Cornell University\\ \texttt{mgurumuk@cs.cornell.edu} \and  Noam Ringach \thanks{Supported by NSF GRFP grant DGE – 2139899,  NSF CAREER Award 2045576, a Sloan Research Fellowship, and NSF Award CCF-2514586.}   \\ Cornell University\\ \texttt{nomir@cs.cornell.edu} \and Rocco A. Servedio\thanks{Supported by NSF Award CCF-2106429 and NSF Award CCF-2211238.} \\ Columbia University \\ \texttt{rocco@cs.columbia.edu}}
 \date{}

 \maketitle
 \pagenumbering{Roman}
 
\begin{abstract}
We investigate the tasks of deterministically condensing and extracting randomness from Online Non-Oblivious Symbol Fixing (oNOSF) sources, a natural model of defective random sources for which it is known that extraction is impossible in many parameter regimes [AORSV, EUROCRYPT'20]. A $(g,\ell)$-oNOSF source is a sequence of $\ell$ blocks $\mathbf{X} = (\mathbf{X}_1, \dots, \mathbf{X}_{\ell})\sim (\{0, 1\}^{n})^{\ell}$, where at least $g$ of the blocks are \emph{good} (are independent and have some min-entropy), and the remaining \emph{bad} blocks are controlled by an \emph{online adversary} where each bad block can be arbitrarily correlated with any block that appears before it. 

The existence of condensers (in regimes where extraction is impossible) was recently studied in [CGR, FOCS'24]. They proved condensing impossibility results for various values of $g$ and $\ell$, and they showed the existence of condensers matching the impossibility results in the special case when $n$ is exponential in $\ell$ (i.e., the setting of few blocks of large length).

In this work, not only do we construct the first explicit condensers matching the existential results of [CGR, FOCS'24], but we make a doubly exponential improvement by handling the case when $n$ is only polylogarithmic in $\ell$. We also obtain a much improved explicit construction for transforming low-entropy oNOSF sources (where the good blocks only have min-entropy, as opposed to being uniform) into uniform oNOSF sources. 

As our next result, we essentially resolve the question of the existence of condensers for oNOSF sources by showing the existence of condensers in almost all parameter regimes, even when $n$ is a large enough constant and $\ell$ is growing.

We find interesting connections and applications of our results on condensers to collective coin flipping and collective sampling, problems that are well-studied in fault-tolerant distributed computing. We use our condensers to provide very simple protocols for these problems. 

Next, we turn to understanding the possibility of extraction from oNOSF sources. For proving lower bounds, we introduce and initiate a systematic study of a new, natural notion of the influence of  functions, which we call \emph{online influence}, and establish tight bounds on the total online influence of  functions, which imply extraction lower bounds. Lastly, we give explicit extractor constructions for oNOSF sources using novel connections to leader election protocols, and we further construct the required leader election protocols. These extractor constructions achieve parameters that go beyond the standard resilient functions of [AL, Combinatorica'93].
\end{abstract}


\maketitle
\newpage


\cleardoublepage
\begingroup
\newgeometry{top=0.8in,bottom=0.8in} 

\setcounter{tocdepth}{2}
\tableofcontents
\clearpage
\restoregeometry
\endgroup

\newpage
\pagenumbering{arabic}


\section{Introduction}

Randomness is extremely useful in computation with wide-ranging applications in algorithm design, cryptography, distributed computing protocols, machine learning, error-correcting codes, and much more \cite{motwani1995randomized, vadhan_pseudorandomness_2012}. Most of these applications require access to high quality randomness. However in a lot of settings, especially arising in practice, algorithms only have access to low quality source of randomness.
This motivates the notion of \emph{condensers}: functions that transform weak random sources into strong random sources that are of \emph{better quality}.

The standard way of measuring the amount of randomness is using min-entropy. Formally, for a source (distribution) $\X$ with support $\Omega$,  define its min-entropy as $\minH(\X) = \min_{x\in \Omega} \log_2(1 / \Pr[\X = x])$. If $\Omega=\zo^n$ and $\minH(\X)\geq k$, we say that $\X$ is an $(n,k)$-source. We will also need the notion of smooth min-entropy, which measures how close a distribution is to having high entropy. Formally, for a source $\X$, its smooth min-entropy with parameter $\eps$ is defined as $\sminH(\X)=\max_{\Y: \abs{\X - \Y}\le \eps}\{\minH(\Y)\}$, where $\abs{\cdot}$ denotes the statistical distance (\cref{defn: statistical distance}).

With this, we are ready to formally define \emph{deterministic condensers}:
\begin{definition}
A function $\Cond: \zo^n \to \zo^m$ is a \emph{$(k_{in}, k_{out}, \eps)$-condenser for a family of distributions $\cX$} if for all $\X\in \cX$ with $\X\sim \zo^n$ and $\minH(\X)\ge k_{in}$, we have that $\sminH(\Cond(\X))\ge k_{out}$.

We say $\frac{k_{in}}{n}$ is the \emph{input entropy rate}, $\frac{k_{out}}{m}$ is the \emph{output entropy rate}, and $m - k_{out}$ is the \emph{entropy gap} of $\Cond$. 
\end{definition}

The task of the condenser is to make the output entropy rate as high as possible compared to the input entropy rate; i.e., to make the output distribution more ``condensed''. 
Related to this, it is also desirable to have as small entropy gap as possible. 
Condensers with entropy gap $0$ are known as \emph{randomness extractors} and have been extensively studied in theoretical computer science.

When $\cX$ is the family of all distributions, it is folklore that no non-trivial condensing is possible.\footnote{Assuming $m\le n$ (wlog this holds since $\abs{\Cond(\zo^n)}\le 2^n$), $m - k_{out}\ge (n - k_{in}) - \log(1 / (1 - \eps))$ and hence the output entropy rate cannot be more than the input entropy rate without incurring extremely large error ($> 0.999$).} So, we additionally assume that $\cX$ is a structured family of sources.\footnote{A different route, that has been widely studied, is to assume access to a short independent seed. In this work, we will limit ourselves to the  \emph{deterministic or seedless setting}.} Since extractors are the highest quality condensers, a significant amount of work has focused on constructing extractors for many interesting families of sources
\cite{tv00, chattopadhyay_explicit_2019, dgw09extractpolynomials, kamp_deterministic_2007}. However, for many natural family of sources, one can provably show that no extractor can exist.

In this work, we focus on one natural family of sources where it is known that extraction is impossible (for many interesting parameter regimes): online non-oblivious symbol fixing sources (\oNOSFs).\footnote{These sources are in contrast to non-oblivious symbol fixing (NOSF) sources where bad blocks can be arbitrary functions of all the good blocks and oblivious symbol fixing sources (OSF) where bad blocks do not depend on any good blocks. NOSF sources were introduced in \cite{kamp_deterministic_2007} and OSF sources were introduced in \cite{chor_bit_1985} with applications in leakage-resilient cryptography, and have been well-studied.} 
Formally:
\begin{definition}
A \emph{\oNOSF[g, \ell, n, k]} $\X = (\X_1, \dots, \X_{\ell})$ with $\ell$ blocks is such that each block $\X_i$ is over $\zo^n$, $g$ of the blocks are independent sources with min-entropy $k$ (``good blocks''), and each ``bad block'' is an arbitrary function of the blocks with an index smaller than it.
When $k = n$, we will call such sources \unioNOSFs[g, \ell, n], and when $n$ is clear from context, we will simply call them \unioNOSFs[g,\ell].
\end{definition}

\paragraph{Our results at a glance} The previous work of \cite{CGR_seedless_condensers} gave a condenser impossibility result for \oNOSFs and showed the existence of a condenser matching that result as long as the block length $n$ was exponential in the number of blocks $\ell$.  We construct explicit condensers for \oNOSFs that match the results of \cite{CGR_seedless_condensers}; in fact, we only require $n$ to be polylogarithmic in $\ell$, providing a doubly exponential improvement over the (existential) result of \cite{CGR_seedless_condensers}.   Next, we essentially resolve the  existence question for \oNOSF condensers by showing that good condensers exist even when $n$ is  a  (large) constant and $\ell$ grows. To go with these results, we obtain an improved construction for transforming low-entropy \oNOSFs into \unioNOSFs. 
Moreover, we find new applications  of our results on condensers to collective coin flipping and collective sampling, and use these connections to provide simple protocols for these problems. In the complementary direction, we construct explicit extractors for \oNOSFs by explicitly constructing coin flipping protocols in our setting; the results of which are summarized in \cref{tab:Non-online results,tab:online results}. Also in the context of extractors for \oNOSFs, we introduce the new, natural notion of \emph{online influence} for Boolean functions and show extraction lower bounds for \oNOSFs by establishing tight bounds on the total online influence of functions.

\subsection*{On the Utility of Condensing for \oNOSFs}
We note that  condensers (and sources with high min-entropy rate) are very useful: the condensed distribution can be used to efficiently simulate randomized algorithms with small overhead, perform one-shot simulations for randomized protocols, cryptography, interactive proofs, and much more.
For details on these applications and more, see \cite{aggarwal_how_2020, doron_almost_2023, CGR_seedless_condensers, DPW14key}.

\paragraph{Practical applications to blockchains and cryptography}
\oNOSFs are inspired by real-time randomness generation settings such as in blockchains where the adversary has some probability of corrupting a block. Moreover, it is known that non-corrupted blocks have some amount of min-entropy \cite{bonneau_bitcoin_2015}. In fact, several works have attempted to use Bitcoin or Ethereum as a source of randomness in cryptographic protocols \cite{bonneau_bitcoin_2015,bentov_bitcoin_2016,pierrot_malleability_2018, bunz_proofs--delay_2017}. However, the authors of \cite{bentov_bitcoin_2016} showed that even when the adversary has a small, constant probability of corrupting a block, randomness extraction is impossible from Bitcoin.\footnote{This mirrors our extraction impossibility result for \oNOSFs in \cref{sec:extracting oNOBF lower bound}} Our results show that in this setting, it is still possible to get a condensed source with a high min-entropy rate. It is known that such sources are still useful for cryptographic protocols, such as hedged public-key enryption \cite{bellare_hedged_2009}. Further, there are natural cryptographic settings, such as creating a Common Reference String, that are widely used in various cryptographic protocols where \oNOSFs naturally arise \cite{aggarwal_how_2020}.

\paragraph{Practical applications to fault-tolerant distributed computing} One common scenario in distributed computing is that of many agents (e.g., servers in a network) attempting to collectively take a decision using several rounds of communication over a common broadcast channel in the presence of computationally unbounded adversarial agents, which render cryptographic primitives ineffective. Protocols for collective coin flipping, leader election, and collective sampling are prime examples of this scenario that have been intensively studied (\cite{ben-or_collective_1989, goldreich_fault-tolerant_1991, dodis2006fault, alonnoar93collective, feige99} and many more), and already connected to the extractor literature \cite{li_extractors_2013, li_new_2013}. In \cref{sec:intro-application coin flipping sampling}, we explain how condensing or extracting from \oNOSFs can be viewed as a variant of these protocols. As a consequence, our new results on condensers provide a new protocol for collective sampling and impossibility results for these protocols can be translated into lower bounds against extractors and condensers for \oNOSFs.

\paragraph{Organization} The remainder of our introduction is structured as follows. We give an overview of previous work in \cref{subsec:intro-previous work} before presenting our main existential and explicit condenser results in \cref{subsec:intro-new condenser constructions}.  In  \cref{subsec:intro-online influence}, we present our results on the limits  of extraction from \oNOSFs. 
Later on, in \cref{sec:intro-application coin flipping sampling}, we explain how our results on condensers have implications for collective coin flipping and sampling protocols.


\subsection{Previous Work}\label{subsec:intro-previous work}

\paragraph{Extractors} The study of extractors for \oNOSFs was initiated by \cite{aggarwal_how_2020}.\footnote{In \cite{aggarwal_how_2020}, these sources were called SHELA (Somewhere Honest Entropic
Look Ahead) sources.} Their results include the following: 
\begin{itemize} 
\item It is impossible to extract from \unioNOSFs when the fraction of good blocks is $0.99$. 
\item An explicit transformation from \oNOSF[g, \ell, n, 0.9n] into a source over $(\zo^{O(n)})^{\ell-1}$ where $g-1$ of the blocks are uniform and independent. 
\item An explicit transformation from \oNOSF[g, \ell, n, 0.1n] into a source over $(\zo^{O(n)})^{100\ell}$ where $g-1$ of the blocks are uniform and independent. 
\end{itemize}
Even though the output entropy rate is only slightly more than the input-entropy rate in the second result and smaller in the third result, the fact that a lot of the blocks are truly uniform is very useful, and they find interesting cryptographic applications of these ``somewhere-extractors''.

Before our work, the best known extractors for \oNOSFs could be obtained by using resilient functions or equivalently, extractors for \NOSFs (non-online version of \oNOSFs) constructed by \cite{ajtai_influence_1993, chattopadhyay_explicit_2019, meka_explicit_2017, IMV23resilient, IV24resilient} ; these require $g \ge \ell - \frac{\ell}{(\log \ell)^2}$.

\paragraph{Condensers} \oNOSFs were further studied by \cite{CGR_seedless_condensers}, where they obtained the following results regarding condensers:
\begin{itemize}
\item When $n\ge k\ge \ell$, there exist functions that can transform a \oNOSF[g, \ell, n, k] into a \unioNOSF[g-1, \ell-1, O(k/\ell)] (this function can be made explicit with slightly worse dependence on output length).
\item When $n\ge 2^{\omega(\ell)}$ and $g > 0.5 \ell$ , there exists condenser $\Cond: (\zo^n)^{\ell}\to \zo^{m = O(n\cdot \ell/g)}$ such that for any \unioNOSF[g, \ell, n] $\X$, $\sminH(\Cond(\X))\ge m - O(\log(n/\eps))$.\footnote{They get a tradeoff for $g\le 0.5\ell$ as well}
Their result is not explicit.
\item It is impossible to condense from $\unioNOSFs[0.5\ell, \ell, n]$ with output entropy rate more than $0.5$.\footnote{They get impossibility for other smaller $g$ as well}
\end{itemize}

We also mention a related family of sources, namely adversarial Chor-Goldreich sources. 
Uniform oNOSF sources can be seen as a special case of adversarial Chor-Goldreich sources where the good blocks are uniform. Constructing condensers where the output entropy rate is $g/ \ell$ for adversarial Chor-Goldreich sources is already a challenging task, although such condensers in various parameter regimes have been recently constructed \cite{doron_almost_2023, GLZ_cg_condenser}. The paper of \cite{doron_online_2025} recently constructed condensers for a related, more general model.

\subsection{This Work: New Condenser Constructions}\label{subsec:intro-new condenser constructions}

Previous works only showed the existence of condensers for \oNOSFs when $n \ge 2^{\omega(\ell)}$. We vastly improve on this result in two ways. First, we construct explicit condensers that work even when $n\geq\polylog(\ell)$ and provide an explicit transformation from low-entropy \oNOSFs to \unioNOSFs that works even when the min-entropy of a block $k$ is only $\polylog(n)$. Second, we show that condensers for \oNOSFs exist when $n$ is just a large constant, only leaving open the question of the existence of such condensers for when $n$ is a very small constant (e.g., $n=1$). We also discover surprising connections between condensers for \oNOSFs and protocols for natural problems in distributed computing, such as collective coin flipping and collective sampling. Lastly, we initiate the study of \emph{online influence} of Boolean functions, a natural generalization of influence that captures the one-sided nature of our online adversary to help us analyze the setting of $n=1$.  We now discuss our results in detail below.



\subsubsection{Explicit Condensers}

We construct the first explicit condensers for \oNOSFs. Our ultimate result is founded on a baseline construction that itself is an explicit condenser for \oNOSFs that matches the existential results of \cite{CGR_seedless_condensers} and works for any block length $n=2^{\Omega(\ell)}\log(1/\varepsilon)$ as long as at least 51\% of blocks are good.

\begin{thm}[\cref{thm overview:condensing 51 with exp} restated]\label{thm:intro-baseline explicit condenser}
    For all $\varepsilon>0$ and $n,\ell\in\N$ where $n\geq 2^{\Omega(\ell)}\log(1/\varepsilon)$, there exists an explicit condenser $\Cond:(\zo^n)^\ell\to\zo^m$ such that for any $(g=0.51\ell,\ell,n)$-oNOSF source $\X$, we have that $\sminH(\Cond(\X))\geq m-2^{O(\ell)}\log(1/\varepsilon)$ where $m=0.0001 \ell n$.
\end{thm}

Surprisingly, we are able to improve upon this baseline to obtain explicit condensers that work for \oNOSFs where the block length $n$ is only at least $\poly(\log(\ell)/\varepsilon)$.

\begin{thm}[Informal version of \cref{thm overview: 51 percent good n polylog ell}]\label{thm: intro explicitly condense uni oNOSF}
For all $\varepsilon>0$ and $n,\ell\in\N$ where $n\geq\poly(\log(\ell)/\varepsilon)$, there exists an explicit condenser $\Cond:(\zo^n)^\ell\to\zo^m$ such that for any $(g=0.51\ell,\ell,n)$-oNOSF source $\X$, we have that $\sminH(\Cond(\X))\geq m-\poly(\log(\ell)/\varepsilon)\cdot\log(n)$ where $m=0.001\ell n-O(\ell\log(\ell)\log(1/\varepsilon))$.
\end{thm}

Since condensing when $g = 0.5\ell$ is impossible, both results are tight. We note that neither result completely subsumes the other. Our baseline construction in \cref{thm:intro-baseline explicit condenser} has an exponential dependence of $n$ on $\ell$ instead of the polylogarithmic dependence achieved in \cref{thm: intro explicitly condense uni oNOSF}; however, the latter result requires the dependence $n\geq\poly(1/\varepsilon)$ compared to  $n\geq\log(1/\varepsilon)$ for the baseline construction.

Using our new results regarding transforming \oNOSFs to \unioNOSFs, we also obtain explicit condensers for \oNOSFs[0.51\ell, \ell, n, k] for the same parameter regime:
\begin{corollary}[\cref{cor:low entropy oNOSF explicit condenser}, simplified]\label{corr: intro explicit condenser oNOSF}
For all $n,\ell,k\in\N$ where $n\geq\poly(\ell)$ and $k\geq\polylog(n)$, there exists an explicit condenser $\Cond:(\zo^n)^\ell\to\zo^m$ such that for any $(g=0.51\ell,\ell,n,k)$-oNOSF source $\X$, we have that $\sminH(\Cond(\X))\geq m-\polylog(\ell)\cdot\log(n)$ where $m=0.001\ell n-O(\ell\log(\ell)\log\log(\ell))$ and $\varepsilon=\poly(1/\log(\ell))$.
\end{corollary}

We can also extend our result to explicitly condense from \unioNOSFs[g, \ell, n] in the same parameter regime so that the output entropy rate is $1 / \floor{\ell/g} - o(1)$, which is tight according to the impossibility result of \cite{CGR_seedless_condensers}.

Previously, \cite{CGR_seedless_condensers} showed how to existentially condense from \unioNOSFs[g, \ell, n] when $n=2^{\Omega(\ell)}$. However, they relied on the existence of a very strong pseudorandom object: ``output-light'' low-error two-source extractors. Such extractors, even without the output-lightness requirement, are extremely hard to construct and it is a major open problem to obtain such extractors.
We are able to construct explicit condensers by creating new tools that allow us to use an \oNOSF to sample indices within an \oNOSF, and stitching them together so that the base pseudorandom object we rely on are seeded extractors that we know how to explicitly construct with near optimal parameters.

\subsubsection{Transforming Low-Entropy \oNOSFs to \unioNOSFs}

We show how to existentially, as well as explicitly, with a slight loss in parameters, transform \oNOSFs[g, \ell, n, k] into \unioNOSFs[0.99g, \ell-1, n]. Formally, we show:

\begin{theorem}[Informal version of \cref{lem:existential low-entropy transformation}]\label{thm:intro-existential low-entropy transformation}
For all $\ell, n, k, \eps$ where $n = \poly(\log(\ell)), k = O(\log(\ell/\eps))$, there exists a function $f$ such that $f$ transforms \oNOSFs[0.51\ell, \ell, n, k] into \unioNOSFs[0.509\ell, \ell, m] with error $\eps$ where $m = \Omega(k)$.
\end{theorem}

Our construction can also be made explicit with slightly worse dependence on $m$ and $\eps$. See \cref{lem:explicit low-entropy transformation} for the full tradeoff.

Previously, \cite{CGR_seedless_condensers} provided such a transformation only for $n\ge k\ge \Omega(\ell)$. Hence, our transformation makes a major improvement on their parameters. Such an improvement allows us to obtain better condensers for low-entropy \oNOSFs in the regime $n = \poly(\log(\ell/\eps))$ (see \cref{theorem: intro low entropy oNOSF condenser}).

\subsubsection{Existential Condensers}

We show how to condense from \unioNOSFs[g, \ell, n] for almost all settings of $\ell$ and $n$ when $g \ge 0.51\ell$. In particular, we show:

\begin{thm}[Informal version of \cref{theorem: simplified large n existential uni onosf condenser}]\label{theorem: intro constant n existential uni onosf}
For all $\ell, \eps$ where $\ell\ge O(\log(1/\eps))$, and $n = 10^4$, there exists a condenser $\Cond: (\zo^n)^{\ell} \to \zo^m$ such that for any \unioNOSF[0.51\ell, \ell, n] $\X$, we have $\sminH(\Cond(\X)) \ge 0.99m$ where $m = \Omega(\ell + \log(1/\eps))$.
Furthermore, when $n = \omega(1)$, the output entropy rate becomes $1 - o(1)$.
\end{thm}

This is tight since \cite{CGR_seedless_condensers} showed it is impossible to condense \unioNOSFs[0.5\ell, \ell, n] beyond output entropy rate $0.5$.

Using our new results regarding transforming \oNOSFs to \unioNOSFs, we also obtain condensers for \oNOSFs[0.51\ell, \ell, n, k] when $n \ge \poly(\log(\ell))$, 
\begin{thm}\label{theorem: intro low entropy oNOSF condenser}
For all $\ell, n, \eps$ where $n = \poly(\log(\ell/\eps)), k = \Omega(\log(\ell/\eps))$, there exists a condenser $\Cond:(\zo^n)^\ell\to\zo^m$ such that for any \oNOSF[0.51\ell, \ell, n, k] $\X$, we have $\sminH(\Cond(\X)) \ge m - O(m / \log (m)) - O(\log(1/\eps))$ where $m = \Omega(k)$.
\end{thm}

We sketch the proof of both of these theorems in \cref{subsec:overview-better-explicit}
We can also extend our result to condense from \unioNOSFs[g, \ell, n] for all $g, \ell$ and constant $n$ where the output entropy rate is $1/\floor{\ell/g} - 0.001$. This is tight since \cite{CGR_seedless_condensers} showed it is impossible to condense such sources beyond output entropy rate $1/\floor{\ell/g}$.

Previously, \cite{CGR_seedless_condensers} showed how to existentially condense from \unioNOSFs[g, \ell, n] when $g\ge 0.51\ell$, provided $n \ge 2^{\omega(\ell)}$. As $n$ gets smaller, condensing becomes harder since a \unioNOSF[g, \ell, n] is also a \unioNOSF[g\cdot n/1000, \ell\cdot n/1000, 1000]. Hence, we greatly improve the parameters while using different and much simpler techniques.

\subsection{\texorpdfstring{Extraction from \oNOSFS}{Extraction from oNOSF Sources}}\label{subsec:intro-online influence}

 Next we discuss our positive and negative results on the limits of extraction from \oNOSFs. Our upper bound results (explicit extractors) are based on a novel connection to leader election and coin-flipping protocols; to instantiate this connection and give explicit extractors, we construct novel protocols for these distributed problems. Our lower bounds are based on a new notion of influence of functions, namely \emph{online influence}, that we introduce and analyze.

\subsubsection*{Extraction Lower bounds via Online Influence}
 For simplicity, we focus on the case of $n = 1$, which leads to interesting new questions about Boolean functions. We refer to such \unioNOSFs[g, \ell, 1] as 
\oNOBFs[g, \ell]; \textrm{oNOBF} stands for online non-oblivious bit-fixing sources.
We ask what is the exact tradeoff between $g$, $\ell$, and $\eps$ for extracting   from \oNOBFs.
Towards this, we introduce the notion of online influence. 

\begin{definition}[Online influence]\label{intro: oI definition}
For a function $f:\bits^{\ell}\to\bits$, the \emph{online influence} of the $i$-th bit is
\begin{align*}
    \oI_i[f]= \E_{x\sim\U_{i-1}}\left[\abs{\E_{y\sim\U_{\ell-i}}[f(x,1,y)]-\E_{y\sim\U_{\ell-i}}[f(x,0,y)]}\right]
\end{align*}
and the \emph{total online influence} is $\oI[f]  =\sum_{i=1}^{\ell}\oI_i[f].$ 
 \end{definition}

We establish new structural results on online influence, including a Poincar\'e-style inequality and use them to obtain the following extraction lower bound.
\begin{theorem}[Informal version of \cref{cor:impossible_nobf_ext}]
For $\eps < 0.01$, there do not exist extractors for \oNOBFs[0.97 \ell,\ell] with error at most $\eps$.
\end{theorem}

A similar extraction lower bound was shown in \cite{aggarwal_how_2020}  using different techniques.

\subsubsection*{Explicit Extractors via Leader Election Protocols}
Here we present our explicit constructions of extractors for oNOBF and oNOSF sources. The following are our main results. 

 \begin{thm}[Informal version of \cref{thm:oNOBF-extractor}] \label{thm:intro main nobf}
There exists an explicit function $\Ext: \zo^{\ell}\to \zo$ such that for any \oNOBF[g, \ell] $\X$ where $g\ge \ell -   \ell / (C \log(\ell))$, we have $\Ext(\X) \approx_{\eps=1/100} \U_1$, where $C$ is a large constant.
\end{thm}

\begin{thm}[Informal version of \cref{thm:oNOSF-extractor}] \label{thm:intro main nosf}
There exists an explicit function $\Ext: (\zo^n)^{\ell}\to \zo^n$ such that for any \oNOSF[g, \ell, n] $\X$ where $g\ge \ell - \ell / (C\log^*(\ell))$ and $n \ge \log(\ell)$, we have $\Ext(\X) \approx_{\eps=1/100} \U_n$, where $C$ is a large constant.
\end{thm}

It is instructive to contrast our results with the non-online setting (where adversarial bits may depend on any good bit), called \NOSFs and \NOBFs. For both these sources, the current best extractors require $g\ge \ell - \frac{\ell}{(\log \ell)^2}$, which is much more than what \cref{thm:intro main nobf} and \cref{thm:intro main nosf} require.

We contrast the results for both settings in \cref{tab:Non-online results,tab:online results}.
In these tables, we are providing known upper and lower bounds on the value of $b(\ell)$, defined as the maximum number of bad symbols for which extraction is still possible with a small constant error --- so lower bounds correspond to best known constructions of such functions and upper bounds refers to the best known limitation of such functions. We write ``$O(\ell)$'' to mean ``$c\ell$ for some small universal constant $c<1$''.

To interpret our results in terms of (online) influence of coalitions, it will be useful  to extend the definition of online influence to subsets of coordinates, which we do formally in \cref{def: Online influence of coalitions}. Intuitively, we're measuring the influence of the exact same adversary as in an \oNOSF. 

In \cref{sec:oNOBF condensing and extraction lower bounds},  we note that online-resilient functions are equivalent to extractors for uniform \oNOSF sources (with one bit output).   
Thus, our explicit extractor results immediately imply explicit online-resilient functions. 


\begin{table}[H]
    \centering
    \renewcommand{\arraystretch}{1.3} 
    \begin{tabular}{c|p{6cm}p{5cm}}
        \toprule
        Source & Lower bound & Upper bound \\
        \midrule
        NOBF & $\Omega\left(\frac{\ell}{\log^2\ell}\right)$, \cite{ajtai_influence_1993} 
             & $O\left(\frac{\ell}{\log\ell}\right)$, \cite{kahn_influence_1988} \\
        NOSF & $\Omega\left(\frac{\ell}{\log^2\ell}\right)$, \cite{ajtai_influence_1993} 
             & $O(\ell)$, \cite{bourgain_influence_1992} \\
        \bottomrule
    \end{tabular}
    \caption{$b(\ell)$ bounds in the non-online setting.}
    \label{tab:Non-online results}
\end{table}

\begin{table}[H]
    \centering
    \renewcommand{\arraystretch}{1.3} 
    \begin{tabular}{c|p{6cm}p{5cm}}
        \toprule
        Source & Lower bound & Upper bound \\
        \midrule
        oNOBF & $\Omega\left(\frac{\ell}{\log\ell}\right)$, [\cref{thm:oNOBF-extractor}] 
              & $O(\ell)$, \cref{cor:impossible_nobf_ext} or \cite{aggarwal_how_2020} \\
        oNOSF & $\Omega\left(\frac{\ell}{\log^*\ell}\right)$, [\cref{thm:oNOSF-extractor}]\footnotemark 
              & $O(\ell)$, \cite{aggarwal_how_2020} \\
        \bottomrule
    \end{tabular}
    \caption{$b(\ell)$ bounds in the online setting.}
    \label{tab:online results}
\end{table}

\footnotetext{Recall that this lower bound is for \oNOSFs[g,\ell,n] with $n\geq\log(\ell)$.}

Our main technique for \Cref{thm:intro main nobf,thm:intro main nosf} is a new generic way to transform leader election and coin flipping protocols (formally defined in \cref{subsec: prelim leader election})  into extractors for oNOBF and oNOSF sources.  This is given in \cref{lemma: leader to extractor}; the general idea of constructing an extractor is to simulate an appropriate leader election protocol with the source at hand (oNOBF or oNOSF), and output according to the chosen leader.\footnote{To clarify why we do not simply output the protocol transcript, see \Cref{rmk:cant just output protocol transcript}.} To instantiate this transformation, we revisit previous leader election   protocols in \cref{sec:new leader election protocols}.   Our leader election protocols provide a slightly stronger than usual guarantee: a good player is elected as the leader with probability close to $1$ (see \cref{lemma: leader 1 bit per round} and \cref{lemma: leader log ell bits per round}). This contrasts with the usual guarantee in leader election protocols, where a good leader is chosen with only a non-trivial (constant) probability.
We give more connections to distributed computing in \cref{sec:intro-application coin flipping sampling} where we delineate applications of our results to collection coin flipping and collective sampling.

\section{Proof Overview}
Our proof overview begins by outlining our new explicit result for condensers in \cref{subsec:overview-better-explicit} that is able to handle polylogarithmic block length. Next, we present our transformation of low-entropy \oNOSFs to \unioNOSFs in \cref{subsec:overview-conversion} before discussing our existential results for condensers that can handle constant block length in \cref{subsec:overview-existence}. We present   the main ideas behind our results regarding online influence and extractor lower bounds in \cref{subsec:overview-fourier}. In \cref{subsec:overview-explicit-extractors}, we overview our extractor constructions for oNOBF and oNOSF sources that are based on a general transformation from leader election protocols. 


\subsection{Explicit Condensers for Uniform oNOSF Sources}\label{subsec:overview-better-explicit}

We  sketch here  our constructions (as well as proof ideas) of explicit condensers for \unioNOSFs[0.51\ell, \ell, n].\footnote{Since all sources are uniform here, we will not explicitly mention this again.} The goal will be to construct explicit condensers that work with as few good sources as possible while minimizing the block length $n$. In particular, we will show:
\begin{theorem}[\cref{thm: 51 percent good n polylog ell}, simplified]\label{thm overview: 51 percent good n polylog ell}
For all $\eps>0$ and $n,\ell\in\N$ where $n \ge \left(\frac{\log(\ell)}{\varepsilon}\right)^{\Omega(1)}$, there exists an explicit condenser $\Cond:(\zo^n)^\ell\to\zo^m$ such that for any \oNOSF[g=0.51\ell,\ell,n] $\X$, we have that $\sminH(\Cond(\X))\geq m - \left(\frac{\log(\ell)}{\varepsilon}\right)^{O(1)} \cdot \log(n)$ where $m = 0.001 \cdot \ell n - O\left(\ell\log(\ell)\log(1/\eps)\right)$.
\end{theorem}
We will require two main tools to show this. The first one  uses ideas from the leader election literature and allows us to sample an $O(\log(\ell))$ sized committee starting from $\ell$ players while essentially maintaining the fraction of bad players. Formally:
\begin{lemma}[\cref{lem:Sampler from reduce and seeded extractor}, simplified]
\label{lem overview:Sampler from reduce and seeded extractor}
For all $\varepsilon_s>0$, $n, \ell\in \N$, and constant $\varepsilon_a>0$ where $n\geq \Omega(\log(\ell)\log(1/\varepsilon_s))$, there exists an explicit function $\NOSFSamp: (\zo^n)^{\ell}\to [\ell]^D$ where $D \le O(\log(\ell/\varepsilon_s))$ with the following property.\footnote{Even though the output domain of $\NOSFSamp$ is a vector, we will abuse notation and often treat it as a set.}
For all $S\subset [\ell]$ and \oNOSFs[g, \ell, n] $\X$,  we have that 
\[
    \Pr_{x\sim \X}\left[\abs{\frac{\abs{\NOSFSamp(x)\cap S}}{D}-\frac{\abs{S}}{\ell}}\geq\varepsilon_a\right]\leq \varepsilon_s
\]
\end{lemma}
At a high level, to construct $\NOSFSamp$, we slightly modify the committee selection procedure from \cite{russellzuckerman01} and instantiate it with a seeded extractor with near optimal dependence on the seed \cite{zuckerman_linear_2007}.

Our second tool is a seeded condenser for general min-entropy sources $\X$ that uses an \oNOSF $\Y$ as the seed, where the bad bits in $\Y$ can depend on $\X$.
\begin{lemma}[\cref{lem:condenser from funky two source extractor}, simplified]
\label{lem overview:condenser from funky two source extractor}
There exists a constant $C_{\TCond}$ such that for all $n_x, k, n_y, t\in\N$ with $\varepsilon>0$ and $n_y\geq (C_{\TCond})^t\log(tn_x/\varepsilon)$, there exists an explicit condenser $\TCond:\zo^{n_x}\times(\zo^{n_y})^t\to\zo^m$ where $m=\frac{1}{3}(k-(C_{\TCond})^t\log(tn_x/\varepsilon))$ so that the following holds: For all $(n_x,k)$-sources $\X$ and \oNOSFs[g=1, \ell=t] $\Y\sim (\zo^{n_y})^t$ such that the good blocks in $\Y$ are independent of $\X$ and the bad blocks in $\Y$ can depend on $\X$, we have that $\sminH(\TCond(\X,\Y))\geq m-(C_{\TCond})^t\log(tn_x/\varepsilon)$.
\end{lemma}
We will sketch how to construct this in \cref{subsubsec overview:condenser from funky two source extractor}.
Using these tools, we are ready to present our explicit condenser. In fact, we will  provide sketches of five different constructions of explicit condensers with increasingly better parameter dependence; the fifth (and final) one is the construction given by \cref{thm overview: 51 percent good n polylog ell}. The parameters they will vary in are the fraction of good blocks (i.e., $g/\ell$) and the block length $n$. As we will see,  each construction will build on  ideas from the previous construction. 

\subsubsection{Construction 1: 51\% good and $n \ge 2^{\Omega(\ell)}$}
We here construct the following condenser: 
\begin{theorem}[\cref{thm:condensing 51 with exp}, simplified]
\label{thm overview:condensing 51 with exp}
For all $\eps>0$ and $n,\ell\in\N$ where $n \ge 2^{\Omega(\ell)} \log(1 / \eps)$, there exists an explicit condenser $\Cond:(\zo^n)^\ell\to\zo^m$ such that for any \oNOSF[g=0.51\ell,\ell,n] $\X$, we have that $\sminH(\Cond(\X))\geq m - 2^{O(\ell)}\log(1 / \eps)$ where $m = 0.0001\cdot n \ell$.
\end{theorem}

\begin{proof}[Proof sketch]
Let $\gamma = 0.01$ and $\ell' = \ell / 2$.
We decompose the input $\X$ into two equal sized parts so that $\X = (\X_1, \X_2)$ where both $\X_1, \X_2\sim (\zo^n)^{\ell/2} \equiv (\zo^n)^{\ell'}$.
Since $\X$ has $(0.5 + \gamma)\ell$ good players, we infer that each of $\X_1, \X_2$ has at least $\gamma \ell = (2\gamma)\cdot \ell'$ good players.
In particular, we use the fact that $\minH(\X_1) \ge (2\gamma)\cdot (\ell n/2)$ and that $\X_2$ is a \oNOSF[g = (2\gamma) \ell/2, \ell/2, n].
With this, we let $\TCond$ be the condenser from our second tool \cref{lem overview:condenser from funky two source extractor} and let our final output be $\TCond(\X_1, \X_2)$.
Note that here $t = \ell$ and so, this requires $n \ge (C_{\TCond})^{\ell} \log(\ell n / \eps)$, an inequality that we indeed satisfy. The guarantees from \cref{lem overview:condenser from funky two source extractor} provide us with the desired claim.
\end{proof}

Each of the subsequent constructions will use a similar construction idea as above.
However, they will try to decrease $t$ as much as possible, where $t$ is the number of players in the \oNOSF when applying $\TCond$ from \cref{lem overview:condenser from funky two source extractor}. Note that any reduction results in a decrease in the block length requirement $n$.

\subsubsection{Construction 2: 67\% good and $n \ge \poly(\ell)$}
Our next construction requires a slightly larger fraction of good blocks. However, the block length required is exponentially improved.
\begin{theorem}[\cref{thm:condensing 67 with poly}, simplified]
\label{thm overview:condensing 67 with poly}
For all $\eps>0$ and $n,\ell\in\N$ where $n \ge \left(\frac{\ell}{\varepsilon}\right)^{\Omega(1)}$, there exists an explicit condenser $\Cond:(\zo^n)^\ell\to\zo^m$ such that for any \oNOSF[g = 0.67\ell,\ell,n] $\X$, we have that $\sminH(\Cond(\X))\geq m - \left(\frac{\ell}{\varepsilon}\right)^{O(1)}\log(n)$ where $m = 0.001\cdot \ell n$.
\end{theorem}


\begin{proof}[Proof sketch]
Let $\gamma = 0.67 - (2 /3)$.    
Let $\ell' = \ell / 3$.
We decompose the input $\X$ into three equal sized parts so that $\X = (\X_1, \X_2, \X_3)$ where all $\X_1, \X_2, \X_3\sim (\zo^n)^{\ell'}$.
Since $\X$ has $((2/3) + \gamma)\ell$ good players, we easily see that each $\X_i$ is a \oNOSF[g = 3\gamma \ell', \ell', n].

We use $\X_1$ to sample a $O(\log(\ell))$ sized committee from $\X_3$.
To do so, we use $\NOSFSamp_{1\to3}: (\zo^n)^{\ell'}\to (\zo^{\ell'})^D$ from \cref{lem overview:Sampler from reduce and seeded extractor} with $S\subset [\ell']$ being the set of good players from $\X_3$, the approximation factor $\eps_a = \gamma$ and sampling error $\eps_s = \eps/3$.
Let $\mC_3\subset [\ell'], \abs{\mC_3}\le D_3 = O(\log(\ell' / \eps)) = O(\log(\ell / \eps))$ be the committee of players thus obtained.
The approximation property of the sampler guarantees us that out of $\ge 3\gamma \ell'$ good players in $\X_3$, at least $2\gamma \abs{\mC_3}$ many good players will be in $\mC_3$ with probability $1 - \eps/3$.
Let $\Y_3$ be the \oNOSF[2\gamma D_3, D_3, n] obtained by restricting the players in $\X_3$ to the committee $\mC_3$.
We finally use $\TCond$ from \cref{lem overview:condenser from funky two source extractor} and output $\TCond(\X_2, \Y_3)$.
Here, the parameter $t$ in \cref{lem overview:condenser from funky two source extractor} will be set to $D\le O(\log(\ell/\eps))$, and hence, \cref{lem overview:condenser from funky two source extractor} would only require that $n \ge (C_{2_{\TCond}})^t\log(\ell n /\eps) = \left(\frac{\ell}{\eps}\right)^{O(1)}$, a condition that we do meet. We carefully compute the remaining parameters to infer the claim.
\end{proof}

\subsubsection{Construction 3: 76\% good and $n \ge \poly(\log(\ell))$}
We build on our previous construction and show how to condense when the block length requirement is again exponentially decreased. This comes at a cost of slightly larger fraction of good blocks.
\begin{theorem}[\cref{thm: condensing 76 with polylog}, simplified]
\label{thm overview: condensing 76 with polylog}
For all $\eps>0$ and $n,\ell\in\N$ where $n \ge \left(\frac{\log(\ell)}{\varepsilon}\right)^{\Omega(1)}$, there exists an explicit condenser $\Cond:(\zo^n)^\ell\to\zo^m$ such that for any \oNOSF[g = 0.76\ell,\ell,n] $\X$, we have that $\sminH(\Cond(\X))\geq m - \left(\frac{\log(\ell)}{\varepsilon}\right)^{O(1)}\log(n)$ where $m = 0.001\cdot \ell n$.
\end{theorem}


\begin{proof}[Proof sketch]
Let $\gamma = 0.01$.    
Let $\ell' = \ell / 4$.
We decompose the input $\X$ into four equal sized parts such that $\X = (\X_1, \X_2, \X_3, \X_4)$ and conclude that each $\X_i$ is a \oNOSF[g = 4\gamma \ell', \ell', n].
However here, we claim something stronger.
Call $i\in [\ell']$ a totally good index if it corresponds to a good player across each of the four blocks.
Since $\X$ has $(3/4+\gamma)\cdot (4\ell')$ good players, there must be $\ge 4\gamma \ell'$ totally good indices.

We first use $\X_1$ to sample a $D_2 = O(\log(\ell / \eps))$ sized committee $\mC_2\subset [\ell']$ using the sampler from \cref{lem overview:Sampler from reduce and seeded extractor} such that $\mC_2$ will have at least $3\gamma$ fraction of totally good indices.
We let $\Y_2$ be the \oNOSF[g=3\gamma D_2, D_2, n] obtained by restricting $\X_2$ to indices from $\mC_2$.

Second, we use $\Y_2$ to sample a $D_4 = O(\log(\log(\ell) / \eps))$ sized committee $\mC_4 \subset \mC_2$ such that $\mC_4$ has at least $2\gamma$ fraction of totally good indices.
We let $\Y_4$ be the \oNOSF[g=2\gamma D_4, D_4, n] obtained by restricting $\X_4$ to indices from $\mC_4$.

Third and last, we use $\TCond$ from \cref{lem overview:condenser from funky two source extractor} and output $\TCond(\X_3, \Y_4)$.
Here, the parameter $t$ in \cref{lem overview:condenser from funky two source extractor} will be set to $D_4\le O(\log(\log(\ell)/\eps))$, and hence, \cref{lem overview:condenser from funky two source extractor} would only require that $n \ge (C_{2_{\TCond}})^t\log(\ell n /\eps) = \left(\frac{\log(\ell)}{\eps}\right)^{O(1)}$, a condition that we do meet. We carefully compute the remaining parameters to infer the claim.
\end{proof}

\subsubsection{Construction 4: 67\% good and $n \ge \poly(\log(\ell))$}
Our next construction maintains a similar guarantee as before on the block length and decreases the requirement on the fraction of good blocks.
\begin{theorem}[\cref{thm: condensing 67 with polylog}, simplified]
\label{thm overview:condensing 67 with polylog}
For all $\eps>0$ and $n,\ell\in\N$ where $n \ge \left(\frac{\log(\ell)}{\varepsilon}\right)^{\Omega(1)}$, there exists an explicit condenser $\Cond:(\zo^n)^\ell\to\zo^m$ such that for any \oNOSF[g = 0.67\ell,\ell,n] $\X$, we have that $\sminH(\Cond(\X))\geq m - \left(\frac{\log(\ell)}{\varepsilon}\right)^{O(1)}\log(n)$ where $m = 0.001\cdot \ell n$.
\end{theorem}


\begin{proof}[Proof sketch]
Let $\gamma = 0.67 - (2/3)$.    
Let $\ell' = \ell / 3$.
We decompose $\X = (\X_1, \X_2, \X_3)$ so that each $\X_i$ is a \oNOSF[g = 3\gamma \ell', \ell', n] with $3\gamma \ell'$ totally good indices.

For the first step, we again use $\X_1$ along with a sampler from \cref{lem overview:Sampler from reduce and seeded extractor} to obtain $\mC_2 \subset [\ell']$ with $D_2 = \abs{\mC_2}\le O(\log(\ell / \eps)$ and a subsource of $\X_2$ restricted to $\mC_2$ - $Y_2$ that is a \oNOSF[g=2\gamma D_2, D_2, n].
For the second step, we again use $\Y_2$ to sample a $D_3 = O(\log(\log(\ell) / \eps))$ sized committee $\mC_3 \subset \mC_2$ such that $\mC_3$ has $\ge \gamma$ fraction of totally good indices. We let $\Y_3$ be the \oNOSF[g=\gamma D_3, D_3, n] obtained by restricting $\X_3$ to indices from $\mC_3$.
Third and last, we use $\TCond$ from \cref{lem overview:condenser from funky two source extractor} and output $\TCond(\X_2, \Y_3)$.
Again the parameter $t$ in \cref{lem overview:condenser from funky two source extractor} will be set to $D_3\le O(\log(\log(\ell)/\eps))$ and would only require that $n \ge  \left(\frac{\log(\ell)}{\eps}\right)^{O(1)}$.

Analyzing this construction requires more care since we use $\X_2$ to both sample from $\X_3$ and as a source for $\TCond$.
We use the chain rule for min-entropy (\cref{lem:min-entropy-chain-rule}) to argue that most fixings of $\Y_2 = y_2$ will leave $\X_2$ with lot of entropy (since sampler requires few random bits) and observe that such a fixing still leaves $\X_3$ as \oNOSF with the same parameters.
Also since for most fixings of $\Y_2 = y_2$, the committee $\mC_3$ has $\gamma$ fraction of good players, we obtain our claim.
\end{proof}

\subsubsection{Construction 5: 51\% good and $n \ge \poly(\log(\ell))$}
We lastly construct the condenser promised in our main result (\cref{thm overview: 51 percent good n polylog ell}).


\begin{proof}[Proof sketch of \cref{thm overview: 51 percent good n polylog ell}]
Let $\gamma = 0.01$.
Let $\ell' = \ell / 2$.
We decompose $\X = (\X_1, \X_2)$ so that each $\X_i$ is a \oNOSF[g = 2\gamma \ell', \ell', n] with $2\gamma \ell'$ totally good indices.

For the first step, we let $\X_1'$ be the \oNOSF[2\gamma \ell', \ell', n_1'] obtained by taking a prefix of length $n_1'$ from each source where we set $n_1' \ll n$ but also $n_1'$ is long enough to be used by the sampler from \cref{lem overview:Sampler from reduce and seeded extractor}. We use $\X_1'$ with such a sampler to obtain a committee $\mC_{1\to2} \subset [\ell']$ with $D_{1\to2} = \abs{\mC_{1\to2}}\le O(\log(\ell / \eps)$ and a subsource of $\X_2$ restricted to $\mC_{1\to 2}$, which we call $\Y_{1\to 2}$, that is a \oNOSF[g=(3/2)\gamma D_{1\to 2}, D_{1\to 2}, n].

We argue that: 1) most fixings of $\X_1'$ are such that they leave $\X_1$ with high entropy, and 2) that the committee $\mC_{1\to 2}$ obtained will have $3\gamma / 2$ fraction of good players. We obtain this using the chain rule for min-entropy and by guarantees of the sampler. We condition on such a fixing from here on.

For the second step, we use $\Y_{1\to 2}$ to sample a $D_{2\to 2} = O(\log(\log(\ell) / \eps))$ sized committee $\mC_{2\to 2} \subset \mC_{1\to 2}$ such that $\mC_{2\to 2}$ has $\ge \gamma$ fraction of totally good indices. We let $\Y_{2\to 2}$ be the \oNOSF[g=\gamma D_{2\to 2}, D_{2\to 2}, n] obtained by restricting $\X_2$ to indices from $\mC_{2\to 2}$.

Lastly, we use $\TCond$ from \cref{lem overview:condenser from funky two source extractor} and output $\TCond(\X_1, \Y_{2\to 2})$.
The parameter $t$ in \cref{lem overview:condenser from funky two source extractor} will be set so that it would only require that $n \ge  \left(\frac{\log(\ell)}{\eps}\right)^{O(1)}$. We must be a bit careful about the error parameter $\eps_{\TCond}$ for $\TCond$ and we will choose it to be extremely small. 

Analyzing this construction requires care since we use $\X_2$ both to sample from within $\X_2$ itself and also as a seed for $\TCond$.
We cannot use the chain rule since a fixing of $\Y_{1\to 2}$ will destroy the structure of the source $\X_2$. 
We first see that the sampler guarantees that no matter how the adversary behaves, with probability $1 - \eps/4$, the sampler will succeed in selecting a committee $\mC_{2\to 2}$ with $\gamma$ fraction of good players.
We pay $\eps/4$ in error and now assume that the sampler always succeeds in doing so.
We then compare two scenarios: one scenario $\Opt$ where the bits in $\Y_{1\to 2}$ are all uniform and independent of all other bits and another scenario $\Adv$ where the bits in $\Y_{1\to 2}$ are all controlled by an adversary (guarantees on this latter scenario suffice for our claim).
Let the total number of bits in $\Y_{1\to 2}$ be equal to $b_{1\to 2}$.
Under scenario $\Opt$, we easily see that we succeed and with error $\eps_{\TCond}$ will have high entropy - say $k$. 
To compare this to scenario $\Adv$, we use \cref{lem:control-few-bits-can-still-condense-again} that lets us conclude that in scenario $\Adv$, with error $\eps_{\TCond}\cdot 2^{b_{1\to 2}}$, the output will have entropy $k - b_{1\to 2}$.
Since we carefully chose $\eps_{\TCond}$ to be small enough and $b_{1\to 2}$ is small since we only use $\Y_{1\to 2}$ as source for the sampler, the output will still have small error and will have high-entropy as desired.
\end{proof}

\subsubsection{Construction of $\TCond$}\label{subsubsec overview:condenser from funky two source extractor}

We now sketch how to construct our desired $\TCond$.

\begin{proof}[Proof sketch of \cref{lem overview:condenser from funky two source extractor}]
Let $\X$ and $\Y = (\Y_1, \dots, \Y_t)$ be the two sources.
For $i\in [t]$, let $n_i \approx C^{t-i+1} \log(t n_x / \eps)$ where $C$ is a large constant.
For $i\in [t]$, let $\ZZ_i$ be the length $n_i$ prefix of the block $\Y_i$.
Our final construction will be the parity of the outputs of seeded extractors applied with source $\X$ and seeds $\ZZ_i$. More formally, we output
\[
    \bigoplus_{i=1}^{t} \sExt_i(\X, \ZZ_i),
\]
where $\sExt_i$ is any explicit near optimal seeded extractor (such as the extractor from \cref{thm:GUV extractor}).

We proceed to sketch the analysis.
We are guaranteed that there exists at least one $j\in [t]$ from $\Y$ that is good.
We first condition on fixing blocks $\ZZ_1, \dots, \ZZ_{j-1}$. 
Since these blocks can depend on $\X$, we apply the chain rule for min-entropy (\cref{lem:min-entropy-chain-rule}) and conclude $\X$ will only lose some small amounts of entropy (the amount will be very small since these blocks are tiny compared to the amount entropy in $\X$). 
Moreover, since the adversary is online, $\ZZ_j$ remains uniform even after this fixing. We now view our construction as
\[
g(\X)\oplus \bigoplus_{i=j}^{t} \sExt_i(\X, \ZZ_i)
\]
where $g$ is the fixed function obtained by fixing $\Y_1, \dots, \Y_{j-1}$.

We now compare two scenarios: (1) Where all of $\ZZ_j, \dots, \ZZ_{t}$ are uniform (2) Only $\ZZ_j$ is uniform and $\ZZ_{j+1}, \dots, \ZZ_{t}$ are arbitrarily controlled by an adversary and can even depend on $\X$:

In the first scenario, we further condition on fixing $\ZZ_{j+1}, \dots, \ZZ_{t}$. Since in this scenario $\ZZ_i$ are independent and random, $\X$ retains the same entropy and $\Y_j$ remains uniform. So our overall output is of the form $h(\X) \oplus \sExt_j(\X, \ZZ_j)$ for some fixed function $h$. We condition on fixing output $h(\X)$. Since the number of output bits $m\ll \minH(\X)$, we apply the chain rule to infer that $\X$ still has lots of entropy when we do this fixing. Now the output is just $z\oplus \sExt_j(\X, \ZZ_j)$ where $z$ is a fixed string, and hence is uniform.

The second scenario is more realistic and, in the worst case, this is what can actually happen. We then use the result that if an adversary controls few bits in the input distribution, then they cannot make the output of the condenser too bad (see \cref{lem:control-few-bits-can-still-condense-again} for the full statement). With this, since we carefully chose geometrically decreasing lengths of $\ZZ_i$ to help control the error, we indeed obtain that the output will be condensed.
\end{proof}

\subsection{Converting Low-Entropy oNOSF Sources to Uniform oNOSF Sources}\label{subsec:overview-conversion}
The key part of our proof for condensing from low-entropy \oNOSFs is a transformation from low-entropy \oNOSFs to \unioNOSFs. Here, we sketch the proof for our transformation in \cref{thm:intro-existential low-entropy transformation} and compare it to that of \cite{CGR_seedless_condensers}. Both these transformations rely on two-source extractors (see \cref{def:two-source extractor} for definition) as a basic primitive.

Given a \oNOSF[g,\ell,n,k] $\X=\X_1,\dots,\X_\ell$, \cite{CGR_seedless_condensers} uses excellent existential two-source extractors (such as from \cref{lem:amazing two-source-extractors exist}) to define output blocks $\O_i=2\Ext(\X_1\circ\cdots\circ\X_{i-1},\X_i)$ for $i\in\{2,\dots,\ell\}$ and define their transformation as $f(\X)=\O_2,\dots,\O_\ell$. They show that $\O_i$ is a good block if: (1) $\X_i$ is a good block and (2) at least one block amongst $\X_1,\dots,\X_{i-1}$ is a good block. They showed that such a good block will be uniform and independent of the blocks $\O_2,\dots,\O_{i-1}$ and argued there will be $g-1$ such good output blocks. This indeed shows their output is a \unioNOSF[g-1, \ell-1, m]. However, each of their output blocks has length $m=O\left(\frac{k}{\ell}\right)\le O\left(\frac{n}{\ell}\right)$, and so they were not able to handle the case of $n=o(\ell)$. We improve on their construction by using a ``sliding window'' based technique to obtain a much better transformation that can even handle $n = \poly(\log(\ell))$.
\begin{theorem}[\cref{lem:existential low-entropy transformation} restated]\label{thm:overview-existential low-entropy transformation}
Let $d, g, g_{out}, \ell, n, m, k, \eps$ be such that 
$ g_{out} \le g - \frac{\ell-g+2}{d},
n\ge k \ge \log(nd-k) + md + 2\log(2g_{out}/\eps)
$.
Then, there exists a function $f: (\zo^n)^{\ell}\to (\zo^m)^{\ell-1}$ such that for any \oNOSF[g, \ell, n, k] $\X$, there exists \unioNOSF[g_{out}, \ell-1, m] $\Y$ for which $\abs{f(\X) - \Y} \le \eps$.
\end{theorem}

The parameter $d$ in our theorem statement above is the width of our sliding window. 
When we set $d=\ell$ we recover the analysis of \cite{CGR_seedless_condensers}.
The true advantage of our transformation emerges when $d$ is very small compared to $\ell$. For instance, when $g = 0.51\ell, n = \poly(\log(\ell))$ and $k = \poly(\log(\ell))$, we set $d$ to be a large constant and conclude that the output distribution is a \uniNOSF[0.509\ell, \ell, \poly(\log(\ell)].

\begin{proof}[Proof sketch of \cref{thm:overview-existential low-entropy transformation}]
Define $\O_i=2\Ext(\X_{i-d}\circ\cdots\circ\X_{i-1},\X_i)$. We call $\O_i$ a good output block when $\X_i$ is good and there is at least one good block amongst $\{\X_{i-d},\dots,\X_{i-1}\}$. 

We first compute the number of good output blocks $g_{out}$. Let $j_1,\dots,j_g$ be the indices of the good input blocks in $\X$ and $d_i=j_{i+1}-j_i$ be the gap between the $i$-th good block and the next ($i+1$)-th good block. If the gap $d_i$ is at most $d$, then $\O_{i+1}$ must be a good output block. So, $g_{out}$ is the number of $i$ such that $d_i\leq d$. Since $g\ge 0.51\ell$, such large gaps cannot appear too often and we compute that $g_{out} \ge g - \frac{\ell-g+2}{d}$ as desired.

Next, we show that the good output blocks are indeed uniform conditioned on all previous output blocks. With this, we will obtain that the output distribution will be \unioNOSF[g_{out}, \ell-1, m] as desired.
Let $i$ be the index of a good output block. We want to show that $\OO_i$ is uniform conditioned on $\OO_1, \dots, \OO_{i-1}$. To do this, we first observe that any input block contributes to at most $d+1$ good output blocks. This means  that$(\X_{i-d}\circ\cdots\circ\X_{i-1})$, which has min-entropy at least $k$, loses at most $d\cdot m$ min-entropy conditioned on fixing $\OO_1, \dots, \OO_{i-1}$. Moreover, $\X_i$ still remains high entropic and independent of $(\X_{i-d}\circ\cdots\circ\X_{i-1})$ when fixing these previous output blocks. Hence, the output of the two-source extractor will indeed be uniform as desired.
\end{proof}

We can make \cref{thm:overview-existential low-entropy transformation} explicit by using the explicit two-source extractors of \cref{thm:explicit two-source-extractors} at a slight cost of dependence on $m$ and $\varepsilon$ as seen in \cref{lem:explicit low-entropy transformation}.

\subsection{Existence of oNOSF Condensers for All \texorpdfstring{$\ell$}{l} and \texorpdfstring{$n$}{n}}\label{subsec:overview-existence}

Here we sketch the proof of \cref{theorem: intro constant n existential uni onosf}. This result states that when $g = 0.51\ell$ and $n = 1000$, there exists a condenser $\Cond$ for \unioNOSFs[g, \ell, n] so that the output entropy rate is $0.99$, the number of output bits is $m = O(\ell + \log(1/\eps))$, and the error of the condenser is $\eps$ where $\eps \le 2^{-\Omega(\ell)}$ is arbitrary.

Our construction uses amazing seeded condensers (see \cref{def:seeded_cond}) with $1\cdot \log(1/\eps)$ dependence on seed length. We slightly modify our source and then apply such seeded condenser. Here is a proof sketch:

\begin{proof}[Proof sketch for \cref{theorem: intro constant n existential uni onosf}]
Let $\X = (\X_1, \dots, \X_{\ell})$ be such a source. Let $\Y_1\sim (\zo^n)^{0.5\ell}$ be the source obtained by concatenating the first $0.5\ell$ blocks of $\X$.
Since $0.51\ell$ blocks are good, there exist at least $0.01\ell$ uniform blocks in $\Y_1$.
We treat $\Y_1$ as a single distribution over $n\ell$ bits with min-entropy $\ge 0.01\ell n$. 
Let $\Y_2\sim \zo^{0.5\ell}$ be the source obtained by concatenating $1$ bit from each of the last $0.5\ell$ blocks of $\X$.
Once again, since $0.51\ell$ blocks are good, there exist at least $0.01\ell$ uniform bits in $\Y_2$.
We will use the following seeded condenser:
\begin{theorem}[\cref{lem:exists great lossless condenser}, simplified]
For all $d, \eps$ such that $d\ge \log(\ell n/\eps) + O(1)$, there exists a seeded condenser $\sCond: \zo^{0.5\ell n} \times \zo^d\to \zo^m$ such that for all $\X\sim \zo^{0.5\ell n}$ with $\minH(\X)\ge 0.01\ell n$, we have $\sminH(\sCond(\X, \U_{d})) \ge 0.01\ell n + d$ where $m = 0.01\ell n + d + \log(1/\eps) + O(1)$.
\end{theorem}

Our condenser $\Cond$ will output $\sCond(\Y_1, \Y_2)$.
Observe that not only is $\Y_2$ not uniform, there could be as many as $0.49\ell$ ``bad bits'' in $\Y_2$ that can depend on $\Y_1$. To remedy this, we use the well known fact that the behavior of such adversarial $\Y_2$ cannot be far worse than the behavior if $\Y_2$ were uniform. Concretely, suppose if $\Y_2$ were uniform and the output entropy and error were $k$ and $\eps$. Then for the actual $\Y_2$, the output entropy will be $k - 0.49\ell$ and error will be $\eps\cdot 2^{0.49\ell}$.
See \cref{lem:control-few-bits-can-still-condense-again} for the formal statement.

For us, it means the following: let $\eps_{\sCond}, k_{\sCond}$ be such that $\minH^{\eps_{\sCond}}(\sCond(\Y_1, \U_{0.5\ell}))\ge k_{\sCond}$.
Then, it must be that $\minH^{2^{0.49\ell}\cdot \eps_{\sCond}}(\sCond(\Y_1, \Y_2))\ge k_{\sCond} - 0.49\ell$.
So, for our final error to be some $\eps$, we need to have $\eps_{\sCond} = \eps\cdot 2^{-0.49\ell}$. For seeded condensers to exist, we need $0.5\ell\ge \log(\ell n/\eps_{\sCond}) + O(1)$ and we check that such an inequality can indeed be satisfied if $\eps\ge 2^{-0.01\ell}$.

Hence, we finally obtain that our seeded condenser will output $0.01\ell n + O(\ell)$ bits and will have output entropy $m - \Delta$ where $\Delta = O(\ell)$. Hence, if $n$ is a large enough constant, our output entropy rate, $\frac{m-\Delta}{m}$,  will be $\ge 0.99$ as desired.
\end{proof}
\begin{remark}
Here (in the inequality $0.5\ell \ge 1\cdot \log(\ell n/\eps_{\sCond})$) we crucially used the fact that there exist seeded condensers with seed length dependence $1\cdot \log(1/\eps)$. Currently, we do not have explicit constructions with this dependence. We also could not have used a seeded extractor since for them, the seed length dependence is $2\cdot \log(1/\eps)$. For that to work, we would need to assume $g\ge 0.76\ell$.
\end{remark}

\subsection{Online Influence and Extractor Lower Bounds}\label{subsec:overview-fourier}
In this subsection, we provide a brief overview of our results regarding online influence and sketch how they imply extractor lower bounds against \oNOBFs. We also contrast this with the established notion of influence for Boolean functions. For any function $f: \zo^n \to \zo$, define the function $e(f)(x) = (-1)^{f(x)}$.

\paragraph{A Poincar\'e inequality and extractor lower bounds} One fundamental inequality about regular influence is  the Poincar\'e inequality  which states that $\Var(f)\leq\I[f]$. We prove a similar result for online influence. 
\begin{theorem}[\cref{thm:oI ineqs} restated]\label{thm:overview oI ineqs}
    For any  $f:\bits^{\ell}\to\bits$, we have $\Var(e(f)) \leq \oI[f]\leq \sqrt{{\ell}\Var(e(f))}$.
\end{theorem}
It is not hard to derive extractor lower bounds for \oNOBFs from the above result. The high level idea is to collect bits with high online influence, which is guaranteed by the first inequality in the above theorem (using an averaging argument) to form a \emph{coalition of coordinates} that has enough online influence to bias the claimed extractor. We refer the reader to \cref{thm:how many bad bits needed to get to beta} for more details.

The proof of \cref{thm:overview oI ineqs} is based on techniques from the Fourier analysis of Boolean functions.\footnote{We give a very brief recap of necessary notions from Fourier analysis of Boolean functions in \cref{sec:Poincare style inequality}.} The following key result implies \cref{thm:overview oI ineqs}. We refer the reader to \cref{sec:extracting oNOBF lower bound} for more details.
\begin{lemma}[\cref{lemma:Fourier bb} restated] 
For any $f: \bits^{\ell} \to \bits$ and $i \in [{\ell}]$, $$\oI_i(f)^2 \le \sum_{\substack{S\subseteq[i]\\S\ni i}}\widehat{f}(S)^2 \le \oI_i(f).$$
\end{lemma}


\paragraph{Influence vs Online Influence} It is not hard to see that  $\oI_i[f]\leq\I_i[f]$ for all $i\in[{\ell}]$,  with equality always holding for $i={\ell}$ as an adversarial online bit in the last index can see every good bit. Moreover, we observe that for monotone functions, the notion of online influence is equivalent to regular influence, so any separation between the two notions must come from non-monotone functions.

We exactly exhibit such a separation via the address function $\Addr_{\ell}:\zo^{\log \ell + \ell}\to\zo$ which considers its first $\log \ell$ bits as an index in $\{1,\dots,\ell\}$ and then outputs the value of the chosen index. It is easy to show (as we do in \cref{lem: address function tight oI}) that the first $\log \ell$ bits of $\Addr_{\ell}$ have no online influence, while the remaining bits have online influence of $O\left(\frac{1}{\ell}\right)$. This is in contrast to the well known result of \cite{kahn_influence_1988} showing that, for a balanced function such as $\Addr_{\ell}$, there must exist a bit with influence at least $\Omega\left(\frac{\log \ell}{\ell}\right)$.

\subsection{Extractors via   Leader Election Protocols}\label{subsec:overview-explicit-extractors}
We sketch our main idea for constructing an extractor for oNOBF sources (\cref{thm:oNOBF-extractor}). Similar ideas work more generally for extracting from oNOSF sources (\cref{thm:oNOSF-extractor}). As mentioned above, we use a novel connection to leader election protocols to construct extractors. We refer the reader to \cref{subsec: prelim leader election} for a quick recap of the leader election protocols.

Suppose $\pi$ is an $(r-1)$-round leader election protocol over $\ell$ players where in each round, each player sends $1$ bit and with the guarantee that if there are at most $\delta \ell$ bad players, then a good player is chosen as leader  with probability $1-\epsilon$. Suppose $\X$ is an \oNOBF[g, \ell r], where $g \ge lr - \delta \ell$. We simply partition the bits of $\X$ into chunks $\X_1,\X_2,\ldots,\X_{r}$, where each $X_i$ is on $\ell$ bits, and simulate the protocol $\pi$ by using the $j$'th bit of $\X_i$ as the message of the $j$'th player in round $i$, for all $1 \le j \le \ell$ and $1 \le i \le r-1$. At the end of this simulation suppose $j^* \in [\ell]$ is the chosen leader. Then we output the $j^*$'th bit of  $\X_{r}$ as the output of the extractor. 

Briefly, the reason that the above is a valid simulation of $\pi$ is the fact that the value of any bad bit in this online setting just depends on bits that appear before it, which is allowed in the leader election protocol (where in round $i$, the message of a bad player can be any function of the messages in the same round or previous rounds). The correctness of the extractor now follows from the fact that since the number of bad players (i.e., bad bits in $\X$) is at most $\delta \ell$, the guarantee of the protocol ensures that the chosen leader $j^* \in [\ell]$ is a good player with probability at least $1-\epsilon$, and in this case  the $j^*$'th bit of $\X_{r}$ must be uniform.

We note here that in the usual definition of leader election protocols, the requirement is to select a good leader   with constant probability, which is a weaker guarantee than what we need to instantiate the above plan. It turns out that we can combine leader election protocols from prior works, in particular from \cite{feige99} and \cite{alonnoar93collective}, to construct protocols with the stronger guarantee we require. We refer the reader to \cref{sec:new leader election protocols} for more details on the construction of our leader election protocols.

\subsection{Organization}
We give some  preliminaries in \cref{sec:preliminaries} before moving on to our core results. \cref{sec:explicit condensers with small block length} provides proofs for our explicit condenser constructions with small block length, and \cref{sec:transformation} shows how to handle converting low-entropy \oNOSFs to \unioNOSF for a wider range of parameters. Next, \cref{section: existence all} details our proof for the existence of seedless condensers for \oNOSFs for all regimes of $\ell$ and $n$. In \cref{sec:extractors for oNOSF/oNOBF via protocols}, we present our explicit constructions of extractors for oNOBF and oNOSF sources using a connection to leader election protocols. Then, in \cref{sec:new leader election protocols}, we explicitly construct the required leader election protocols. Finally, we introduce the notion of online influence in \cref{sec:extracting oNOBF lower bound} and use it to provide an extraction lower bound for \oNOBFs. We discuss some open questions in \cref{sec:open problems}. 

In \cref{sec:local oNOSFs}, we consider a natural local variant of oNOSF sources and show that it is straightforward to extract from such sources using existing extractors for small-space sources.

\dobib

\section{Application to Collective Coin Flipping and Collective Sampling}\label{sec:intro-application coin flipping sampling}

We now discuss applications of our results on condensers for \oNOSFs to fault-tolerant distributed computing. Condensing from \oNOSFs can be viewed as a special case of coin flipping and collective sampling protocols in the full information model that arise in fault-tolerant distributed computing.

\subsubsection{Background}

Say there are $\ell$ players who have a common broadcast channel and want to jointly perform a task such as collectively flipping a coin. Some $b$ players out of them are ``bad'' and want to deter the task. We assume the bad players are computationally unbounded so cryptographic primitives are of no use. We further assume  that each player has private access to uniform randomness. \cite{ben-or_collective_1989} initiated the study of this model and aptly termed this task as ``collective coin flipping.''

The simplest way to collectively flip a coin would be for all the players to initially agree on a function $f: \zo^{\ell}\to \zo$, then synchronously broadcast one random bit $r_i$, and to finally agree on the output being $f(r_1, \dots, r_{\ell})$. 
However, synchronizing broadcasts is hard, and it could be that the bad players set their output as function of the bits of the good players. 
\cite{kahn_influence_1988} showed that no function $f$ can handle more than $O\left(\frac{\ell}{\log \ell}\right)$ corruptions.

One way to allow for more corruptions (almost linear) among  players is to consider ``protocols'' that allow more rounds of communication. In particular, a protocol can be thought of as a tree where each vertex represents a ``round'' where in every round the following happens: all good players sends their bits, then all bad players send their bits as a function of the bits of the good players, and they jointly compute a function of these bits. Depending on the outcome of the function, everyone branches on one branch in this tree. Furthermore, every leaf is labeled with final outcomes (say $0$ or $1$) and, once a leaf is reached, that is the outcome that everybody agrees on. 
\cite{ggl98} initiated the study of protocols where the outcomes are from a larger range and where the bad players are trying to minimize the largest probability of any outcome. They called this problem ``collective sampling.'' For a formal definition, see \cref{subsec: prelim leader election}.

\subsubsection{Known Results}

\cite{ben-or_collective_1989} showed that for protocols with outcomes $\zo$, $b$ bad players can always ensure that some outcome occurs with probability at least $\frac{1}{2} + \frac{b}{2\ell}$. 
\cite{alonnoar93collective} first constructed a protocol that can handle a linear number of corruptions. Follow-up works tried to reduce the number of rounds in this protocol where, in some settings, players were allowed to send more than one bit per round \cite{russellzuckerman01, feige99}.

\cite{ggl98} showed that for all collective sampling protocols and all outcomes, there exists a way for $b$ bad players to coordinate and ensure that an outcome that happens without corruption with probability $p$, now happens with probability $p^{1 - (b/n)} \ge p\left(1 + \frac{b}{n}\log(1/p)\right)$. Nearly matching collective sampling protocols were constructed by \cite{ggl98, SV08selection, GVZ06selection}. For an overview of further results and bounds, see \cite{dodis2006fault}.

\subsubsection{Connection to oNOSF Sources}

The problem of extracting or condensing from $\oNOSFs$ can be seen as special cases or variants of collective coin flipping and collective sampling that provide very simple protocols. For instance, suppose one has an extractor or condenser $f$ for $\unioNOSFs[g, \ell, n]$. Then, consider a protocol where all $\ell$ players take turns and output $n$ random bits. The agreed final outcome is $f$ applied on these $\ell n$ bits. This leads to protocols that are structurally much simpler since players don't have to carefully compute whose turn it is to go in various rounds and can obliviously prepare for their turn. 

The above protocol can also be viewed as a relaxed version of a $1$-round protocol where instead of everyone providing their output asynchronously, they take turns and provide outputs one after another in a simple sequential manner.

\subsubsection{Previous Results Interpreted in oNOSF source context}

Previous impossibility results can be interpreted in the context of extracting / condensing from \unioNOSFs. For instance, collective coin flipping impossibility results of \cite{ben-or_collective_1989} imply extraction impossibility results for \unioNOSFs[g, \ell, n] when $n=1$. They imply:
\begin{corollary}
    There does not exist an $\frac{b}{2\ell}$-extractor for \unioNOSFs[g, \ell, 1].
\end{corollary}

Similarly, we observe that the notion of collective sampling is equivalent to $0$-error condensing. Hence, lower bounds of \cite{ggl98} imply $0$-error condensing lower bounds for \unioNOSFs[g, \ell, n] when $n = 1$. Formally:
\begin{corollary}
There does not exist a condenser $\Cond: \zo^{\ell}\to \zo^m$ for \unioNOSFs[g, \ell, 1] that can guarantee output smooth min-entropy (with parameter $\eps = 0$) more than $k = \frac{g}{\ell}\cdot m$.
\end{corollary}

\subsubsection{$\eps$-Collective Sampling}

Since collective sampling lower bounds show that for any protocol, $0$-error condensing beyond rate $g/\ell$ is impossible, one can naturally ask whether condensing with small error $\eps$ is possible.
We call this problem \emph{$\eps$-collective sampling}, where the goal is to output a distribution which is $\eps$-close to a distribution where every output has small probability.

Interpreted this way, this is exactly what protocols arising out of our condensers for \unioNOSFs provide: Using \cref{theorem: intro constant n existential uni onosf}, when each player has access to $10^4$ random bits, there exists a simple protocol that can handle $0.49\ell$ corrupt players such that the players can collectively sample a distribution over $m = O(\ell)$ bits which is $2^{-\Omega(\ell)}$-close to having entropy $0.99 m$. As far as we are aware, such a protocol is not implied by any other previous protocol. Most previous protocols are obtained through \emph{leader election} protocols, which do not seem useful here since the leader has access to only constant number of bits.

We similarly obtain explicit protocols using \cref{thm: intro explicitly condense uni oNOSF} for the case when each player has access to $n \ge \poly(\log(\ell) / \eps)$ many bits. 

\subsubsection{Collective Coin Flipping and Sampling with Weak Random Sources}

A natural extension to collective coin flipping and sampling in the full information model is when all players only have access to weak source of randomness (that are independent from each other) instead of true uniform randomness. 
This question was first studied by \cite{gsv05}. 
\cite{klrz08network} used network extractor protocol to transform weak random sources of each player into independent private random sources. This way, after using the network extraction protocol, players can follow the usual collective coin flipping / sampling protocol. \cite{gsz21network} improved the network extraction protocol using two-source non-malleable extractors. 

Using our \oNOSF[g, \ell, n, k] condensers, we obtain alternative, simple $\eps$-collective sampling protocols in the setting where players have access to weak sources of randomness. We obtain such an existential protocol using \cref{theorem: intro low entropy oNOSF condenser}, and explicit protocol using \cref{corr: intro explicit condenser oNOSF}.

\dobib

\section{Preliminaries}\label{sec:preliminaries}

In this section we give some basic background and facts used throughout our paper. We use boldfaced font to indicate a random variable such as $\X$. Often we will use $\circ$ or $,$ to indicate concatenation of blocks. So if $\X_1\sim \zo^n$ and $\X_2\sim \zo^n$, then $\X_1,\X_2$ will be the concatenated random variable over $\zo^{2n}$. We will use the notation $[n]$ as shorthand for $\{1, \dots, n\}$. All logs in this paper will have base $2$ unless stated otherwise.

\subsection{Basic Probability Notions}

We measure the distance between two distributions via statistical distance:

\begin{definition}[Statistical Distance]\label{defn: statistical distance}
For any two distributions $\X, \Y$ over $ \Omega$, we define the statistical distance or total-variation distance (TV) distance as:
\[
\abs{\X - \Y} = \max_{S\subset \Omega} \abs{\Pr[\X\in S] - \Pr[\Y\in S]} = \frac{1}{2}\sum_{s\in \Omega} \abs{\Pr[\X = s] - \Pr[\Y = s]}
\]

We use the notation $\X \approx_{\varepsilon} \Y$ to denote the fact that $|\X- \Y| \le \varepsilon$.
\end{definition}


We also state the useful folklore result of the data processing inequality.
\begin{fact}\label{fact:data processing}
    For any two distributions $\X,\Y$ over $\Omega$ and function $f:\Omega\to\R$,
    \begin{align*}
        \abs{\X-\Y}\geq\abs{f(\X)-f(\Y)}.
    \end{align*}
\end{fact}

We will utilize the very useful min-entropy chain rule in our constructions.

\begin{lemma}[Min-entropy chain rule,  \cite{maurer1997privacy}]\label{lem:min-entropy-chain-rule}
For any random variables $\X\sim X$ and $\Y\sim Y$ and $\eps>0$,
\[
    \Pr_{y\sim\Y}[H_\infty(\X\mid\Y=y)\geq H_\infty(\X)-\log|\supp(\Y)|-\log(1/\eps)]\geq1-\eps.
\]
\end{lemma}

\subsection{Condensers and Extractors}

We recall the definition of a seeded condenser.
\begin{definition}\label{def:seeded_cond}
A $(k_{in}, k_{out}, \eps)$-seeded condenser $\sCond:\zo^n \times \zo^d \to\zo^m$ satisfies the following: for every source $\X\sim \zo^n$ with $\minH(\X)\ge k_{in}$, and $\Y=\U_d$,
\[
\sminH(\sCond(\X, \Y))\ge k_{out}.
\]
Here, $d$ is called the \emph{seed length} of $\sCond$.

\end{definition}

A seeded extractor is the special case of seeded condenser where $k_{out} = m$. We record the full definition for completeness sake:
\begin{definition}\label{def:seeded_ext}
A $(k,\eps)$-seeded extractor $\sExt:\zo^n \times\zo^d \to\zo^m$ satisfies the following:   for every source $\X\sim \zo^n$ with $\minH(\X)\ge k$, and $\Y=\U_d$,
$$\sExt(\X,\Y)\approx_{\eps}\U_m.$$
Here, $d$ is called the \emph{seed length} of $\sExt$. 
$\sExt$ is called strong if
$$\sExt(\X,\Y), \Y \approx_{\eps}\U_m, \Y.$$
\end{definition}


We will use the following near optimal explicit construction of seeded extractors:

\begin{theorem}[Theorem 1.5 in \cite{guruswami_unbalanced_2009}]
\label{thm:GUV extractor}
For all constant $0 < \alpha < 1$, there exists a constant $C$ such that for all $n, k, \eps$, there exists an explicit $(k, \eps)$-seeded extractor $\sExt: \zo^n \times \zo^d \rightarrow \zo^m$ with $d = C\log(n / \eps)$ and $m \ge (1 - \alpha)k$.
\end{theorem} 

Next, we recall the definition of two-source extractors.
\begin{definition}\label{def:two-source extractor}
A function $2\Ext:\zo^{n_1}\times\zo^{n_2}\to\zo^m$ is a \emph{$(k_1,k_2,\varepsilon)$-two-source extractor} if for every source $\X_1\sim \zo^{n_1}$ with $\minH(\X_1)\ge k_1$ and $\X_2\sim \zo^{n_2}$ with $\minH(\X_2)\ge k_2$ where $\X_1$ and $\X_2$ are independent of each other, we have
\[
    2\Ext(\X_1,\X_2)\approx_\varepsilon \U_m.
\]
It is said to be \emph{strong in the first argument} if 
\[
    2\Ext(\X_1,\X_2),\X_1\approx_\varepsilon \U_{m}, \X_1.
\]
\end{definition}


\subsection{Averaging Samplers}
Recall the definition of an averaging sampler
\begin{definition}
    A $(k,\delta,\varepsilon)$-averaging sampler is a function $\Samp:\zo^n\to(\zo^m)^D$ such that for any function $f:\zo^m\to[0,1]$ and any $(n,k)$-source $\X$, we have that
    \begin{align*}
        \Pr_{(x_1,\dots,x_D)\sim\Samp(\X)}\left[\abs{\frac{1}{D}\sum_{i=1}^Df(x_i)-\E_{x\sim\U_m}[f(x)]}\geq\varepsilon\right]\leq \delta.
    \end{align*}
\end{definition}

It was shown in \cite{zuckerman_randomness-optimal_1997} that strong extractors and averaging samplers are equivalent. We reproduce the proof here for completeness.
\begin{lemma}\label{lem:sampler extractor equivalence}
    Let $\Ext:\zo^n\times\zo^d\to\zo^m$ be a $(k,\varepsilon)$-seeded extractor. Define $\Samp:\zo^n\to(\zo^m)^D$, where $D=2^d$ as $\Samp(x)=(\Ext(x,1),\dots, \Ext(x, D))$ where we identify $[D]$ with $\zo^d$. Then $\Samp$ is a $(k+\log(1/\delta),\delta,\varepsilon)$-averaging sampler for all $\delta>0$.
\end{lemma}
\begin{proof}
    Let $\X$ be an $(n,k+\log(1/\delta))$-source. For $x\in\supp(\X)$, call $x$ \emph{bad} if $\abs{\Ext(x,\U_d)-\U_m}>\varepsilon$ and let $B\subseteq\zo^n$ be the subset of bad $x$'s. Now, suppose for the sake of contradiction that $\Pr[\X\in B]>\delta$. Then, letting $\X_B=\X\mid(\X\in B)$, we can compute the min-entropy of $\X_B$ as
    \begin{align*}
        \minH(\X_B)&\geq\minH(\X)-\log\left(\frac{1}{\Pr[\X\in B]}\right)\\
        &\geq k+\log(1/\delta)-\log(1/\delta)\\
        &=k.
    \end{align*}
    Therefore, we can apply $\Ext$ to $\X_B$ and obtain that $\abs{\Ext(\X_B,\U_d)-\U_n}\leq\varepsilon$. However, by assumption we know that for all $x\in B$, $\abs{\Ext(x,\U_d)-\U_m}>\varepsilon$, meaning that $\abs{\Ext(\X_B,\U_d)-\U_m}>\varepsilon$, giving us a contradiction. Thus, we have that $\Pr[\X\in B]\leq\delta$.

    Now, we turn our attention to the good $x\notin B$. Using \cref{fact:data processing}, we know that for any $x\notin B$,
    \begin{align*}
        \abs{f(\Ext(x,\U_d))-f(\U_m)}&\leq\abs{\Ext(x,\U_d)-\U_m}\\
        &\leq\varepsilon.
    \end{align*}
    This is equivalent to saying that, for all good $x\in\zo^n$, $\abs{\frac{1}{D}\sum_{s\in\zo^d}f(\Ext(x,s))-f(\U_m)}\leq\varepsilon$. This is exactly the requirement of our sampler, and we have shown that this happens with probability $\Pr[\X\notin B]\geq1-\delta$, as required.
\end{proof}

In particular, we will use the following strong extractor from \cite{zuckerman_linear_2007} to instantiate an averaging sampler.

\begin{theorem}[\cite{zuckerman_linear_2007}]\label{lem:zuckerman linear seeded extractor base}
    For all constant $\alpha,\delta,\varepsilon>0$, there is an efficient family of strong $(k=\delta n,\varepsilon)$-extractors $\Ext:\zo^n\times\zo^d\to\zo^m$ with $m\leq(1-\alpha)\delta n$ and $D=2^d=O(n)$.
\end{theorem}

Using \cref{lem:sampler extractor equivalence}, we get the following averaging sampler that we shall use later.

\begin{lemma}
\label{lem:zuckerman linear seeded extractor}
    For all constant $\alpha,\delta,\varepsilon>0$, we can construct an explicit sampler $\Samp:\zo^t\to(\zo^m)^D$ with $m\leq(1-\alpha)\delta t$ and $D=O(t)$ such that for all sets $S\subseteq[M]$, where $M=2^m$, and all $(t,k)$-sources $X$, we have that 
    \begin{align*}
        \Pr_{x\sim X}\left[\abs{\frac{\abs{\Samp(x)\cap S}}{D}-\frac{\abs{S}}{M}}\geq\varepsilon\right]\leq 2^{\delta t-k}.
    \end{align*}
\end{lemma}
\begin{proof}
    Simply apply \cref{lem:sampler extractor equivalence} to \cref{lem:zuckerman linear seeded extractor base} and consider the indicator function $f(x)=1_{x\in S}$ of $S$ for the resulting sampler.
\end{proof}

\subsection{Leader Election, Collective Coin Flipping, and Sampling Protocols}
\label{subsec: prelim leader election}

We formalize the definition of protocols in the full information model. Collective coin flipping protocols, leader election protocols, and collective sampling protocols are special cases of such protocols where the output domain is $[\ell]$, $\zo$ and $\zo^m$ for some $m$ respectively. 

\begin{definition}[Protocol in the full information model]
A \emph{$k$-round protocol with output domain $Y$ over $\ell$ players where each player sends $n$ random bits per round} is a function 
\[
\pi: \left(\left(\zo^n\right)^{\ell}\right)^k \to Y
\]
that takes in the input of each of the players during each round and outputs an element from set $Y$ which is the outcome of the protocol.

Here is how the protocol operates in the presence of a set $B\subset [\ell]$ of bad players:
In round $i$, each of the players from $[\ell]\setminus B$ independently output a uniformly random element from $\zo^n$. Let their collective outputs be $\alpha_i\in \left(\zo^n\right)^{[\ell]\setminus B}$. Then, depending on $\alpha_1, \dots, \alpha_i$, the players in $B$ together output an element of $\left(\zo^n\right)^{B}$. Hence, we model the strategy of the bad players as a sequence of functions $\sigma = (\sigma_1, \dots, \sigma_k)$, where 
\[
\sigma_i: \left(\left(\zo^n\right)^{[\ell]\setminus B}\right)^{i} \to \left(\zo^n\right)^{B},
\]
where $\sigma_i$ takes in the inputs of the good players from the first $i$ rounds and maps it to the output of the bad players for round $i$.
For a fixed strategy $\sigma$, the outcome of the protocol can be modeled as follows: uniform random strings $\alpha_1, \dots, \alpha_k\in \left(\zo^n\right)^{[\ell]\setminus B}$ are chosen, and the outcome of the protocol is
\[
\pi(\alpha_1: \sigma_1(\alpha_1), \alpha_2: \sigma_2(\alpha_1, \alpha_2), \dots, \alpha_k: \sigma_k(\alpha_1, \dots, \alpha_k)).
\]
To denote the presence of the bad players in $B$, we may also write $\pi|_B$.
\end{definition}

We now specialize this definition to define collective coin flipping protocols

\begin{definition}[Collective coin flipping protocol]
A \emph{collective coin flipping protocol} $\pi$ is a protocol in the full information model with output domain $Y = \zo$.
Furthermore, we say $\pi$ is $(b, \gamma)$ resilient if in the presence of any set $B$ of bad players with $\abs{B}\le b$, we have that $\max_{o\in \zo} \Pr[\pi|_B = o] \le 1-\gamma$.
\end{definition}

Note that when $k=1$, the protocol $\pi$ just becomes a function over $\zo^{\ell}$; such 1-round coin flipping protocols which cannot be biased by any small set of bad players are also known as \emph{resilient functions}.

We also specialize the definition of protocols to define leader election protocols:

\begin{definition}[Leader election protocol]
A \emph{leader election protocol} $\pi$ is a protocol in the full information model with output domain $Y = [\ell]$, the number of players the protocol is operating on.
Furthermore, we say $\pi$ is $(b, \gamma)$ resilient if in the presence of any set $B$ of bad players with $\abs{B}\le b$, we have that $\Pr[\pi|_B \in B]\le 1-\gamma$.
\end{definition}

\begin{remark}
The definition of resilience that we use, which is standard in the leader election and collective coin flipping literature, requires only that bad players can be elected as a leader with probability at most $1-\gamma$.    
Our leader election protocols satisfy (and need) the stronger measure of quality that is standard in the pseudorandomness literature:  that bad players are chosen with probability at most $\eps$ for small $\eps$.
\end{remark}

We lastly define collective sampling protocols:

\begin{definition}[Collective sampling protocol]
A \emph{collective sampling protocol} $\pi$ is a protocol in the full information model, typically with output domain $Y = \zo^m$ for some $m$ which is a function of $\ell$ and $n$.
The goal of collective sampling protocols is to ensure that for every output set $S\subset \zo^m$ with density $\mu$, in the presence of $b$ bad players, the probability that the output lies in $S$ is at most $\eps$, with the goal to make $\eps$ as close to $\mu$ as possible.
\end{definition}

\dobib

\section{Explicit Condensers for oNOSFs with Small Block Length}\label{sec:explicit condensers with small block length}
In this section we will give an explicit construction of condensers for \unioNOSFs with small block length. Our main theorem shows that, given an \oNOSF \footnote{Unless stated otherwise, in this section we will use \oNOSF to refer to \unioNOSF} with slightly more than a half fraction of good players, the block length only needs to be polylogarithmic in the length of the source to condense from it. This nearly matches the lower bound that it is impossible to condense above entropy rate $1/2$ when the fraction of good players is at most $1/2$.

\begin{theorem}\label{thm: 51 percent good n polylog ell}
There exists a universal constant $C$ such that for any constant $\gamma>0$ the following holds.
For all $0 < \eps < 1/2$ and $n,\ell\in\N$ where $n \ge \left(\frac{\log(\ell)}{\varepsilon}\right)^{C}$, there exists an explicit condenser $\Cond:(\zo^n)^\ell\to\zo^m$ satisfying: 
For any \oNOSF[g=(1/2+\gamma)\ell,\ell,n] $\X$, we have that $\sminH(\Cond(\X))\geq m - \left(\frac{\log(\ell)}{\varepsilon}\right)^{C}\log(n)$ where $m = \frac{1}{3}\cdot \gamma\ell n - C\ell\log(\ell)\log(1/\eps)$.
\end{theorem}
We will prove this result in \cref{sec:condensing 51 with polylog}.

Using the transformation of low-entropy \oNOSFs to \unioNOSFs from \cref{lem:explicit low-entropy transformation}, we get an explicit condenser for low-entropy \oNOSFs.
\begin{corollary}\label{cor:low entropy oNOSF explicit condenser}
There exists universal constants $C, C'$ such that for any constant $\gamma>0$ the following holds.
For all $n,\ell, k\in\N$ where $n \ge \ell^{C}, k \ge (\log(n))^C$, there exists an explicit condenser $\Cond:(\zo^n)^\ell\to\zo^m$ satisfying: 
For any \oNOSF[g=(1/2+\gamma)\ell,\ell,n, k] $\X$, we have that $\sminH(\Cond(\X))\geq m - \left(\log(\ell)\right)^C\log(n)$ where $m = \frac{1}{4}\cdot \gamma\ell n - C\ell\log(\ell)\log(\log(\ell))$ and $\eps = \frac{1}{\left(\log(\ell)\right)^{C'}}$.
\end{corollary}

\begin{proof}
To do this, we apply \cref{lem:explicit low-entropy transformation} with $d$ as a very large universal constant, and its number of output bits $m = (\log(\ell))^{C_0}$ for some large universal constant $C_0$.
We obtain that with error at most $\eps/2$, the resultant distribution is 
\unioNOSF[g=(1/2+0.99\gamma)\ell',\ell', (\log(\ell'))^{C_1}] where $\ell' = \ell-1$ and $C_1$ is a large universal constant.
On that resultant source, we apply \cref{thm: 51 percent good n polylog ell} with error $\eps/2$ to obtain the desired result.
\end{proof}





We will need two main tools to prove \cref{thm: 51 percent good n polylog ell}. The first one will allow us to use an \oNOSF to sample a logarithmically sized ``committee'' from any given subset of players that still has approximately the same fraction of good players.
\begin{lemma}
\label{lem:Sampler from reduce and seeded extractor}
There exists a universal constant $C$ such that for all constant $0 < \gamma, \eps_a < 1$, the following holds.
For all $0 < \eps_s < 1/2$, and $n, \ell\in \N$ where $n\geq 6\log(\ell)\log(1/\varepsilon_s)/\gamma$, there exists an explicit function $\NOSFSamp: (\zo^n)^{\ell}\to [\ell]^D$ where $D \le C\log(\ell/\varepsilon_s)$ with the following property.\footnote{Even though the output domain of $\NOSFSamp$ is a vector, we will abuse notation and often treat it as a set}
For all $S\subset [\ell]$ and \oNOSFs[\gamma \ell, \ell, n] $\X$,  we have that 
\[
    \Pr_{x\sim \X}\left[\abs{\frac{\abs{\NOSFSamp(x)\cap S}}{D}-\frac{\abs{S}}{\ell}}\geq\varepsilon_a\right]\leq \varepsilon_s
\]
\end{lemma}
We will prove this in \cref{subsec:Sampler from reduce and seeded extractor}.

Our second tool is a seeded condenser that works even when the seed is an \oNOSF.
In fact,  the bad bits in the seed are allowed to depend on the general min-entropy source that the condenser is acting on. 
\begin{lemma}\label{lem:condenser from funky two source extractor}
    There exists a universal constant $C$ such that for all $n_x, k, n_y, t\in\N$ with $\varepsilon>0$ and $n_y\geq (C)^t\log(tn_x/\varepsilon)$, there exists an explicit condenser $\TCond:\zo^{n_x}\times(\zo^{n_y})^t\to\zo^m$ where $m=\frac{1}{3}(k-(C)^t\log(tn_x/\varepsilon))$ so that the following holds: For all $(n_x,k)$-sources $\X$ and \oNOSFs[g=1, \ell=t] $\Y\sim (\zo^{n_y})^t$ such that the good blocks in $\Y$ are independent of $\X$ and the bad blocks in $\Y$ can depend on $\X$, we have that $\sminH(\TCond(\X,\Y))\geq m-(C)^t\log(tn_x/\varepsilon)$.
\end{lemma}
We prove this in \cref{subsec:funky two source extractor exists}.
\begin{remark}
In \cref{lem:condenser from funky two source extractor}, we have the requirement that $n_y \ge (C)^t\log(tn_x/\varepsilon)$ so that both $m>0$ and $\TCond$ outputs non-zero min-entropy. 
Therefore, decreasing $t$ will give a smaller bound on the length of each block of $\Y$ and, ultimately, the block length of the \oNOSF we are condensing from. Our trick here will be to get $t$ down to $t\approx O(\log\log(\ell))$ when starting from a \oNOSF[g,\ell,n] so that the block length of this source only needs to be at least $\polylog(\ell)$.
\end{remark}

To give some intuition behind our explicit construction in \cref{thm: 51 percent good n polylog ell}, this section is organized in  parts that each provide a step towards the proof. 
For base line construction, we observe that \cref{lem:funky two source extractor exists} already yields an explicit condenser for \oNOSFs[g, \ell] where $g > \ell / 2$ with block length $n \ge \exp(\ell)$. This is formally proven in \cref{sec:condensing 51 with exp}.
In \cref{sec:condensing 67 with poly}, we will first see how we use both of these main tools from above to handle the easier setting of condensing from a \oNOSF[g, \ell] with the number of good blocks $g>\frac{2}{3}\ell$ and we have polynomial block length $n\geq\poly(\ell)$, an exponential improvement over the base line construction. We build upon these ideas to further exponentially decrease our block length requirement in \cref{sec:condensing 76 with polylog} to handle \oNOSFs with $g>\frac{3}{4}\ell$ and $n\geq\polylog(\ell)$. To then decrease the fraction of good blocks that we require, we introduce a correlated sampling trick in \cref{sec:condensing 67 with polylog} so that we only require $g>\frac{2}{3}\ell$ while retaining the requirement $n\ge \poly(\log(\ell))$. Finally, we obtain \cref{thm: 51 percent good n polylog ell}, which only requires $g>\frac{1}{2}\ell$ and $n\ge \poly(\log(\ell))$ in \cref{sec:condensing 51 with polylog} by repeating such a correlated sampling trick, carefully handling the fact that our sources lose some structure each time we do so. Each of these 4 sections is self contained and only depends only on \cref{lem:Sampler from reduce and seeded extractor} and \cref{lem:condenser from funky two source extractor}.

\subsection{\texorpdfstring{Condensing from 51\% good \oNOSFs with $n\geq\exp(\Omega(\ell))$}{Condensing from 51\% good oNOSF sources with n >= exp(Omega(l))}}\label{sec:condensing 51 with exp}
We first construction our baseline condenser that requires at least 51\% good blocks but requires the block length $n \ge \exp(\Omega(\ell))$. This construction solely relies on \cref{lem:condenser from funky two source extractor}:
\begin{theorem}\label{thm:condensing 51 with exp}
There exists a universal constant $C$ such that for any constant $\gamma > 0$, the following holds.
For all $n, \ell\in \N$, and $0 < \eps < 1/2$ where $n\ge (C)^{\ell}\log(1 / \eps)$, there exists an explicit condenser $\Cond:(\zo^n)^\ell\to\zo^m$ satisfying: for any \unioNOSF[g = (1/2 + \gamma)\ell,\ell] $\X$, we have $\sminH(\Cond(\X))\geq m - (C)^{\ell}\log(\ell n/\eps)$ where $m = \frac{1}{6} \cdot \gamma \ell n$.
\end{theorem}

\begin{proof}
If $\ell$ is odd, then we split each block of $\X$ into two contiguous blocks of length $n/2$ each and view $\X$ as a \oNOSF[(1/2+\gamma)2\ell, 2\ell, n/2]. This allows us to without loss of generality assume $\ell$ is even since this transformation preserves the output guarantees required by our condenser.

We begin by decomposing $\X = (\X_1, \X_2)$ where in this decomposition we are simply splitting $\X$ into two parts, so that each $\X_i$ is a \oNOSF[g=(2\gamma)\cdot (\ell/2), (\ell/2), n]. 
We use $\TCond$ from \cref{lem:condenser from funky two source extractor} with $n_x = \ell n / 2, k = \gamma \ell n, n_y = n, t=\ell/2$, error parameter equal to $\eps$ and output $\TCond(\X_1, \X_2)$.
Let $C_{\TCond}$ be the universal constant from \cref{lem:condenser from funky two source extractor}.
We let our universal constant $C$ be much larger than $C_{\TCond}$ so that we satisfy $n \ge (C_{\TCond})^{\ell} \log(\ell^2 n / \eps)$ as required by \cref{lem:condenser from funky two source extractor} and also so that in odd $\ell$ cases, $n/2$ is also sufficiently large.
\end{proof}

\subsection{\texorpdfstring{Condensing from 67\% good \oNOSFs with $n\geq\poly(\ell)$}{Condensing from 67\% good oNOSF sources with n >= poly(l)}}\label{sec:condensing 67 with poly}
We begin by constructing condenser that requires at least 67\% good blocks instead of just 51\%, but allows for the block length $n$ to be polynomial instead of super-exponential in $\ell$. 
Formally, we will show that:
\begin{theorem}
\label{thm:condensing 67 with poly}
    There exists a universal constant $C$ such that for any constant $\gamma>0$ the following holds.
    For all $0 < \eps < 1/2$ and $n,\ell\in\N$ where $n \ge \left(\frac{\ell}{\varepsilon}\right)^{C}$, there exists an explicit condenser $\Cond:(\zo^n)^\ell\to\zo^m$ satisfying: 
    For any \oNOSF[g=(2/3+\gamma)\ell,\ell,n] $\X$, we have that $\sminH(\Cond(\X))\geq m - \left(\frac{\ell}{\varepsilon}\right)^{C}\log(n)$ where $m = \frac{1}{3}\cdot \gamma\ell n$.
\end{theorem}

The key idea here will be to sample a few blocks in the last third of the source to take short prefixes from and then use these as the seed of a particular two source condenser. With this, we present the proof.
\begin{proof}[Proof of \cref{thm:condensing 67 with poly}]
    Let $\ell' = \ell / 3$.
    We begin by decomposing $\X = (\X_1, \X_2,\X_3)$ where in this decomposition we are simply splitting $\X$ into thirds, so each $\X_i$ is a \oNOSF[g=3\gamma\ell',\ell',n]. 

    Our first step is to use $\X_1$ to sample a logarithmically sized committee of players from $\X_3$. We do so by using $\NOSFSamp:(\zo^{n})^{\ell'}\to[\ell']^D$ from \cref{lem:Sampler from reduce and seeded extractor} with $\varepsilon_s=\varepsilon/2$, $\varepsilon_a=\gamma$, the corresponding $\gamma$ equal to $3\gamma$, and the $S$ of \cref{lem:Sampler from reduce and seeded extractor} corresponding to the indices $B\subseteq[\ell]$ of the bad players in $\X_3$, so $\frac{\abs{B}}{\ell'}\leq 1-3\gamma$. Note that $D\le C_0\log(\ell/\varepsilon_s) \le C_1\log(\ell/\varepsilon)$ where $C_0, C_1$ are some universal constants. \cref{lem:Sampler from reduce and seeded extractor} then allows us to conclude that 
    \begin{align*}
        \Pr_{x\sim \X_1}\left[\abs{\frac{\abs{\NOSFSamp(x)\cap B}}{D}-\frac{\abs{B}}{\ell'}}\geq\varepsilon_a\right]&\leq\varepsilon_s\\
        \implies \Pr_{x\sim \X_1}\left[\frac{\abs{\NOSFSamp(x)\cap B}}{D}\leq\varepsilon_a+\frac{\abs{B}}{\ell'}\right]&\geq1-\varepsilon_s\\
        \implies \Pr_{x\sim \X_1}\left[\abs{\NOSFSamp(x)\cap \overline{B}}\geq 2\gamma\cdot D\right]&\geq1-\frac{\varepsilon}{2}.
    \end{align*}
    Let $\Y=(\X_3)_{\NOSFSamp(\X_1)}$ be the $D \le C_1\log(\ell'/\varepsilon)$-sized committee of players from $\X_3$ chosen by $\NOSFSamp(\X_1)$. The above can then be interpreted as saying that, with at least a $1-\varepsilon/2$ probability over $\X_1$, we have that $\Y$ is a \oNOSF[g=2\gamma D,D,n] (we will only need the fact that $\Y$ contains at least one good block, though).

    Our second step is to apply the condenser from \cref{lem:condenser from funky two source extractor} to $\X_2$ and $\Y$. We instantiate \cref{lem:condenser from funky two source extractor} with $n_x= \ell' n$, $n_y = \left(\frac{\ell}{\varepsilon}\right)^{C'}\log(n)$, $t = D \le C_1\log(\ell'/\varepsilon)$, $k = 3\gamma\ell' n$, and error equal to $\frac{\varepsilon}{2}$ where $C'$ is a large enough constant. One can verify that this setting of parameters along with our assumption that $n \ge \left(\frac{\ell}{\varepsilon}\right)^{C}$ for a universal constant $C$ (by setting it to be large enough) satisfies the requirements of \cref{lem:condenser from funky two source extractor}.
    This yields a condenser that we shall call $\TCond:(\zo^n)^{\ell'}\times(\zo^{n})^D\to\zo^m$ (so as to avoid confusion with our ultimate condenser $\Cond$). 

    Lastly, we let $\ZZ = \TCond\left(\X_2,(\X_3)_{\NOSFSamp(\X_1)}\right)$ and let $m_z$ be the length of $\ZZ$.
    If $m_z \le m = \frac{\gamma \ell n}{3}$, then we output $\ZZ$ followed by $m_z - m$ many zeros.
    Otherwise, we output a prefix of $\ZZ$ of length $m = \frac{\gamma \ell n}{3}$.

    We now analyze the guarantees of this condenser.
    We first observe that 
    \begin{align*}
    (C_{\TCond})^t\log(2tn_x/\varepsilon) 
    & \le (C_{\TCond})^{C_1(\log(\ell/3\varepsilon))}\log(4\cdot C_1\log(\ell/\varepsilon)\ell n /\varepsilon)\\
    & \le \left(\frac{\ell}{\varepsilon}\right)^{C_2}\log(n) & \tag{$\ast$} \label{eq: condensing 67 with poly ell *}
    \end{align*}
    where $C_{\TCond}$ is the universal constant from \cref{lem:condenser from funky two source extractor} (we will use this notation in subsequent proofs) and $C_2$ is a large enough universal constant.
    With this, we are guaranteed that the length of $\ZZ = m_z$ is such that
    \begin{align*}
        m_z
        & = \frac{1}{3}\left(k-(C_{\TCond})^t\log(2tn_x/\varepsilon)\right)\\
        & = \frac{1}{3}\left(\gamma \ell n - (C_{\TCond})^t\log(2tn_x/\varepsilon)\right)\\
        & \ge \frac{\gamma \ell n}{3} - \left(\frac{\ell}{\varepsilon}\right)^{C_2}\log(n) & \textrm{(by \cref{eq: condensing 67 with poly ell *})}\\
        & = m - \left(\frac{\ell}{\varepsilon}\right)^{C_2}\log(n)
    \end{align*}
    Moreover, conditioned on the fact that $\Y$ has at least one good block, \cref{lem:condenser from funky two source extractor} guarantees that
    \begin{align*}
        \minH^{\eps/2}(\ZZ)
        & = \minH^{\eps/2}(\TCond(\X_2,\Y))\\
        &\geq m_z - (C_{\TCond})^t\log(2tn_x/\varepsilon)\\
        &\geq m_z - \left(\frac{\ell}{\varepsilon}\right)^{C_2}\log(n) & \textrm{(by \cref{eq: condensing 67 with poly ell *})}
    \end{align*}
    Thus, adding up the two $\varepsilon/2$ errors from both our steps, we see that $\sminH(\ZZ)\geq m_z - \left(\frac{\ell}{\varepsilon}\right)^{C_2}\log(n)$. 

    If $m_z > m$, then our final output inherits the smooth min-entropy gap of $\ZZ$ which is $\left(\frac{\ell}{\varepsilon}\right)^{C_2}\log(n)$.
    If $m_z \le m$, then our output inherits not only the entropy gap of $\ZZ$ but also an entropy gap of $m - m_z$. Since $m_z \ge m - \left(\frac{\ell}{\varepsilon}\right)^{C_2}\log(n)$, our output will have smooth min-entropy gap at most $2\left(\frac{\ell}{\varepsilon}\right)^{C_2}\log(n)$.
    In either case, our gap will be at most $2\left(\frac{\ell}{\varepsilon}\right)^{C_2}\log(n)$. 
    We let our final universal constant $C$ be much larger than $C_2$ to obtain our claim.
\end{proof}

\begin{remark}
The padding or truncating trick at the end of all our steps to meet the desired output length is standard and for next subsections and proofs, we will omit it and use it implicitly.    
\end{remark}

\subsection{\texorpdfstring{Condensing from 76\% good \oNOSFs with $n\geq\polylog(\ell)$}{Condensing from 76\% good oNOSF sources with n >= poly(log(l))}}\label{sec:condensing 76 with polylog}
To decrease our block length requirement all the way down to $\polylog(\ell)$, we simply apply the idea in the previous section twice. We split up our \oNOSF $\X$ into four blocks $\X=\X_1,\X_2,\X_3,\X_4$ and use $\X_1$ to sample a logarithmically sized committee from $\X_2$, use this committee to sample a doubly logarithmically sized committee from $\X_4$, and finally apply the condenser from \cref{lem:condenser from funky two source extractor} to $\X_3$ and this final committee.
\begin{theorem}\label{thm: condensing 76 with polylog}
    There exists a universal constant $C$ such that for any constant $\gamma>0$ the following holds.
    For all $0 < \eps < 1/2$ and $n,\ell\in\N$ where $n \ge \left(\frac{\log(\ell)}{\varepsilon}\right)^{C}$, there exists an explicit condenser $\Cond:(\zo^n)^\ell\to\zo^m$ satisfying: 
    For any \oNOSF[g=(3/4+\gamma)\ell,\ell,n] $\X$, we have that $\sminH(\Cond(\X))\geq m - \left(\frac{\log(\ell)}{\varepsilon}\right)^{C}\log(n)$ where $m = \frac{1}{3}\cdot \gamma\ell n$.
\end{theorem}
\begin{proof}
    Let $\ell' = \ell / 4$.
    We decompose $\X$ into quarters as $\X= (\X_1,\X_2,\X_3,\X_4)$, so each $\X_i$ is a \oNOSF[g=4\gamma\ell',\ell',n]. We in fact claim something stronger. Call an index $i\in [\ell']$ totally good if it is good in each of $\X_1, \X_2, \X_3, \X_4$. 
    For the rest of the $i\in [\ell']$ that are not totally good, we refer to them as somewhat bad. Since $\X$ has $(3+4\gamma)\ell'$ good indices out of $4\ell'$, we see that there are must be at least $4\gamma\ell'$ totally good indices, i.e., indices that are good across all of the 4 blocks. 

    Our first step is to use $\X_1$ to sample a logarithmically sized committee of players from $\X_2$. We obtain $\NOSFSamp_2:(\zo^{n})^{\ell'}\to[\ell']^{D_2}$ from \cref{lem:Sampler from reduce and seeded extractor} with $\varepsilon_s=\varepsilon/3$, $\varepsilon_a=\gamma$, the corresponding $\gamma$ equal to $4\gamma$, and the set $S$ of \cref{lem:Sampler from reduce and seeded extractor} corresponding to the indices $B_2\subset [\ell]$ of the somewhat bad players in $\X_2$, so $\frac{\abs{B_2}}{\ell'}\leq 1-4\gamma$. Note that $D_2\le C_0\log(\ell/\varepsilon_s) \le C_1(\log(\ell/\varepsilon))$ where $C_0$ and $C_1$ are some universal constants. \cref{lem:Sampler from reduce and seeded extractor} then guarantees that
    \begin{align*}
        \Pr_{x\sim \X_1}\left[\abs{\frac{\abs{\NOSFSamp_2(x)\cap B_2}}{D_2}-\frac{\abs{B_2}}{\ell'}}\geq\varepsilon_a\right]&\leq\varepsilon_s\\
        \implies \Pr_{x\sim \X_1}\left[\frac{\abs{\NOSFSamp_2(x)\cap B_2}}{D_2}\leq\varepsilon_a+\frac{\abs{B_2}}{\ell'}\right]&\geq1-\varepsilon_s\\
        \implies \Pr_{x\sim \X_1}\left[\abs{\NOSFSamp_2(x)\cap \overline{B_2}}\geq 3\gamma\cdot D_2\right]&\geq1-\frac{\varepsilon}{3}.
    \end{align*}
    Let $\mC_2\subset [\ell']$ be the $D_2$ sized-committee of indices such sampled and let $\Y_2=(\X_2)_{\NOSFSamp_2(\X_1)}$ be the induced source by restricting $\X_2$ to indices from  $\mC_2$. The above then says that with at least a $1-\frac{\varepsilon}{3}$ probability over $\X_1$, we have that $\Y_2$ is a \oNOSF[g=3\gamma D_2,D_2,n] and $\mC_2$ contains at least $ 3\gamma D_2$ totally good indices.

    Our second step is to use $\Y_2$ to sample from $\mC_2$ and obtain a subsource over those indices from $\X_4$. We again do this by using $\NOSFSamp_4:(\zo^{n})^{D_2}\to[D_2]^{D_4}$ from \cref{lem:Sampler from reduce and seeded extractor} with $\eps_s = \eps/3$, $\eps_a = \gamma$, the corresponding $\gamma$ equal to $3\gamma$, and the set $S$ of \cref{lem:Sampler from reduce and seeded extractor} corresponding to the indices $B_{\mC_2}\subseteq \mC_2$ of the weakly bad indices in $\mC_2$ so that $\frac{\abs{B_{\mC_2}}}{D_2}\leq 1 - 3\gamma$. Here, $D_4 \le C_3\log(D_2/\varepsilon_s)\le C_4(\log(\log(\ell)/\varepsilon))$ where $C_3, C_4$ are some universal constants. From \cref{lem:Sampler from reduce and seeded extractor}, once again we are guaranteed that
    \begin{align*}
        \Pr_{y\sim \Y_2}\left[\abs{\frac{\abs{\NOSFSamp_4(y)\cap B_{\mC_2}}}{D_4}-\frac{\abs{B_{\mC_2}}}{D_2}}\geq\varepsilon_a\right]&\leq\varepsilon_s\\ 
        \implies \Pr_{y\sim \Y_2}\left[\frac{\abs{\NOSFSamp_4(y)\cap B_{\mC_2}}}{D_4}\leq\varepsilon_a+\frac{\abs{B_{\mC_2}}}{D_2}\right]&\geq1-\varepsilon_s\\
        \implies \Pr_{y\sim \Y_2}\left[\abs{\NOSFSamp_4(y)\cap \overline{B_{\mC_2}}}\geq 2\gamma\cdot D_4\right]&\geq1-\frac{\varepsilon}{3}.
    \end{align*}
    If we define $\Y_4=(\X_4)_{\NOSFSamp_4(\Y_2)}$, then the above guarantees that, with probability $1-\frac{\varepsilon}{3}$ over $\Y_2$ (conditioned on $\mC_2$ containing $\ge 3\gamma D_2$ totally good indices), we have that $\Y_4$ is a \oNOSF[g=2\gamma D_4,D_4,n].

    In our third and final step, we will use \cref{lem:condenser from funky two source extractor} to condense from $\X_3$ and $\Y_4$. We instantiate \cref{lem:condenser from funky two source extractor} with $n_x = \ell' n$, $n_y = \left(\frac{\log(\ell)}{\eps}\right)^{C'}\log(n)$, $t=D_4 \le C_4\log(\log(\ell)/\varepsilon)$, $k= 4\gamma\ell' n$, and error equal to $\frac{\eps}{3}$ where $C'$ is a large enough universal constant. Given these parameters and our assumption that $n \ge (\log(\ell)/\varepsilon)^{C}$ for a universal constant $C$ (that is large enough), as well as the fact that $\Y_4$ contains at least one good index, the requirements of \cref{lem:condenser from funky two source extractor} are satisfied. Consequently, we obtain the function $\TCond:(\zo^{n})^{\ell'}\times(\zo^{n})^{D_4}\to\zo^{m_{\TCond}}$. 
    Finally the overall output of our explicit condenser is $\ZZ = \TCond\left(\X_3,\Y_4\right)$.

    We now analyze the guarantees of this condenser. 
    We first observe that 
    \begin{align*}
    (C_{\TCond})^t\log(3tn_x/\varepsilon) 
    & \le (C_{\TCond})^{C_4(\log(\log(\ell)/\varepsilon))}\log(3\cdot \ell n\cdot \left(\ell / \eps\right)^{C'} /\varepsilon)\\
    & \le \left(\frac{\log(\ell)}{\varepsilon}\right)^{C_5}\log(n) \tag{$\ast$} \label{eq: condensing 76 with polylog ell *}
    \end{align*}
    where $C_5$ is a large enough universal constant.
    With this, we are guaranteed that the length of $\ZZ = m_z$ above is
    \begin{align*}
        m_z
        & = \frac{1}{3}\left(k-(C_{\TCond})^t\log(3tn_x/\varepsilon)\right)\\
        & = \frac{1}{3}\left(\gamma \ell n - (C_{\TCond})^t\log(3tn_x/\varepsilon)\right)\\
        & \ge \frac{\gamma \ell n}{3} - \left(\frac{\log(\ell)}{\varepsilon}\right)^{C_5}\log(n) & \textrm{(by \cref{eq: condensing 76 with polylog ell *})}\\
    \end{align*}
    Moreover, conditioned on the fact that $\Y_4$ has at least one good block, \cref{lem:condenser from funky two source extractor} guarantees that
    \begin{align*}
        \minH^{\eps/3}(\ZZ)  
        & = \minH^{\eps/3}(\TCond(\X_3,\Y_4))\\
        &\geq m_z - (C_{\TCond})^t\log(3tn_x/\varepsilon)\\
        &\geq m_z - \left(\frac{\log(\ell)}{\varepsilon}\right)^{C_5}\log(n) & \textrm{(by \cref{eq: condensing 76 with polylog ell *})}
    \end{align*}
    Thus, adding up the three $\varepsilon/3$ errors from both our steps, we see that $\sminH(\ZZ) \ge m_z - \left(\frac{\log(\ell)}{\varepsilon}\right)^{C_5}\log(n)$. 

    We let our final universal constant $C$ be much larger than $C_5$ to obtain our desired claim.
\end{proof}

\subsection{\texorpdfstring{Condensing from 67\% good \oNOSFs with $n\geq\polylog(\ell)$}{Condensing from 67\% good oNOSF sources with n >= poly(log(l))}}\label{sec:condensing 67 with polylog}

We now decrease not only our block length requirement all the way down to $\polylog(\ell)$ but also require that only 67\% of the blocks are good. To do this, we apply the idea in the previous section twice; but this time at the end, we reuse one of the blocks we previously used to sample as a source.
We show this by observing that sampling requires the usage of very few bits. By using the chain rule for min-entropy, we conclude that fixing those bits still leaves the source with lots of entropy.
\begin{theorem}\label{thm: condensing 67 with polylog}
    There exists a universal constant $C$ such that for any constant $\gamma>0$ the following holds.
    For all $0 < \eps < 1/2$ and $n,\ell\in\N$ where $n \ge \left(\frac{\log(\ell)}{\varepsilon}\right)^{C}$, there exists an explicit condenser $\Cond:(\zo^n)^\ell\to\zo^m$ satisfying: 
    For any \oNOSF[g=(2/3+\gamma)\ell,\ell,n] $\X$, we have that $\sminH(\Cond(\X))\geq m - \left(\frac{\log(\ell)}{\varepsilon}\right)^{C}\log(n)$ where $m = \frac{1}{3}\cdot \gamma\ell n$.
\end{theorem}
\begin{proof}
    Let $\ell' = \ell / 3$.
    We decompose $\X$ into three parts as $\X= (\X_1,\X_2,\X_3)$, so each $\X_i$ is a \oNOSF[g=3\gamma\ell',\ell',n]. We in fact claim something stronger. Call an index $i\in [\ell']$ totally good if it is good in each of $\X_1, \X_2, \X_3$. 
    For the rest of the indices $i\in [\ell']$ that are not totally good, we refer to them as somewhat bad.
    Since $\X$ has $(2+3\gamma)\ell'$ good indices out of $3\ell'$, we see that there are must be at least $3\gamma\ell'$ totally good indices, i.e., indices that are good across each of the 3 blocks. 

    Our first step is to use $\X_1$ to sample a logarithmically sized committee of players from $\X_2$. We use $\NOSFSamp_2:(\zo^{n})^{\ell'}\to[\ell']^{D_2}$ from \cref{lem:Sampler from reduce and seeded extractor} with $\varepsilon_s=\varepsilon/4$, $\varepsilon_a=\gamma$, the corresponding $\gamma$ equal to $3\gamma$, and the set $S$ of \cref{lem:Sampler from reduce and seeded extractor} corresponding to the indices $B_2\subset [\ell']$ of the somewhat bad indices in $\X_2$, so $\frac{\abs{B_2}}{\ell'}\leq 1-3\gamma$. Note that $D_2\le C_0\log(\ell/\varepsilon_s) \le C_1(\log(\ell/\varepsilon))$ where $C_0$ and $C_1$ are some universal constants. \cref{lem:Sampler from reduce and seeded extractor} then guarantees that
    \begin{align*}
        \Pr_{x\sim \X_1}\left[\abs{\frac{\abs{\NOSFSamp_2(x)\cap B_2}}{D_2}-\frac{\abs{B_2}}{\ell'}}\geq\varepsilon_a\right]&\leq\varepsilon_s\\
        \implies \Pr_{x\sim \X_1}\left[\frac{\abs{\NOSFSamp_2(x)\cap B_2}}{D_2}\leq\varepsilon_a+\frac{\abs{B_2}}{\ell'}\right]&\geq1-\varepsilon_s\\
        \implies \Pr_{x\sim \X_1}\left[\abs{\NOSFSamp_2(x)\cap \overline{B_2}}\geq 2\gamma\cdot D_2\right]&\geq1-\frac{\varepsilon}{4}.
    \end{align*}
    Let $\mC_2\subset [\ell']$ be the $D_2\le C_1(\log(\ell/\eps))$ sized-committee of indices thus sampled and let $\Y_2=(\X_2)_{\NOSFSamp_2(\X_1)}$ be the source obtained by restricting $\X_2$ to indices from $\mC_2$.
    However, when we do this, instead of each player in $\Y_2$ holding $n$ bits, we take a prefix of length $n_2 = C_2\cdot \log(\log(\ell)/\eps)\cdot \log(1/\eps)$ from each where $C_2$ is a sufficiently large universal constant.
    The above then says that with at least a $1-\frac{\varepsilon}{4}$ probability over $\X_1$, we have that $\Y_2$ is a \oNOSF[g=2\gamma D_2,D_2,n_2] and $\mC_2$ contains $\ge 2\gamma D_2$ totally good indices.

    Our second step is to use $\Y_2$ to sample from $\mC_2$ and obtain a subsource over those indices from $\X_3$. We again do this by using $\NOSFSamp_3:(\zo^{n})^{D_2}\to[D_2]^{D_3}$ from \cref{lem:Sampler from reduce and seeded extractor} with $\eps_s = \eps/4$, $\eps_a = \gamma$, the corresponding $\gamma$ equal to $2\gamma$, and the set $S$ of \cref{lem:Sampler from reduce and seeded extractor} corresponding to the indices $B_{\mC_2}\subseteq \mC_2$ of the weakly bad indices in $\mC_2$ so that  $\frac{\abs{B_{\mC_2}}}{D_2}\leq 1 - 2\gamma$. Here, $D_3 \le C_3\log(D_2/\varepsilon_s)\le C_4(\log(\log(\ell)/\varepsilon))$ where $C_3, C_4$ are some universal constants. From \cref{lem:Sampler from reduce and seeded extractor}, once again we are guaranteed that
    \begin{align*}
        \Pr_{y\sim \Y_2}\left[\abs{\frac{\abs{\NOSFSamp_3(y)\cap B_{\mC_2}}}{D_3}-\frac{\abs{B_{\mC_2}}}{D_2}}\geq\varepsilon_a\right]&\leq\varepsilon_s\\
        \implies \Pr_{y\sim \Y_2}\left[\frac{\abs{\NOSFSamp_3(y)\cap B_{\mC_2}}}{D_3}\leq\varepsilon_a+\frac{\abs{B_{\mC_2}}}{D_2}\right]&\geq1-\varepsilon_s\\
        \implies \Pr_{y\sim \Y_2}\left[\abs{\NOSFSamp_3(y)\cap \overline{B_{\mC_2}}}\geq \gamma\cdot D_3\right]&\geq1-\frac{\varepsilon}{4}.
    \end{align*}
    If we define $\Y_3=(\X_3)_{\NOSFSamp_3(\Y_2)}$, then the above guarantees that, with probability $1-\frac{\varepsilon}{4}$ over $\Y_2$ (conditioned on $\mC_2$ containing $\ge 2\gamma D_2$ totally good indices), we have that $\Y_3$ is a \oNOSF[g=\gamma D_3,D_3,n].

    We will show that $\X_2$ has entropy conditioned on most fixings of $\Y_2$.
    Recall that $\X_2$ is a \oNOSF[g=3\gamma\ell',\ell',n].
    We use the min-entropy chain rule (\cref{lem:min-entropy-chain-rule}) to conclude that with probability $1 - \eps/4$ over $y\sim \Y_2$, we have that 
    \begin{align*}
    \minH(\X_2 | (\Y_2 = y) ) 
    & \ge 3\gamma \ell' n - \log(4 / \eps) - D_2\cdot n_2\\
    & \ge 3\gamma \ell' n - \log(4 / \eps) - C_1\log(\ell/\eps)\cdot C_2\log(\log(\ell) / \eps)\cdot \log(1 / \eps)\\
    & \ge 3\gamma \ell' n - C_5\left(\log(\ell / \eps)\right)^{3}
    \end{align*}
    where $C_5$ is a large enough universal constant.
    
    With this, we apply union bound to conclude that conditioned on $\mC_2$ containing $\ge 2\gamma D_2$ totally good indices, with probability $1 - \eps / 2$ over $y\sim \Y_2$, we have that $\Y_3$ is a \oNOSF[g=\gamma D_3,D_3,n] and $\minH(\X_2) \ge \gamma \ell n - C_5\left(\log(\ell / \eps)\right)^{3}$.
    We refer to such a fixing of $\Y_2 = y_2$ as `good.'

    In our third and final step,  we use \cref{lem:condenser from funky two source extractor} to condense from $\X_2$ and $\Y_3$. 
    We instantiate \cref{lem:condenser from funky two source extractor} with $n_x = \ell' n$, $n_y = \left(\frac{\log(\ell)}{\eps}\right)^{C'}\log(n)$, $t = D_3 \le C_6\log(\log(\ell)/\varepsilon)$, $k = 3\gamma\ell' n - C_5\left(\log(\ell / \eps)\right)^{3}$, and error equal to $\frac{\eps}{4}$ where $C'$ is a large enough universal constant. 
    Using our assumption that $n \ge (\log(\ell)/\varepsilon)^{C}$ for a universal constant $C$ (that is large enough), these parameters satisfy the requirements of \cref{lem:condenser from funky two source extractor},, and the lemma gives us the explicit condenser $\TCond:(\zo^{n})^{\ell'}\times(\zo^{n_y})^{D_3}\to\zo^{m_{\TCond}}$. 
    Let $\ZZ = \TCond\left(\X_2,\Y_3\right)$ and let this be the final output of our own condenser.

    We now analyze the guarantees of this condenser. 
    We first observe that 
    \begin{align*}
    (C_{\TCond})^t\log(4tn_x/\varepsilon) 
    & \le (C_{\TCond})^{C_6(\log(\log(\ell)/\varepsilon))}\log(4\cdot C_6(\log(\log(\ell)/\varepsilon))\cdot \ell n /\varepsilon)\\
    & \le \left(\frac{\log(\ell)}{\varepsilon}\right)^{C_7}\log(n) \tag{$\ast$} \label{eq: 67 with polylog}
    \end{align*}
    where $C_7$ is a large enough universal constant.
    
    Let $m_z$ be the length of the source $\ZZ$.
    With this, we are guaranteed from \cref{lem:condenser from funky two source extractor} the following lower bound on $m_z$:
    \begin{align*}
        m_z
        & = \frac{1}{3}\left(k-(C_{\TCond})^t\log(4tn_x/\varepsilon)\right)\\
        & = \frac{1}{3}\left(\gamma \ell n - C_5(\log(\ell / \eps))^3 - (C_{\TCond})^t\log(4tn_x/\varepsilon)\right)\\
        & \ge \frac{1}{3}\left(\gamma \ell n - C_5(\log(\ell / \eps))^3 - \left(\frac{\log(\ell)}{\varepsilon}\right)^{C_7}\log(n)\right) & \textrm{(by \cref{eq: 67 with polylog})}\\
        & \ge \frac{\gamma \ell n}{3} - \left(\frac{\log(\ell)}{\varepsilon}\right)^{C_8}\log(n)
    \end{align*}
    where $C_8$ is a large enough universal constant.
    
    We condition on both $\mC_2$ containing $2\gamma D_2$ totally good indices and a `good' fixing of $\Y_2$. Under this conditioning, \cref{lem:condenser from funky two source extractor} guarantees that
    \begin{align*}
        \minH^{\eps/4}(\ZZ)  
        & = \minH^{\eps/4}(\TCond(\X_2,\Y_3))\\
        &\geq m_z - (C_{\TCond})^t\log(4tn_x/\varepsilon)\\
        &\geq m_z - \left(\frac{\log(\ell)}{\varepsilon}\right)^{C_7}\log(n) & \textrm{(by \cref{eq: 67 with polylog})}\\
        &\geq m_z - \left(\frac{\log(\ell)}{\varepsilon}\right)^{C_8}\log(n).
    \end{align*}
    Thus, adding up the four $\varepsilon/4$ errors from both our steps, we see that $\sminH(\ZZ) \ge m_z - \left(\frac{\log(\ell)}{\varepsilon}\right)^{C_8}\log(n)$. 

    We let our final universal constant $C$ be much larger than $C_8$ to obtain our final claim.
\end{proof}

\subsection{\texorpdfstring{Condensing from 51\% good \oNOSFs with $n\geq\polylog(\ell)$}{Condensing from 51\% good oNOSF sources with n >= poly(log(l))}}\label{sec:condensing 51 with polylog}

We now decrease not only our block length requirement all the way down to $\polylog(\ell)$ but also require that only 51\% of the blocks are good. To do this, we build upon the previous ideas with one more `self-sampling' idea - where we sample from within the blocks in the same source. This introduces correlations between the bits that are being used to sample and the source itself, in a way that makes us lose the structure of our sources. Nevertheless we rely on the fact that very few bits are required to do sampling, and that sampling succeeds regardless of the behavior of the bad players. To handle this situation, we use \cref{lem:control-few-bits-can-still-condense-again} that states if an adversary is allowed to arbitrarily control few bits of the source (that were previously uniform), then the damage they can do is not too much.

\begin{proof}[Proof of \cref{thm: 51 percent good n polylog ell}]
Let $\ell' = \ell / 2$.
We begin by decomposing $\X$ into two parts as $\X= (\X_1,\X_2)$, so each $\X_i$ is a \oNOSF[g=2\gamma\ell',\ell',n]. We in fact claim something stronger. Call an index $i\in [\ell']$ totally good if it is good in both of $\X_1, \X_2$. 
For the rest of the indices $i\in [\ell']$ that are not totally good, we refer to them as somewhat bad.
Since $\X$ has $(1+2\gamma)\ell'$ good indices out of $2\ell'$, we see that there are must be at least $2\gamma\ell'$ totally good indices, i.e., indices that are good across both the blocks.

Let $\X_1'$ be the subsource obtained from $\X_1$ by taking prefixes of all blocks of length $n_1' = C_1\log(\ell)\log(1 / \eps)$ where $C_1$ is a large enough universal constant.
Hence, $\X_1'$ is a \oNOSF[g=2\gamma \ell', \ell', n_1'].

Our first step is to use $\X_1'$ to sample a logarithmically sized committee of players from $\X_2$. We obtain $\NOSFSamp_{1\to 2}:(\zo^{n_1'})^{\ell'}\to[\ell']^{D_2}$ from \cref{lem:Sampler from reduce and seeded extractor} with $\varepsilon_s=\varepsilon/4$, $\varepsilon_a = \gamma / 2$, the corresponding $\gamma$ equal to $2\gamma$, and the set $S$ of \cref{lem:Sampler from reduce and seeded extractor} corresponding to the indices $B_2\subset [\ell']$ of the somewhat bad indices in $\X_2$, so $\frac{\abs{B_2}}{\ell'}\leq 1-2\gamma$. Note that 
\begin{align}\label{eq: condensing 51 with polylog D_2}
D_2\le C_0\log(\ell/\varepsilon_s) \le C_1(\log(\ell/\varepsilon))
\end{align}
where $C_0$ and $C_1$ are some universal constants. \cref{lem:Sampler from reduce and seeded extractor} then guarantees that
\begin{align*}
    \Pr_{x\sim \X_1'}\left[\abs{\frac{\abs{\NOSFSamp_{1\to 2}(x)\cap B_2}}{D_2}-\frac{\abs{B_2}}{\ell'}}\geq\varepsilon_a\right]&\leq\varepsilon_s\\
    \implies \Pr_{x\sim \X_1'}\left[\frac{\abs{\NOSFSamp_{1\to 2}(x)\cap B_2}}{D_2}\leq\varepsilon_a+\frac{\abs{B_2}}{\ell'}\right]&\geq1-\varepsilon_s\\
    \implies \Pr_{x\sim \X_1'}\left[\abs{\NOSFSamp_{1\to 2}(x)\cap \overline{B_2}}\geq (3\gamma/2) \cdot D_2\right]&\geq1-\frac{\varepsilon}{4}.
\end{align*}
Let $\mC_2\subset [\ell']$ be the $D_2\le C_1(\log(\ell/\eps))$ sized committee of indices thus sampled and let $\Y_2=(\X_2)_{\NOSFSamp_2(\X_1)}$ be the source obtained by restricting $\X_2$ to indices from $\mC_2$.
Let $E_{1\to 2}$ be the event that the sampler $\NOSFSamp_{1\to 2}$ above succeeds.
We have that $\Pr[E_{1\to 2}] \ge 1 - \eps/4$ with the probability being over sampling from $\X_1'$.
We see that when $E_{1\to 2}$ occurs,  $\Y_2$ will be an \oNOSF[g= (3\gamma/2) D_2, D_2, n].

We will show that $\X_1$ has entropy conditioned on most fixings of $\X_1'$. We use the min-entropy chain rule (\cref{lem:min-entropy-chain-rule}) to conclude that with probability $1 - \eps/4$ over $x\sim \X_1'$, we have that 
\begin{align*}
\minH(\X_1 | (\X_{1}' = x) ) 
& \ge 2\gamma \ell' n - \ell'n_1' - \log(4 / \eps)\\
& = 2\gamma \ell' n - \ell'\cdot C_1\log(\ell)\log(1/\eps) - \log(4 / \eps)\\
& \ge 2\gamma \ell' n - C_3\ell'\log(\ell')\log(1/\eps)
\end{align*}
where $C_3$ is a large enough universal constant.
Let 
\begin{align}\label{eq: condensing 51 with polylog k_1}
k_1 = 2\gamma \ell' n - C_3\ell'\log(\ell')\log(1/\eps)
\end{align}
Let $E_1$ be the event that $x\sim \X_1'$ is such that $\minH(\X_1 | (\X_1' = x_1)) \ge k_1$.
Then, we have that $\Pr[E_1]\ge 1 - (\eps / 4)$ with the probability being over sampling from $\X_1'$.

By a union bound, we have that both $E_1$ and $E_{1\to2}$ happen together with probability at least $1 - \eps/2$.
Note that conditioning on both $E_1$ and $E_{1\to2}$, the online structure of the source still remains intact, i.e., $\Y_2$ still remains an \oNOSF and the good bits in $\Y_2$ still are independent of $\X_1$. So, for instance we also satisfy the required independence conditions of \cref{lem:funky two source extractor exists} and if we also satisfied the parameter conditions for it, we could apply it. However, we do not satisfy the parameter conditions since our guarantees on $n$ are too small. To remedy this, we will use a subsource of $\Y_2$ to sample from within itself. Doing so will shrink our source and let us satisfy the parameter conditions from \cref{lem:funky two source extractor exists}. However, we will then no longer satisfy the independence requirements to apply it. Nevertheless we do this anyways and argue that we can so, while only sacrificing the final guarantees of the condenser by a tiny amount.

Let $\Y_{2,\Samp}$ be the subsource obtained from $\Y_2$ by taking prefixes of all blocks of length 
\begin{align}\label{eq: condensing 51 with polylog n_2'}
n_2' = C_2\cdot \log(\log(\ell)/\eps)\cdot \log(1/\eps) 
\end{align}
where $C_2$ is a large enough universal constant.
So, when $E_{1\to 2}$ occurs, we have that $\Y_{2, \Samp}$ is \oNOSF[g= (3\gamma/2)D_2, D_2, n_2'] 

In the second step, we will use $\Y_{2, \Samp}$ to sample from $\mC_2$ and obtain a subsource over those indices from $\Y_2$. We again do this by using $\NOSFSamp_{2\to 2}:(\zo^{n})^{D_2}\to[D_2]^{D_{2, \Cond}}$ from \cref{lem:Sampler from reduce and seeded extractor} with $\eps_s = \eps/4$, $\eps_a = \gamma/2$, the corresponding $\gamma$ equal to $3\gamma/2$, and the set $S$ of \cref{lem:Sampler from reduce and seeded extractor} corresponding to the indices $B_{\mC_2}\subseteq \mC_2$ of the weakly bad indices in $\mC_2$ so that  $\frac{\abs{B_{\mC_2}}}{D_2}\leq 1 - (3\gamma/2)$. Here, $D_{2, \Cond} \le C_4\log(D_2/\varepsilon_s)\le C_5(\log(\log(\ell)/\varepsilon))$ where $C_4, C_5$ are some universal constants. From \cref{lem:Sampler from reduce and seeded extractor}, once again we are guaranteed that
\begin{align*}
    \Pr_{y\sim \Y_{2, \Samp}}\left[\abs{\frac{\abs{\NOSFSamp_{2\to 2}(y)\cap B_{\mC_2}}}{D_{2, \Cond}}-\frac{\abs{B_{\mC_2}}}{D_2}}\geq\varepsilon_a\right]&\leq\varepsilon_s\\
    \implies \Pr_{y\sim \Y_{2, \Samp}}\left[\frac{\abs{\NOSFSamp_{2\to 2}(y)\cap B_{\mC_2}}}{D_{2, \Cond}}\leq\varepsilon_a+\frac{\abs{B_{\mC_2}}}{D_2}\right]&\geq1-\varepsilon_s\\
    \implies \Pr_{y\sim \Y_{2, \Samp}}\left[\abs{\NOSFSamp_{2\to 2}(y)\cap \overline{B_{\mC_2}}}\geq \gamma\cdot D_{2, \Cond}\right]&\geq1-\frac{\varepsilon}{4}. & \tag{$\ast\ast$}\label{eq: condensing 51 with polylog second sampler guarantee}
\end{align*}
For $y_{2, \Samp}\in (\zo^{n_2'})^{D_2}$, let $\mC_{2, \Cond}(y_{2, \Samp}) = \NOSFSamp_{2\to 2}(y_{2, \Samp})$ so that the number of players in the resultant committee is $\abs{\mC_{2, \Cond}(y_{2, \Samp})} = D_{2, \Cond} \le C_5(\log(\log(\ell)/\varepsilon))$.
We let $\Y_{2, \Cond}(y_{2, \Samp})$ be the subsource obtained from $\Y_2$ by taking the suffix of all blocks of length $n - n_2'$ where $n_2'$ is as above.
Then \cref{eq: condensing 51 with polylog second sampler guarantee} guarantees that, with probability $1-\frac{\varepsilon}{4}$ over sampling $y_{2, \Samp}\sim\Y_{2, \Samp}$ (conditioned on $E_1$ and $E_{1\to 2}$), we have that $\Y_{2, \Cond}(y_{2, \Samp})$ is a \oNOSF[g=\gamma D_{2, \Cond},D_{2, \Cond},n-n_2'].
We refer to such a $y_{2, \Samp}$ as `good.'

In our third and final step,  we use \cref{lem:condenser from funky two source extractor} to condense from $\X_1$ and $\Y_{2, \Cond}(y_{2, \Samp})$. We condition on events $E_1$ and $E_{1\to2}$ here.
We also pay additional $\eps/4$ in error and assume that all $y_{2, \Samp}$ are good, i.e., the sampler always succeeds. This brings the total error we have incurred so far to $3\eps/4$.
Note that since we used $y_{2, \Samp}\sim\Y_{2, \Samp}$ to obtain $\Y_{2, \Cond}$, any fixing of the output of $\Y_{2, \Samp} = y_{2, \Samp}$ can create correlations between $\X_1$ and $\Y_{2, \Cond}$ and it may not even preserve the structure of $\Y_{2, \Cond}$.
Formally for any fixed $y_{2, \Samp}$, conditioned on $\Y_{2, \Samp} = y_{2, \Samp}$, 1) it is not necessarily true that $\Y_{2, \Cond}$ still remains an \oNOSF, and 2) the good bits in $\Y_{2, \Cond}$ may not necessarily be independent of $\X_1$.
We address these concerns by using \cref{lem:control-few-bits-can-still-condense-again} and paying with more error and more entropy gap at the end. 

Let $\Opt$ (short for optimistic) be the assumption that all the bits (including the bad ones) in $\Y_{2, \Samp}$ were truly uniform and independent of $\X_1$ and independent of all length $n - n_2'$ suffixes of the bits of good players in $\X_2$ (these bits in the suffixes are the ones that potentially can be used to form $\Y_{2, \Cond}$ above).
We use this to assume we do meet the preconditions of \cref{lem:condenser from funky two source extractor}. Let $\Actual$ be the realistic scenario where the above does not happen and $\Y_{2, \Samp}$ is allowed to have bad bits.

Let $\eps_{\Cond} = 2^{-C_6 (\log(\ell/\eps))^3}$ where $C_6$ is a large universal constant.
We then instantiate \cref{lem:condenser from funky two source extractor} with $n_x = \ell' n$, $n_y = \left(\frac{\log(\ell)}{\eps}\right)^{C'}\log(n)$, $t = D_3 \le C_6\log(\log(\ell)/\varepsilon)$, $k = k_1$ (from \cref{eq: condensing 51 with polylog k_1}), and error equal to $\eps_{\Cond}$ where $C'$ is a large enough universal constant. 
Given these parameters and our assumption that $n \ge (\log(\ell)/\varepsilon)^{C}$ for a universal constant $C$ (that is large enough), we indeed satisfy the requirements of \cref{lem:condenser from funky two source extractor} under $\Opt$. 
Consequently, we obtain the function $\TCond:(\zo^{n})^{\ell'}\times(\zo^{n})^{D_{2, \Cond}}\to\zo^{m_{\TCond}}$ and our final output will be $\TCond(\X_1, \Y_{2, \Cond})$. 

Let $\ZZ_{\Opt} = \TCond(\X_1, \Y_{2, \Cond})$ be the distribution under the assumption $\Opt$.
Let $\ZZ_{\Actual}$ be the actual output distribution that we obtain. i.e., $\ZZ_{\Actual}:= \Cond(\X)$, the output distribution of our condenser.

We first analyze the guarantees of this condenser under the assumption $\Opt$.
We first observe that 
\begin{align*}
(C_{\TCond})^t\log(tn_x/\varepsilon_{\Cond})
& \le (C_{\TCond})^{C_6(\log(\log(\ell)/\varepsilon))}\log(C_6(\log(\log(\ell)/\varepsilon))\cdot \ell n / \varepsilon_{\Cond})\\
& = (C_{\TCond})^{C_6(\log(\log(\ell)/\varepsilon))}\log(C_6(\log(\log(\ell)/\varepsilon))\cdot \ell n\cdot 2^{C_6 (\log(\ell / \eps))^3})\\
& \le \left(\frac{\log(\ell)}{\varepsilon}\right)^{C_7}\log(n) & \tag{$\ast$} \label{eq: condensing 51 with polylog *}
\end{align*}  
where $C_7$ is a large enough universal constant.

Since $n \ge (\log(\ell)/\varepsilon)^{C}$ for a very large universal constant $C$, our parameter setting indeed satisfies all the requirements of \cref{lem:condenser from funky two source extractor} (and the independence based requirements hold because we are arguing under the assumption $\ZZ_{\Opt}$.

With this, \cref{lem:condenser from funky two source extractor} provides us with the following guarantee on the length $m_z$ of our final output (recall that $k_1$ from \cref{eq: condensing 51 with polylog k_1} is the entropy of $\X_1$ conditioned on event $E_1$ occurring):
\begin{align*}
m_z
& = \frac{1}{3}\left(k_1 - (C_{\TCond})^t\log(tn_x/\varepsilon_{\Cond})\right)\\
& = \frac{1}{3}\left(2\gamma \ell'n - C_3\ell'\log(\ell')\log(1 / \eps) - (C_{\TCond})^t\log(tn_x/\varepsilon_{\Cond})\right) & \textrm{(by \cref{eq: condensing 51 with polylog k_1})}\\
& \ge \frac{1}{3}\left(2\gamma \ell'n - C_3\ell'\log(\ell')\log(1 / \eps) - \left(\frac{\log(\ell)}{\varepsilon}\right)^{C_7}\log(n)\right) & \textrm{(by \cref{eq: condensing 51 with polylog *})}\\
& = \frac{1}{3}\left(\gamma \ell n - C_3(\ell / 2)\log(\ell / 2)\log(1 / \eps) - \left(\frac{\log(\ell)}{\varepsilon}\right)^{C_7}\log(n)\right) & \textrm{(by definition of $\ell'$)}\\
&\ge \frac{\gamma \ell n}{3} - C_8\left(\ell\log(\ell)\log(1 / \eps)\right) - \left(\frac{\log(\ell)}{\eps}\right)^{C_7}\log(n)
\end{align*}
where $C_8$ is some large enough universal constant.

Then, \cref{lem:condenser from funky two source extractor} (under assumption $\Opt$) guarantees that
\begin{align*}
    \minH^{\eps_{\Cond}}(\ZZ_{\Opt})  
    &\geq m_z - (C_{\TCond})^t\log(tn_x/\varepsilon_{\Cond})\\
    &\geq m_z - \left(\frac{\log(\ell)}{\eps}\right)^{C_7}\log(n) & \textrm{(by \cref{eq: condensing 51 with polylog *})}
\end{align*}

We are not yet done since in reality, the assumption $\Opt$ does not hold and we need to argue under $\Actual$.
To handle this situation, instead of using $\Y_{2, \Samp}$ above, we consider a distribution $\X_{\Adv}\sim (\zo^{n})^{\ell}$ which is same as the distribution $\X$ but the $n_2'\cdot D_2$ many bits in $\Y_{2, \Samp}$ are instead controlled by an adversary ; we allow those bits to depend on any other bits from $\X$.
Let $\ZZ_{\Adv}$ be the resulting output distribution when we do this.
Since the adversary $\Adv$ is arbitrary, this adversarial assumption is stronger than $\Actual$ scenario, so it suffices to argue about $\ZZ_{\Adv}$.
To argue regarding $\ZZ_{\Adv}$, we apply \cref{lem:control-few-bits-can-still-condense-again} to infer that 
\begin{align*}
\minH^{\eps_{\Cond}\cdot 2^{n_2'\cdot D_2}}(\ZZ_{\Adv}) 
& \ge \minH^{\eps_{\Cond}}(\ZZ_{\Opt}) - n_2'\cdot D_2\\
& \ge m_z - \left(\frac{\log(\ell)}{\eps}\right)^{C_7}\log(n) - \left(C_2\cdot \log(\log(\ell) / \eps)\cdot \log(1 / \eps)\right)\cdot \left(C_1\log(\ell / \eps)\right) \\
& \textrm{(by \cref{eq: condensing 51 with polylog n_2'} and \cref{eq: condensing 51 with polylog D_2})}\\
& \ge m_z - \left(\frac{\log(\ell)}{\eps}\right)^{C_9}\log(n)
\end{align*}
where $C_9$ is a large enough universal constant.
We also see that
\begin{align*}
\eps_{\Cond}\cdot 2^{n_2'\cdot D_2}
& = 2^{-C_6(\log(\ell / \eps))^3}\cdot 2^{\left(C_2\cdot \log(\log(\ell) / \eps)\cdot \log(1 / \eps)\right)\cdot \left(C_1\log(\ell / \eps)\right)} & \textrm{(by \cref{eq: condensing 51 with polylog n_2'} and \cref{eq: condensing 51 with polylog D_2})}\\
& \le \eps / 4
\end{align*}
The last inequality follows since we will pick $C_6$ to be much larger than $C_1$ and $C_2$.

We note that $\X$, even after conditioning on $E_1, E_{1\to 2}$ can be assumed to be a flat source. This is because we can express $\X_1$ as a convex combination of sources with the same min-entropy and doing so retains the structure of our source $\X$. Hence, we do meet all the preconditions of \cref{lem:control-few-bits-can-still-condense-again}.

Lastly, since we incurred $3\eps/4$ error at the beginning of the fourth step, and we incur additional $\eps/4$ here, we obtain that our final output distribution will be $\eps$-close to having min-entropy $m_z - \left(\frac{\log(\ell)}{\eps}\right)^{C_9}\log(n)$.
By taking $C$ to be a large enough constant for our actual claimed parameters, we infer the claim.

\end{proof}

\subsection{\texorpdfstring{Constructing $\NOSFSamp$}{Constructing oNOSFSamp}}\label{subsec:Sampler from reduce and seeded extractor}

In this subsection, we will construct $\NOSFSamp$ and prove \cref{lem:Sampler from reduce and seeded extractor}.
Our construction of $\NOSFSamp$ itself requires two ingredients: (1) A $\Reduce$ function that reduces an \oNOSF of length $\ell$ to an $O(\log(\ell))$ length source, and (2) a good averaging sampler with linear sample complexity from \cref{lem:zuckerman linear seeded extractor}. 

Let's formally define this $\Reduce$ function:

\begin{lemma} 
\label{lem:RZ reduce}
    There exists a universal constant $C$ such that the following holds.
    For all $0 < \gamma\le 1, 0 < \eps < 1/2$ and all $\ell, n\in \N$ such that $n\geq 6\log(\ell)\log(1/\varepsilon)/\gamma$, there exists an explicit function $\Reduce: (\zo^n)^{\ell} \to \zo^t$ such that for all \oNOSFs[\gamma \ell, \ell] $\X$, we have that $\Reduce(\X)$ is a $(t, k)$-source where $t\leq C\log(\ell/\varepsilon)$ and $k\geq 3\log(\ell/\varepsilon)$.
\end{lemma}

We construct this function in \cref{subsubsec:RZ reduce}. Let's see how using it we can construct $\NOSFSamp$.

\begin{proof}[Proof of \cref{lem:Sampler from reduce and seeded extractor}]
    We use the given parameters $\gamma,n,\ell$ to instantiate $\Reduce:[N]^\ell\to\zo^t$ from \cref{lem:RZ reduce} with $\varepsilon=\varepsilon_s$. This gives us a constant $C_0$ such that $\Reduce(\X)$ is a $(t\leq C_0\log(\ell/\varepsilon_s),k\geq6\log(\ell/\varepsilon_s))$-source. We then let $\delta=\frac{k}{2t}=\frac{3}{C_0}$ and $\alpha=\frac{1}{3}$ which we use to instantiate \cref{lem:zuckerman linear seeded extractor} with $\varepsilon=\varepsilon_a$ to get $\Samp:\zo^t\to(\zo^m)^D$.

    Define $\NOSFSamp(\X)=\Samp(\Reduce(X)):(\zo^n)^\ell\to(\zo^m)^D$. Since $(1-\alpha)\delta t=\frac{2}{3}\cdot\frac{3}{2}\log(\ell/\varepsilon_s)\geq\log(\ell)$, \cref{lem:zuckerman linear seeded extractor} allows us to take $m=\log(\ell)$. Moreover, because $k-\delta t=k/2\geq 3\log(\ell/\varepsilon_s)\geq\log(1/\varepsilon_s)$, we have that $2^{\delta t-k}\leq\varepsilon_s$, giving us the desired error bound. Finally, \cref{lem:zuckerman linear seeded extractor} also gives us that $D=O(t)=O(\log(\ell/\varepsilon_s))$, as claimed.
\end{proof}

\subsubsection{Constructing $\Reduce$}\label{subsubsec:RZ reduce}

We here construct $\Reduce$ function as required by \cref{lem:RZ reduce}. Our construction is based on the construction from \cite{russellzuckerman01} that utilizes hitting sets for combinatorial rectangles. We call their general constructed function as $\Reduce'$. 

This function $\Reduce'$ has the following guarantee:
\begin{lemma}[\cite{russellzuckerman01}]\label{lem:RZ reduce base}
There exists a universal constant $C'$  such that for any $\gamma>0$ and $a,d\in\N$, there exists an efficient function $\Reduce':[a]^d\to\zo^t$ such that for any \oNOSF[g=\gamma d,d,\log(a)] $\X$, we have that $\Reduce'(\X)$ is a $(t,k)$-source with $t\leq C'(\log(a)+\log\log(d)+d/a)$ and $k\geq \gamma d/a$.
\end{lemma}

We construct our desired function $\Reduce'$ in \cref{subsubsec:RZ reduce base}.
Let's see first how by carefully choosing $a$ and $d$ in \cref{lem:RZ reduce base}, we get the $\Reduce$ function we require.
\begin{proof}[Proof of \cref{lem:RZ reduce}]
    We will consider two cases for the parameters of our \oNOSF and apply \cref{lem:RZ reduce base} with different parameters in each case.
    Recall that, in \cref{lem:RZ reduce base}, $d$ represents the number of blocks in our \oNOSF and $\log(a)$ represents the number of bits in each block. However, our given \oNOSF[g=\gamma\ell,\ell, n] $\X$ is on $[N]^\ell$, so we must make these parameters match. We take cases on the relative size of $\log(\ell)$ and $\log(1/\varepsilon)$.

    \paragraph{Case 1}
    First, if $\log(1/\varepsilon)\leq\log(\ell)$, then let $d=\ell$ and $a=\frac{\gamma\ell}{6\log(\ell)}$ in \cref{lem:RZ reduce base}. In this setting, since we are guaranteed that $n\geq6\log(\ell)\log(1/\varepsilon)/\gamma$, so $N\geq\ell>a$, we can simply truncate each block to $\log(a)$ bits and take $\X$ to be a \oNOSF[g=\gamma\ell,\ell] on $[a]^d$. \cref{lem:RZ reduce base} gives us that there exists some $C'$ such that
    \begin{align*}
        t&\leq C'(\log(a)+\log\log(d)+d/a)\\
        &= C'(\log(\gamma)+\log(\ell)-\log(6)-\log\log(\ell)+\log\log(\ell)+6\log(\ell)/\gamma)\\
        &\leq C_2\log(\ell)\\
        &\leq C_2\log(\ell/\varepsilon)
    \end{align*}
    for some constant $C_2$. Then, we compute the min-entropy of $\Reduce(\X)$ as
    \begin{align*}
        k&\geq\gamma d/a\\
        &=\gamma \ell\cdot 6\log(\ell)/(\gamma\ell))\\
        &=6\log(\ell).
    \end{align*}Recall that the assumption in this case is that $\log(1/\varepsilon)\leq\log(\ell)$, which we can rearrange into $\log(\ell)\geq\log(\ell/\varepsilon)/2$. Applying this yields that $k\geq 6\log(\ell)\geq3\log(\ell/\varepsilon)$, as desired.

    \paragraph{Case 2}
    Second, if $\log(1/\varepsilon)>\log(\ell)$, then let $d=6\ell\log(1/\varepsilon)/\gamma$ and $a=\ell$ in \cref{lem:RZ reduce base}. In order to convert $\X$ to a source over $[a]^d$, we split each $n$ length block of $\X$ into length $n'=\frac{\gamma n}{6\log(1/\varepsilon)}$ blocks. This gives us $\ell'=\ell\log(1/\varepsilon)\cdot\frac{6}{\gamma}=d$ total blocks with $g'=\gamma\ell'$ total good blocks. Thus, we now view $\X$ as a source $\X'$ over $[N']^{\ell'}$ where $N' = 2^{n'}$. To finish the conversion, we recall that $n\geq6\log(\ell)\log(1/\varepsilon)/\gamma$, so $n'\geq\log(\ell)=\log(a)$, allowing us to just take a length $\log(a)$ prefix of each length $n'$ block to create a new source $\X''$ over $[a]^d$, as required. Finally, we can analyze $t$ and $k$ in this setting. We begin with $t$ using \cref{lem:RZ reduce base} to infer that there exists some $C'$ such that
    \begin{align*}
        t&\leq C'(\log(a)+\log\log(d)+d/a)\\
        &= C'(\log(\ell)+\log\log(6\ell\log(1/\varepsilon)/\gamma)+6\log(1/\varepsilon)/\gamma)\\
        &\leq C_3\log(\ell/\varepsilon)
    \end{align*}
    for some constant $C_3$. 
    We compute $k$ as
    \begin{align*}
        k&\geq\gamma d/a\\
        &=\gamma(6\ell\log(1/\varepsilon)/\gamma)/\ell\\
        &=6\log(1/\varepsilon).
    \end{align*}
    showing that $k\geq 6\log(1/\varepsilon)$. Finally, recall that in this case $\log(1/\varepsilon)>\log(\ell)$, so $\log(1/\varepsilon)>\log(\ell/\varepsilon)/2$, which we can apply to get that $k\geq6\log(1/\varepsilon)>3\log(\ell/\varepsilon)$.

    Let $C=\max(C_2,C_3)$. In either case, we have that the number of output bits is $t\leq C(\log(\ell/\varepsilon))$ and the min-entropy $k$ of $\Reduce(\X)$ is $\geq 3\log(\ell/\varepsilon)$, as claimed.
\end{proof}

\subsection{\texorpdfstring{Constructing $\TCond$}{Constructing 2Cond}}\label{subsec:funky two source extractor exists}
In this subsection we will prove our remaining helper lemma - \cref{lem:condenser from funky two source extractor}. First, we will require the following result:


\begin{lemma}\label{lem:funky two source extractor exists}
There exists universal constant $C'$ such that for all $n_x, k_x, n_{y,1}, \dots, n_{y,t}, m, 0 < \eps_1\le \dots \le \eps_t < 1$ satisfying $n_{y, i} \ge C'\log(2n_x / \eps_i)$ and $m = \frac{k_x - \log(2/\eps_1)}{3}$, the following holds:
There exists an explicit extractor $\Ext: \zo^{n_x}\times \zo^{n_{y, i}} \times \dots \times \zo^{n_{y, t}}\to \zo^m$ satisfying:
For all $1\le j\le t$ and all independent sources $\X\sim \zo^{n_x}, \Y_1\sim \zo^{n_{y,1}}, \dots, \Y_t\sim \zo^{n_{y,t}}$ where $\minH(\X) = k_x$, each of $\Y_1, \dots, \Y_{j-1}$ are fixed constants and all $\Y_j, \dots, \Y_t$ are uniform, we have that $\Ext(\X, \Y_1, \dots, \Y_t)$ is $\eps_j$-close to $\U_m$.
\end{lemma}
\begin{proof}[Proof of \cref{lem:funky two source extractor exists}]
For $1\le i\le t$, let $\sExt_{i}: \zo^{n_x} \times \zo^{n_{y, i}} \to \zo^m$ be the explicit $(k_x,\eps_i/2)$-seeded-extractor guaranteed by \cref{thm:GUV extractor}---where we assume the universal constant from \cref{thm:GUV extractor} is $C'$ and check that our parameters meet the requirements.
Our extractor construction is:
\[
\Ext(x, y_1, \dots, y_t) = \bigoplus_{i=1}^t \sExt_{i}(x, y_i).
\]

Let $\ZZ_{good} = \sExt_j(\X, \Y_j)$ and let $\ZZ_{rest} = \bigoplus_{1\le i\le t, i\ne j} \sExt_{i}(\X, \Y_i)$.
Notice that our final output distribution is $\ZZ_{good}\oplus \ZZ_{rest}$.
We will argue that on most fixings of $\ZZ_{rest}$, the output will be  close to uniform.

By \cref{lem:min-entropy-chain-rule}, we have the following (where the probability below is over sampling from $\ZZ_{rest}$)
\[
\Pr[\minH(\X | \ZZ_{rest} = z_{rest}) \ge k_x - m - \log(2/\eps_j)] \ge 1 - \eps_j / 2.
\]
Call the fixings $z_{rest}$ of $\ZZ_{rest}$ that satisfy the above property of leaving $\X$ with a lot of entropy when conditioning on them, as the ``good fixings.''
As $\ZZ_{rest}$ is independent of $\Y_j$ and $\X$ is left with a lot of entropy conditioning on a good fixing $z_{rest}$, we have that 
\[
\sExt_{j}((\X | \ZZ_{rest} = z_{rest}), (\Y_j | \ZZ_{rest} = z_{rest})) \approx_{\eps_j/2} \U_m.
\]
As $1 - \eps_j/2$ fraction of fixings of $\ZZ_{rest}$ are good, we conclude that $\Ext(\X, \Y_1, \dots, \Y_{t}) \approx_{\eps_j} \U_m$ as desired.
\end{proof}
With this, we finally proved the proof of our lemma:
\begin{proof}[Proof of \cref{lem:condenser from funky two source extractor}]
Let $C'$ be a universal constant that we set later.
For $1\le i\le t$, let $n_{z, i} = 2C'(3C')^{t-i}\log(2t n_x/\eps)$ and let $n_z = \sum_{i=1}^{t} n_{z, i}$.
Let $\ZZ_i$ be the length $n_{z, i}$ prefix of the block $\Y_i$, and let $\ZZ=\ZZ_1,\dots,\ZZ_t$ be the concatenation of these prefixes. Note that by our lower bound guarantee on $n_y$, each block is long enough to take such prefixes.
We use the extractor $\Ext$ from \cref{lem:funky two source extractor exists} with $k_x = k - n_z - \log(2/\eps)$, $m=\frac{1}{3}(k-(3C')^t\log(2tn_x/\varepsilon))$ as in the lemma statement, and for $1\le i\le t$, we set $\eps_i = \left(\frac{\eps}{2t n_x}\right)^{(3C')^{t-i}}$.
With this, we define our condenser as:
\[
\TCond(\X,\Y) =\Ext(\X,\ZZ)= \Ext(\X, \ZZ_1,\dots,\ZZ_t).
\]

We easily compute and check that our parameter settings satisfy the requirements of \cref{lem:funky two source extractor exists}. We will show that the output entropy (with error $\eps$) is at least $m - n_z$. We compute that $n_z \le (3C')^{t}\log(2t n_x/\eps)$, the output entropy gap. Hence if we show this, then our condenser will indeed have the claimed property.

We now show that our condenser construction is correct. Since $\Y$ is guaranteed to have at least one good block by assumption, let $j\in [t]$ be the index of this good block. Now, let $\A = \ZZ_1, \dots, \ZZ_{j-1}$ and let $\B = \ZZ_{j+1}, \dots, \ZZ_{t}$ so that $\ZZ = (\A, \ZZ_j, \B)$.
We will show that $\sminH(\TCond(\X,\Y)) \ge m - n_z$.

We will now consider fixings of $\A$.
We say a fixing of $\A = a$ is good if $\minH(\X | \A = a) \ge k - n_z - \log(2/\eps) = k_x$.
By the min-entropy chain rule (\cref{lem:min-entropy-chain-rule}), at least $1-\eps/2$ fraction of fixings of $\A$ are good.
Since $\Y$ is an \oNOSF, $\ZZ_j$ remains independent and uniform of $\X$ for every fixing of $\A$.

We will show that, conditioned on a good fixing $a$ of $\A$, we have $\minH^{\eps/2}(\TCond(\X,\Y)) \ge m - \sum_{i=j+1}^t n_{z, i} \ge m - n_z$.
This will prove our result as our total error will be $\eps/2+\eps/2=\eps$ and the min-entropy guarantee will be $m - n_z$, as desired.

Consider the best case scenario when $(\B | \A = a) = \U_{|\B|}$. This is unrealistic since it is possible that all bits in $\B$ are bad and arbitrarily depend on the remaining bits. Nevertheless, it is instructive to see what happens in this scenario.
In this case, $\X, \Y$ are independent distributions, and we can infer that $\TCond(\X,\Y)=\Ext(\X, \ZZ)\approx_{\eps_j} \U_m$. 
However, as alluded before, all bits in $\B$ can be adversarially set. To overcome this, we invoke \cref{lem:control-few-bits-can-still-condense-again} that allows us to compare how worse off our output distribution can be compared to the best case scenario. We conclude that even when $\B$ is completely adversarially controlled, $\minH^{\epspr}(\TCond(\X, \Y)) \ge m - \abs{\B} = m - \sum_{i=j+1}^t n_{z, i}$ where
\begin{align*}
    \epspr &= \eps_{j}\cdot 2^{\abs{\B}}\\
           &= \left(\frac{\eps}{2t n_x}\right)^{(3C')^{t-j}}\cdot 2^{\sum_{i=j+1}^{t} n_{z, i}}\\
           &= \left(\frac{\eps}{2t n_x}\right)^{(3C')^{t-j}}\cdot 2^{2C'\log(2t n_x/\eps)\sum_{i=j+1}^{t} (3C')^{t-i}}\\
           &= \left(\frac{\eps}{2t n_x}\right)^{(3C')^{t-j}}\cdot \left(\frac{2t n_x}{\eps}\right)^{2C' \frac{(3C')^{t-j}-1}{3C'-1}}\\
           &\le \left(\frac{\eps}{2t n_x}\right)^{(3C')^{t-j}}\cdot \left(\frac{2t n_x}{\eps}\right)^{(3C')^{t-j}-1}\\
           &\le \frac{\eps}{2t n_x}\\
           &\le \eps/2\\
\end{align*}
This proves our claim, showing that for all good fixings, our output is highly condensed.
We set our final universal constant $C$ to be $4\cdot C'$ and see that doing so only weakens the promise of our condenser.

We also need to be careful when invoking \cref{lem:control-few-bits-can-still-condense-again} since it requires that $(\X, \A, \ZZ_j, \U_{|\B|})$ should be a flat distribution. While that may not be true, we can express $\X$ as a convex combination of flat sources with the same min-entropy and since $\A$ is fixed and $\ZZ_j$ and $\U_{|\B|}$ are independent and uniform, we can express the joint distribution as a convex combination of flat sources, for each of them invoke the lemma, and conclude that the original distribution will be condensed as well.
\end{proof}

\dobib

\section{Transforming Low-Entropy oNOSF Sources to Uniform oNOSF Sources}\label{sec:transformation}

In this section, we show how to transform low-entropy \oNOSFs into \unioNOSFs.
Such a transformation was also provided in \cite{CGR_seedless_condensers}. Here, we obtain improved bounds using a generalized construction that allows us to obtain better tradeoffs and parameters in many more regimes of $n,\ell$. Our main theorem is:

\begin{theorem}\label{lem:existential low-entropy transformation}
Let $d, g, g_{out}, \ell, n, m, k, \eps$ be such that 
$ g_{out} \le g - \frac{\ell-g+2}{d},
n\ge k \ge \log(nd-k) + md + 2\log(2g_{out}/\eps)+O(1)
$.
Then, there exists a function $f: (\zo^n)^{\ell}\to (\zo^m)^{\ell-1}$ such that for any \oNOSF[g, \ell, n, k] $\X$, there exists \unioNOSF[g_{out}, \ell-1, m] $\Y$ for which $\abs{f(\X) - \Y} \le \eps$.
\end{theorem}

The flexibility of setting $d$ to any desired value allows us to obtain stronger results. 
For instance by setting $d$ to be a large constant, we can get the following transformation that works even when $n$ is very small compared to $\ell$:

\begin{corollary}[Transformation for small $n$]\label{cor: small n low entropy transformation}
Let $g, \ell, n, m, k, \eps, \delta$ be such that 
$
\delta \le 0.99,
g = \delta \ell,
n = \poly(\log(\delta \ell/\eps)),
k = \Omega(\log(\delta \ell/\eps)),
m = \Omega(k)
$.
Then, we can construct a function $f: (\zo^n)^{\ell}\to (\zo^m)^{\ell-1}$ such that: for any \oNOSF[g, \ell, n, k] $\X$, there exists \unioNOSF[0.99\delta\ell, \ell-1, m] $\Y$ such that $\abs{f(\X) - \Y} \le \eps$.
\end{corollary}

We additionally note that when we set $d = \ell$, we recover the same construction as in \cite{CGR_seedless_condensers}, matching its parameters. This is most interesting in the regime when say $\ell = O(1)$ and $n$ is arbitrarily growing.

\begin{corollary}[similar parameters as Theorem 5.2 from \cite{CGR_seedless_condensers}]
Let $g, \ell, n, m, k, \eps$ be such that 
$
k \ge 1.01(\log(n\ell) + 2\log(2(g-1)/\eps)),
m = k/200\ell
$.
Then, we can construct a function $f: (\zo^n)^{\ell}\to (\zo^m)^{\ell-1}$ such that for any \oNOSF[g, \ell, n, k] $\X$, there exists \unioNOSF[g-1, \ell-1, m] $\Y$ such that $\abs{f(\X) - \Y} \le \eps$.
\end{corollary}

To obtain these transformations, we will use two-source extractors.
In fact, using explicit construction of two-source-extractors, we also obtain an explicit transformation:

\begin{corollary} [Explicit Transformation]\label{lem:explicit low-entropy transformation}
There exists a universal constant $C$ such that for all $d, g, g_{out}, \ell, n, m, k, \eps$ satisfying
$
g_{out} \le g - \frac{\ell-g+2}{d},
k\ge \poly(\log(n)) + md + 2\log(2g_{out}/\eps) + O(1),
m\le \poly(\log n),
\eps\ge n^{-\Omega(1)} / 2g_{out}
$.
the following holds:
There exists an explicit function $f: (\zo^n)^{\ell}\to (\zo^m)^{\ell-1}$ such that for any \oNOSF[g, \ell, n, k] $\X$, there exists \unioNOSF[g_{out}, \ell-1, m] $\Y$ for which $\abs{f(\X) - \Y} \le \eps$.
\end{corollary}

We can instantiate this lemma even in the case of constant $d$ and get an explicit transformation similar to \cref{cor: small n low entropy transformation} with fewer output bits per block.

We will use the following main technical lemma that shows how to use two-source extractors to obtain these transformations:

\begin{lemma}[Main Lemma]\label{lem:two-source-extractor to low-entropy transformation}
Let $d, g, g_{out}, \ell, n, m, k_{2\Ext}, k, \eps_{2\Ext}$ be such that $k \ge k_{2\Ext} + m\cdot d + \log(1/\eps_{2\Ext}), g_{out} \le \frac{g(d+1)-\ell-2}{d}$.
Let $2\Ext: \zo^{d\cdot n}\times \zo^{n} \to \zo^m$ be an $(k_{2\Ext},k_{2\Ext}, \eps_{2\Ext})$-average-case-strong two-source extractor.\footnote{As defined in \Cref{def:avg case strong two source extractor}.}
Then, we can construct a function $f: (\zo^n)^{\ell}\to (\zo^m)^{\ell-1}$ such that for any \oNOSF[g, \ell, n, k] $\X$, there exists \oNOSF[g_{out}, \ell-1, m] $\Y$ such that $\abs{f(\X) - \Y} \le \eps$ where $\eps = 2g_{out}\cdot \eps_{2\Ext}$.
\end{lemma}

Existentially, two-source-extractors with following parameters exist:

\begin{lemma}[Lemma 5.4 from \cite{CGR_seedless_condensers}]\label{lem:amazing two-source-extractors exist}
Let $n_1, n_2, k_1, k_2, m, \eps$ be such that $k_1 \le n_1, k_2\le n_2, m = k_1 + k_2 - 2\log(1/\eps) - O(1)$, $k_2 \ge \log(n_1 - k_1) + 2\log(1/\eps) + O(1)$, and $k_1 \ge \log(n_2 - k_2) + 2\log(1/\eps) + O(1)$.
Then, a random function $2\Ext:\zo^{n_1}\times \zo^{n_2}\to \zo^m$ is a $(k_1, k_2, \eps)$-two source extractor with probability $1 - o(1)$.
\end{lemma} 

Using this, our main result follows:

\begin{proof}[Proof of \cref{lem:existential low-entropy transformation}]
We use the two-source-extractors from \cref{lem:amazing two-source-extractors exist} and apply it in \cref{lem:two-source-extractor to low-entropy transformation}.
\end{proof}

To make this transformation explicit, we can use the following construction of a two-source-extractor:
\begin{theorem}[\cite{chattopadhyay_explicit_2019, meka_explicit_2017, li_improved_2016}]\label{thm:explicit two-source-extractors}
There exists a universal constant $C\ge 1$ such that for all $n, k, m, \eps$ with $k\ge \log^C(n), m\le n^{1/C}, \eps \ge n^{-1/C}$, the following holds:
There exists an explicit $(k, k, \eps)$ two-source-extractor $2\Ext: \zo^n\times \zo^n\to \zo^m$.
\end{theorem}

With this our explicit transformation follows:

\begin{proof}[Proof of \cref{lem:explicit low-entropy transformation}]
We use the explicit two-source-extractors from \cref{thm:explicit two-source-extractors} and apply it in \cref{lem:two-source-extractor to low-entropy transformation}.
\end{proof}

\subsection{Low-Entropy oNOSF Source  to Uniform Using Two-Source-Extractors}

In this subsection, we will prove \cref{lem:two-source-extractor to low-entropy transformation}. To do this, we will use two-source-extractors and average-case two-source-extractors. Let's first define them:

\begin{definition}\label{def:avg case strong two source extractor}
We say that $2\Ext$ is $(k_1, k_2, \eps)$ \emph{average-case} strong if 
\[
2\Ext(\X_1,\X_2),\W \approx_\varepsilon \U_m,\W
\]
for every $\X_1$ and $\W$ such that $\avgcondminH(\X_1\mid\W)\geq k_1$ with $\X_2$ independent of $\X_1$ and $\minH(\X_2)\ge k_2$ and $\W$.
\end{definition}

Here we use the notion of average conditional min-entropy from \cite{dodis_fuzzy_2008}.
\begin{definition}
    For two distributions $\mathbf{A},\mathbf{B}$ over $\Omega$, the \emph{average conditional min-entropy} of $\mathbf{A}$ given $\mathbf{B}$ is
    \begin{align*}
        \avgcondminH(\mathbf{A}\mid\mathbf{B})&=\log\left(\E_{b\sim\mathbf{B}}\left[\max_a\Pr[\mathbf{A}=a\mid\mathbf{B}=b]\right]\right)\\
        &=-\log\left(\E_{b\sim\mathbf{B}}\left[2^{-\minH(\mathbf{A}\mid\mathbf{B}=b)}\right]\right).
    \end{align*}
\end{definition}

This notion of average-case two-source-extractors allows us obtain a simpler chain rule:
\begin{lemma}\cite{dodis_fuzzy_2008}\label{lem:avg cond min H chain rule}
    Let $\mathbf{A}$, $\mathbf{B}$, and $\mathbf{C}$ be distributions such that $\Supp(\mathbf{B})\leq 2^\lambda$. Then $\avgcondminH(\mathbf{A}\mid \mathbf{B},\mathbf{C})\geq\avgcondminH(\mathbf{A},\mathbf{B}\mid\mathbf{C})-\lambda\geq\avgcondminH(\mathbf{A}\mid\mathbf{C})-\lambda$.
\end{lemma}

Lemma 2.3 of \cite{dodis_fuzzy_2008} shows that all two-source extractors are average-case-two-source extractors with similar parameters.
\begin{lemma}\cite{dodis_fuzzy_2008}\label{lem:strong two source is average case strong}
    For any $\eta>0$, if $2\Ext$ is a $(k_1,k_2,\varepsilon)$-two-source extractor, then $2\Ext$ is a $(k_1+\log(1/\eta),k_2,\varepsilon+\eta))$-average-case-two-source extractor.
\end{lemma}

With this, we will finally prove our main lemma that shows how to use two-source-extractors to obtain our transformation:

\begin{proof}[Proof of \cref{lem:two-source-extractor to low-entropy transformation}]
For $-d\le i\le 0$, define $\X_i$ to be the random variable that always outputs $0^n$.
For $2\le i\le \ell$, we output $\OO_i = 2\Ext(\X_{i-d}\circ \dots \circ \X_{i-1}, \X_i)$.

For $2\le i\le \ell$, we say that $\OO_i$ is good if (1) $\X_i$ is good and (2) there exists a block amongst $\X_{i-d}, \dots, \X_{i-1}$ that is good.
We observe that if $\OO_i$ is good, then $\abs{\OO_i - \U_m}\le \eps_{2\Ext}$.
Let $g'$ be the number of such good $\OO_i$.
Let $j_1, \dots, j_g$ be the indices of the good blocks in $\X$.
For $1\le i\le g-1$, let $d_i = j_{i+1} - j_i$.
We observe that $g'$ equals number of $i$ such that $d_i\le d$.
As $\sum_{i=1}^{g-1} d_i \le \ell$ and $d_i \ge 1$, we infer that $g' \ge \frac{(g-1)(d+1) - \ell}{d}$.
Hence, as long as $g_{out} \le \ceil{g'}$, we can guarantee the desired number of good blocks in the output.
This holds as long as $g_{out} \le \frac{g(d+1)-\ell-2}{d}$.

Using \cref{lem:strong two source is average case strong}, we infer that $2\Ext$ is a $(k_{2\Ext} + \log(1/\eps_{2\Ext}),k_{2\Ext}, 2\eps_{2\Ext})$-average-case-two-source extractor. We will use this property below.

Now, using a hybrid argument we will show that 
\[
(\OO_2, \dots, \OO_{\ell}) \approx_{2g_{out}\cdot \eps_{2\Ext}} (\Y_2, \dots, \Y_{\ell}) 
\]
where $\Y = (\Y_2, \dots, \Y_{\ell})$ is a \unioNOSF[g_{out}, \ell, m] that we will define as the proof goes.
Let $\Y^{(1)} = (\OO_2, \dots, \OO_{\ell})$ and for $2\le i\le \ell$, let $\Y^{(i)} = (\OO_2, \dots, \OO_{\ell-i+1}, \Y_{\ell-i+2}, \dots, \Y_{\ell})$. Hence, $\Y^{(\ell)} = \Y$.
We proceed by induction. We will show that for $2\le i\le \ell$, 
\[
\abs{\Y^{(i)} - \Y^{(i-1)}}\le 2\eps_{2\Ext}
\]
whenever $\OO_i$ is good and
\[
\Y^{(i)} = \Y^{(i-1)}
\]
whenever $\OO_i$ is bad.
By repeated applications of the triangle inequality, we will have shown that our output is indeed close to some $\unioNOSF$ with desired parameters.

We proceed by induction and let $i\ge 2$ be arbitrary.
If $\OO_i$ is bad, then we let $\Y_{i} = \OO_i$.
Then, we indeed have that $\Y^{(i)} = \Y^{(i-1)}$ as desired.
Otherwise, we assume $\OO_i$ is good.
Then, it must be that $\X_i$ is good.
Let $i_{prev}$ be the index of the good block before $\X_i$ in $\X$.
Then, we know that $i - i_{prev}\le d$.
We first claim that 
\[
    \avgcondminH(\X_{i_{prev}} | \OO_1, \dots, \OO_{i-1}) \ge k_{2\Ext} = k - m\cdot d
\]
Firstly, by construction, blocks $\OO_{2},\dots, \OO_{i_{prev}-1}$ are functions of blocks $\X_1, \dots, \X_{i_{prev}-1}$. As $\X_{i_{prev}}$ is independent of $\X_1, \dots, \X_{i_{prev}-1}$, we infer that 
$\X_{i_{prev}}$ is independent of $\OO_{2}, \dots,\OO_{i_{prev}-1}$.
As $2\Ext$ is average-case-strong, we apply \cref{lem:avg cond min H chain rule} to get that
\[
    \avgcondminH(\X_{i_{prev}} | \OO_2, \dots, \OO_{i-1}) \ge k - m\cdot (i - i_{prev}) \ge k - m\cdot d = k_{2\Ext} + \log(1/\eps)
\]
where for the last inequality, we used the fact that $i - i_{prev}\le d$.
Moreover, as $\X_i$ is independent of $\X_1, \dots, \X_{i-1}$ and $\OO_2, \dots, \OO_{i-1}$ are solely functions of $\X_1, \dots, \X_{i-1}$, we infer that $\X_i$ is independent of $\OO_2, \dots, \OO_{i-1}$.
Hence, conditioned on fixing $\OO_2, \dots, \OO_{i-1}$, $\OO_i$ will be $2\eps_{2\Ext}$ close to $\U_m$.
This implies $\Y^{(i-1)}\approx_{2\eps_{2\Ext}} \Y^{(i)}$ as desired.
This shows that a good block in $\Y$ is uniform conditioned on all previous blocks---i.e., it is independent of all the blocks before it. This shows all bad blocks can only depend on good blocks appearing before them and that good blocks are independent of each other. This implies $\Y$ is indeed a $\unioNOSF$ as desired.
\end{proof}

\dobib

\section{Existence of Condensers for All Values of \texorpdfstring{$\ell, n$}{l, n}}\label{section: existence all}



We will show that there exist condensers for \unioNOSFs[g, \ell, n] for almost all settings of $\ell, n$, provided $g > 0.5\ell$. 
Observe that a \unioNOSF[g, \ell, n] is also a \unioNOSF[g\cdot s, \ell\cdot s, n/s] by simply dividing up all blocks into $s$ parts.
This implies that as $n$ becomes smaller (relative to $\ell$), it gets harder to condense with the hardest case being $n = 1$. Our condenser will also be able to handle the case of $n = O(1)$ and $\ell$ arbitrarily growing:



\begin{theorem}[Simplified version of \cref{cor: large n existential uni onosf condenser}]\label{theorem: simplified large n existential uni onosf condenser}
For all $g, \ell, n, \eps, \delta$ where $g = 0.51\ell$, and $0.01\ell n \ge 2\log(\ell n/2\eps) + O(1)$, there exists a condenser $\Cond: (\zo^n)^{\ell} \to \zo^m$ such that for any \unioNOSF[g, \ell, n] $\X$, we have 
$\sminH(\Cond(\X)) \ge m - \Delta$ where 
$m = 0.005\ell n + 200(\ell + \log(\ell n / 2\eps)) + O(1)$ and 
$\Delta = 200(\ell + \log(\ell n / 2\eps)) + O(1)$.
\end{theorem}
Note that when $n$ is a large enough constant, $m\ge 100\Delta$ and hence, the output entropy rate is at least $0.99$.

In fact, we obtain a general result for all values of $n, \ell$ and when $g = 0.5\ell + e$ where $e\in \N$ is arbitrary. See \cref{lem: existential uni onosf condenser general} for the full tradeoff; to get slightly better parameters for small $n$, see \cref{cor: small n existential uni onosf condenser}.

We combine the above condenser for \unioNOSFs with the transformation for low-entropy \oNOSFs to \unioNOSFs from \cref{cor: small n low entropy transformation} to obtain the following condenser for low-entropy \oNOSFs:

\begin{corollary}
Let $g, \ell, n, m, k, \eps$ be such that 
$
g = 0.51\ell,
n = \poly(\log(\ell/\eps)),
k = \Omega(\log(\ell/\eps)),
m = \Omega(\ell \log(\ell/\eps))
$.
Then, we can construct condenser $\Cond: (\zo^n)^{\ell}\to \zo^m$ such that for any \oNOSF[g, \ell, n, k] $\X$, we have $\sminH(\Cond(\X)) \ge m - \Delta$ where $\Delta = O(\ell + \log(1/\eps))$.
\end{corollary}

\begin{remark}
Previous condensers from \cite{CGR_seedless_condensers} could only show that condensers exist for \unioNOSFs when $\ell = o(\log n)$. They relied on existence of low-error two source extractors equipped with an additional ``regularity'' property. Our constructions are much simpler, recover all their results with even better parameters, and work for all values of $n$ and $\ell$, including the hardest case of $n = O(1)$.
\end{remark}

We provide our general construction of condensers in \cref{subsec: construct condenser unionosf}. To do that, we will require another type of condenser for two \unioNOSFs where the bad bits of the second block are allowed to depend on the bits of the first block. We provide this construction in \cref{subsec: condenser for two uni onosfs}.

\subsection{Constructing Condensers for  Uniform  oNOSF Sources}\label{subsec: construct condenser unionosf}

In this subsection, we will construct the following general condenser for $\unioNOSFs$:

\begin{lemma}[General \unioNOSF condensing]\label{lem: existential uni onosf condenser general}
For all $g, \ell, n, \eps, e$ where $g \ge (\ell/2) + e$, and $en \ge 2\log(\ell n/2\eps) + O(1)$, there exists a condenser $\Cond: (\zo^n)^{\ell} \to \zo^m$ such that for any \unioNOSF[g, \ell, n] $\X$, we have $\sminH(\Cond(\X)) \ge m - \Delta$ where 
$m = \frac{en}{2} + (2\ell-e)\ceil{\frac{\log(\ell n / 2\eps) + O(1)}{e}} + \log(1/\eps) + O(1)$ and 
$\Delta = (2\ell-2e)\ceil{\frac{\log(\ell n / 2\eps) + O(1)}{e}} + \log(1/\eps) + O(1)$.
\end{lemma}

To do this, we will use a condenser for two distinct \unioNOSFs where one source can depend on the other:

\begin{lemma}\label{lem:general tradeoff length-good rate-block size condenser}
For all $g, \ell, n_x, n_y, \eps$ where $n_x\ge n_y$ and $gn_y \ge \log(\ell n_x/\eps) + O(1)$, there exists a condenser $\Cond: (\zo^{n_x})^{\ell} \times (\zo^{n_y})^{\ell} \to \zo^{m}$ such that:
For any \unioNOSF[g, \ell, n_x] $\X$ and \unioNOSF[g, \ell, n_y] $\Y$ with the additional property that bad blocks in $\Y$ can depend on $\X$ as well, we have that $\sminH(\Cond(\X, \Y)) \ge m - \Delta$ where
$m = g n_x + (2\ell - g) n_y + \log(1/\eps) + O(1)$ and $\Delta = (2\ell-2g)n_y + \log(1/\eps) + O(1)$.
\end{lemma}

We construct this condenser in \cref{subsec: condenser for two uni onosfs}. Using this, our main general condenser can be constructed as follows:

\begin{proof}[Proof of \cref{lem: existential uni onosf condenser general}]
We split each block in $\X$ into $2$ parts to obtain a \unioNOSF[2g, 2\ell, n/2]. We call this resultant source $\X$ as well since it is the same distribution, just viewed differently.
Let $\U = (\U_1, \dots, \U_{\ell})$ and where for $1\le i\le \ell$, $\U_i = \X_i$.
Let $\V = (\V_1,\dots, \V_{\ell})$ where for $1\le i\le \ell$, we define $\V_i$ to be prefix of length $n_v$ of $\X_{\ell+i}$ where $n_v = \ceil{\frac{\log(\ell n / 2\eps) + O(1)}{e}}$.

We observe that $\U$ is a \unioNOSF[e, \ell, n/2] and $\V$ is a \unioNOSF[e, \ell, n_v] where bad bits in $\V$ can depend on $\U$ and the good bits in both sources are independent.
We now define our condenser $\Cond$ to be the condenser from \cref{lem:general tradeoff length-good rate-block size condenser} applied to sources $\U, \V$.
Hence, we will have that $\sminH(\Cond(\U, \V)) \ge m - \Delta$ where $m = en/2 + (2\ell-e)n_y + \log(1/\eps) + O(1)$ and $\Delta = (2\ell-2e)n_y + \log(1/\eps) + O(1)$ as desired.
\end{proof}

Our first corollary will apply to the regime that is the hardest to condense from, namely when $n$ is very small compared to $\ell$, even when $n = O(1)$ and $\ell$ is arbitrarily growing:

\begin{corollary}[Small $n$]\label{cor: small n existential uni onosf condenser}
For all $g, \ell, n, \eps, \delta$ where $g \ge (0.5 + \delta)\ell, \eps\ge 2^{-\delta \ell + O(1)}$, and $n\le 2^{\delta\ell/2}$, there exists a condenser $\Cond: (\zo^n)^{\ell} \to \zo^m$ such that for any \unioNOSF[g, \ell, n] $\X$, we have $\sminH(\Cond(\X)) \ge m - \Delta$ where 
$m = \delta \ell n / 2 + (2-\delta)\ell + \log(1/\eps) + O(1)$ and 
$\Delta = (2-\delta)\ell + \log(1/\eps) + O(1)$.
\end{corollary}

\begin{proof}
With $e=\delta\ell$, we observe that $\ceil{\frac{\log(\ell n / 2\eps) + O(1)}{e}} = 1$ and directly apply \cref{lem: existential uni onosf condenser general}.
\end{proof}

We also obtain the following general tradeoff for larger $n$ that may be growing with $\ell$ or even when $\ell = O(1)$ and $n$ growing alone (this applies to all $n$ but is most interesting when $n$ is large since \cref{cor: small n existential uni onosf condenser} provides better tradeoff for small $n$). 

\begin{corollary}[Larger $n$]\label{cor: large n existential uni onosf condenser}
For all $g, \ell, n, \eps, \delta$ where $g \ge (0.5+\delta)\ell$, and $\delta \ell n \ge 2\log(\ell n/2\eps) + O(1)$, there exists a condenser $\Cond: (\zo^n)^{\ell} \to \zo^m$ such that for any \unioNOSF[g, \ell, n] $\X$, we have $\sminH(\Cond(\X)) \ge m - \Delta$ where 
$m = \frac{\delta \ell n}{2} + (2/\delta-1)(\log(\ell n / 2\eps) + O(1)) + (2-\delta)\ell + \log(1/\eps) + O(1)$ and 
$\Delta = (2/\delta-1)(\log(\ell n / 2\eps) + O(1)) + 2(2-\delta)\ell + \log(1/\eps) + O(1)$.
\end{corollary}

\begin{proof}
With $e=\delta\ell$, we observe that $\ceil{\frac{\log(\ell n / 2\eps) + O(1)}{e}} \le 1 + \frac{\log(\ell n / 2\eps) + O(1)}{e}$ and apply that to the condenser from \cref{lem: existential uni onosf condenser general}.
\end{proof}

\subsection{Condenser for Two Uniform oNOSF Sources }\label{subsec: condenser for two uni onosfs}

In this subsection, we will prove \cref{lem:general tradeoff length-good rate-block size condenser}. To construct the claimed condenser, we will use the following folklore result regarding existence of excellent seeded condensers (e.g., see Corollary 3 of \cite{GLZ_cg_condenser}).

\begin{theorem} \label{lem:exists great lossless condenser}
For all $n, k, d, \eps$ such that $d \ge \log(n/\eps) + O(1)$, there exists a seeded condenser $\sCond:\zo^n\times \zo^d\to \zo^m$ such that for all $\X\sim \zo^n$ with $\minH(\X) = k$, we have $\sminH(\Cond(\X)) \ge k + d$ where $m = k + d + \log(1/\eps) + O(1)$.
\end{theorem}

We will also use the following result from \cite{CGR_seedless_condensers} that states an adversary cannot make things too bad if it controls very few bits. We note that similar lemmas have been useful in previous construction of condensers \cite{ben-aroya_two-source_2019, ball_randomness_2022, GLZ_cg_condenser}: 

\begin{lemma}[Lemma 6.18 in \cite{CGR_seedless_condensers}]\label{lem:control-few-bits-can-still-condense-again}
Let $\X \sim\zo^n$ be an arbitrary flat distribution and let $\Cond: \zo^n\to\zo^m$ be such that $\sminH(\Cond(\X)) \ge k$.
Let $G\subset [n]$ with $|G| = n - b$ be arbitrary.
Let $\X_G\sim \zo^{n-b}$ be the projection of $\X$ onto $G$.
Let $\Xpr\sim \zo^n$ be the distribution where the output bits defined by $G$ equal $\X_G$ and remaining $b$ bits are deterministic functions of the $n-b$ bits defined by $G$ under the restriction that $\supp(\Xpr)\subset \supp(\X)$.
Then, $\minH^{\epspr}(\Cond(\Xpr)) \ge k - b$ where $\epspr = \eps\cdot 2^b$.
\end{lemma}

With this, we are ready to provide the construction of condensers for two \unioNOSFs:

\begin{proof}[Proof of \cref{lem:general tradeoff length-good rate-block size condenser}]
Let $\sCond: (\zo^{n_x})^{\ell} \times (\zo^{n_y})^{\ell} \to \zo^{m}$ be the condenser guaranteed from \cref{lem:exists great lossless condenser} with $\eps_{\sCond} = \eps\cdot 2^{-(\ell - g) n_y}$.
We define $\Cond(x, y) = \sCond(x, y)$.

Let $\OO_{unif} = \Cond(\X, \U_{\ell n_y})$ and $\OO_{adv} = \Cond(\X, \Y)$.
We argue that $\OO_{unif}$ will be highly condensed and since the adversary controls so few bits in $\Y$, $\OO_{adv}$ will be condensed as well.

We first see that by the property of the seeded condenser, $\minH^{\eps_{\sCond}}(\OO_{unif}) \ge gn_x + \ell n_y$.
Next we observe that $\OO_{adv}$ can be obtained from $\OO_{unif}$ by an adversary controlling $b = (\ell - g)n_y$ bits from $(\X, \U_{\ell n_y})$ to obtain $(\X, \Y)$ and considering the output of $\sCond$.
We apply \cref{lem:control-few-bits-can-still-condense-again} which allows us to compare output entropy in such scenarios and obtain that
\[
\minH^{\eps_{\sCond}\cdot 2^b}(\OO_{adv}) \ge \minH^{\eps}(\OO_{unif}) - b\ge (gn_x + \ell n_y) - ((\ell - g)n_y) = m - \Delta.
\]
As $\eps_{\sCond}\cdot 2^b = \eps$, we indeed have that $\sminH(\OO_{adv}) \ge m - \Delta$ as desired.
\end{proof}

\dobib

\section{Extractors for oNOSF and oNOBF Sources via Leader Election Protocols}\label{sec:extractors for oNOSF/oNOBF via protocols}

In this section, we provide a generic way to transform leader election and coin flipping protocols into extractors for \oNOSFs and \oNOBFs. To do so, we must formally define the online influence of coalitions.

\begin{definition}\label{def: Online influence of coalitions}
    For any function $f: \Sigma^{\ell} \to \bits$, and any $B \subset [{\ell}]$, where $B = \{ i_1 < i_2 < \ldots < i_k\}$, define $\oI_B(f)$ as follows: an online adversary $\mathcal{A}$ samples a distribution $\X$ in online manner. It starts by sampling the variables $x_{1},x_2,\ldots,x_{i_1-1}$ independently and uniformly from $\Sigma$, then picking the value of $x_{i_1}$ depending on $x_{<i_1}$. Next, the variables $x_{i_1+1},\ldots,x_{i_2-1}$ are sampled independently and uniformly from $\Sigma$, and $\mathcal{A}$ sets the value of $x_{i_2}$ based on all variables set so far, and so on. Define the advantage of $\mathcal{A}$ to be $\textrm{adv}_{f,B}(\mathcal{A}) = |\E[f(\X)] - \E[f(\U_{\ell})]|$. Then $\oI_B(f)$  is defined to be $\max_{\mathcal{A}} \{\textrm{adv}_{f,B}(\mathcal{A})\}$, where the maximum is taken over all online adversaries $\mathcal{A}$ that control the bits in~$B$.

    We say a function $f$ is $(b,\varepsilon)$-online-resilient if $\oI_B(f) \le \varepsilon$ for every set $B \subset [\ell]$ of size at most~$b$. 
\end{definition}

We note that \cref{intro: oI definition} is a special case of \cref{def: Online influence of coalitions}, for $\Sigma = \{0,1 \}$ and $|B|=1$.

Now we return to our transformation from leader election and coin flipping protocols into extractors for \oNOSFs and \oNOBFs. Conceptually, given a leader election protocol, we can use an \oNOSF to simulate the protocol and then have the elected leader output its last block. We formalize this below.

\begin{lemma}\label{lemma: leader to extractor}
For any integers $r> 1,\ell > 0$ and any $\delta>0$, let $\pi$ be an $(r-1)$-round protocol over $\ell$ players that send $n$ bits per round such that for any $\delta \ell$ bad players, the protocol elects a good leader with probability $1-\eps$.

Then, there exists an explicit function $\Ext: (\zo^n)^{\ell r}\to \zo^n$ such that for any \oNOSF[g, \ell r, n] $\X$ where $g\ge \ell r - \delta \ell$, we have $\Ext(\X) \approx_{\eps} \U_n$.
\end{lemma}

\begin{remark}\label{rmk:cant just output protocol transcript}
    It may initially seem like a good idea to use the fact that a $(1-\delta)$-fraction of players are good and just directly output the input to $\Ext$ to get many more bits out than we do. However, the input only has entropy rate at least $1-\delta$ but is \emph{not} $\delta$-close to the uniform distribution, which is what we are achieving here.
\end{remark}

    

Instantiating our lemmas with the leader election protocols from \cref{sec:new leader election protocols}, we construct explicit extractors for \oNOBFs and uniform \oNOSFs:
 
\begin{theorem}\label{thm:oNOBF-extractor}
    There exists an explicit function $\Ext: \zo^{\ell}\to \zo$ such that for any $\delta$ and any \oNOBF[g, \ell] $\X$ where $g\ge \ell - \delta \ell / \log(\ell)$, we have $\Ext(\X) \approx_{\eps} \U_1$ where $\eps = C\delta + 12\left(C\delta\right)^{3/2} + \log(\ell)^{-1/3}$ where $C$ is a large universal constant.
\end{theorem}
\begin{proof}
This directly follows by instantiating \cref{lemma: leader to extractor} with the protocol guaranteed from \cref{lemma: leader 1 bit per round}.
\end{proof}

By using a the leader election protocol of \cref{lemma: leader log ell bits per round} with multiple bits per round, we construct extractors for \oNOSFs:

\begin{theorem}\label{thm:oNOSF-extractor}
    There exists an explicit function $\Ext: (\zo^n)^{\ell}\to \zo^n$ such that for any constant $\delta$ and any \oNOSF[g, \ell, n] $\X$ where $g\ge \ell - \delta \ell / \log^*(\ell)$ and $n \ge \log(\ell)$, we have $\Ext(\X) \approx_{\eps} \U_n$ where $\eps = C\delta + 13\left(C\delta\right)^{3/2}$.
\end{theorem}
\begin{proof}
This directly follows by instantiating \cref{lemma: leader to extractor} with the protocol guaranteed from \cref{lemma: leader log ell bits per round}.
\end{proof}

We finally prove our lemma regarding obtaining extractors for \oNOSFS from leader election protocols:

\begin{proof}[Proof of \cref{lemma: leader to extractor}]
Define function $\Ext$ as follows:    
On input $(y_1, \dots, y_r)$ where $y_i\in (\zo^n)^{\ell}$, let $y_{i,j} \in \zo^n$ denote the $j$'th block of $y_i$. Simulate the protocol $\pi$ with the message of the $j$'th player  in round $i$ being $y_{i,j}$, where $1 \le i \le r-1$ and $1 \le j \le \ell$. Let $j^* \in [\ell]$ be the leader that is elected by $\pi$; then output $y_{r,j^*}$

Let us analyze $\Ext$ on some source $\Y\sim (\zo^n)^{\ell r}$.
Let the bad symbols in $\Y$ be given by $A\subset [\ell] \times [r]$ where $\abs{A}\le \delta \ell$.
Let $\X\sim ((\zo^n)^{\ell})^{r}$ be the exact same source as $\Y$.
We write $\X = \{\X_{i, j}\}_{1\le i\le r, 1\le j\le \ell}$ and interpret it as the distribution where $\X_{i, j}$ denotes the random bits of player $j$ in round $i$.
Call $\X_{i, j}$ a bad block if the corresponding index $(i,j)$ is in $A$, i.e., the block is bad in $\Y$.
Since a bad block in $\Y$ can only depend on blocks before it, the corresponding bad block in $\X$ satisfies the criteria for being bad in $\X$; this is because a bad block in the protocol setting is allowed to depend on all blocks in the same or previous rounds.
Thus $\X$  has at most $\delta \ell$ bad blocks as well. 
By declaring the player corresponding to the bad block in $\X$ as bad, we obtain that the distribution $\X$ can be simulated by at most $\delta \ell$ bad players.
Formally, for $1\le i\le r$, let $B_i \subset [\ell]$ be the set of bad blocks in $\X$ among all blocks in round $i$.  
Let $B = \cup_{i=1}^r B_i$.
We declare all players in $B$ as bad players.
Finally, observe that
\[
\abs{B}\le \sum_{i=1}^r \abs{B_i} = \abs{A} = \delta \ell
\]
as desired. Thus the correctness of $\pi$ implies that after $(r-1)$ rounds, the chosen leader $j^*$ does not belong to $B$ with probability at least $1-\varepsilon$. By construction, it follows that $(r,j^*) \notin A$ whenever $j^*\notin B$. Thus, the output of the extractor, $\Y_{r,j*}$ is uniform on $n$ bits,  with probability at least $1-\varepsilon$.
\end{proof}
\dobib

\section{High Probability Leader Election Protocols}\label{sec:new leader election protocols}

We use this section to provide the leader election  protocols that are used in \cref{sec:extractors for oNOSF/oNOBF via protocols}. In \cref{subsec: one bit}, we present leader protocols where each player is allowed to send one bit per round.
We tackle the case where players can send multiple bits per round in \cref{subsec:many bits}.

\subsection{One Bit per Round} \label{subsec: one bit}

We will construct leader election protocols with the following guarantees:
\begin{lemma}\label{lemma: leader 1 bit per round}
There exists a universal constant $C$ and an explicit protocol over $\ell$ players, where each player sends $n = 1$ bit per round, that lasts for $C\log(\ell)$ rounds such that for any $\delta>0$, if $\delta \ell$ players are bad, then a good leader is chosen with probability $\ge 1 - \eps$ where $\eps = \delta + 12\delta^{3/2} + \log(\ell)^{-1/3}$.
\end{lemma}

We will use the following protocol from \cite{alonnoar93collective}:
\begin{lemma}\label{lemma: leader alon naor}
There exists a protocol $\pi$ over $\ell$ players where each player sends at most $1$ bit per round, that lasts for $O(\ell)$ rounds such that if $\delta \ell$ players are bad for $\delta \le 1/4$, then a good leader is chosen with probability $\ge 1 - \eps$ where $\eps = \delta + 12\delta^{3/2}$.
Furthermore, this protocol can be explicitly constructed in time $2^{O(\ell)}$.
\end{lemma}

We will also need the Chernoff bound:
\begin{lemma}\label{lemma: chernoff}
Let $\X_1, \dots, \X_n$ be independent random variables taking values in $\{0, 1\}$. 
Let $\X = \sum_{i=1}^n \X_i$ and let $\mu = \E[\X]$. 
Then, $\Pr[\X \le (1-\delta)\mu] \le e^{-\delta^2 \mu/2}$.
\end{lemma}

\begin{proof}[Proof of \cref{lemma: leader 1 bit per round}]
Our protocol will have two stages. In the first stage, we will use the lightest bin protocol from \cite{feige99} until the number of players is small enough, and then in the second stage we use the protocol from \cref{lemma: leader alon naor}.
Let $C_0$ be a large constant that we set later.
In particular, our final protocol will be:
\begin{enumerate}
\item 
Let $P_1 = [\ell]$.
\item
In round $i$ of stage $1$, all players in $P_i$ will present their value in $\zo$ and based on that, they will be divided into $P_i^0, P_i^1$.
\item
Set $P_{i+1}$ equal to the smaller set among $P_i^0, P_i^1$ (breaking ties arbitrarily).
\item
Repeat this until the number of players becomes at most $C_0 \log \ell$. 
Let this happens after $r$ rounds. This marks the end of the first stage.

\item
In the second stage, apply the protocol from \cref{lemma: leader alon naor} to $P_{r+1}$ and output the leader from that protocol.
\end{enumerate}

We now analyze this protocol. 
We argue that at the end of the first stage, with high probability, the fraction of good players in $P_{r+1}$ will be at least $(1-\delta) - o(1)$. For the second stage, the correctness of the protocol follows from \cref{lemma: leader alon naor}.

For $1\le i\le r+1$, let $g_i$ be the number of good players in $P_i$ and let $p_i = \abs{P_i}$.
As we always choose the lightest bin at each stage, $p_{i+1} \le p_i / 2$.
Hence, we infer that $p_i \le 2^{-i+1}\cdot \ell$.
Let $g_1 = g$. We next lower bound $g_i$:

\begin{claim}\label{claim: lower bound gi}
With probability at least $1 - \exp(-(1/10)\cdot (g/2^r))$, it holds that for all $1\le i\le r+1$, $g_i \ge \frac{g}{2^i} - 5\left(\frac{g}{2^i}\right)^{2/3}$.
\end{claim}

We prove this claim using concentration bounds later.
Using this claim, we see that in $P_{r+1}$, the number of good players will be at least
\[
\frac{(1-\delta)\ell}{2^r} - 5\left(\frac{(1-\delta)\ell}{2^r}\right)^{2/3}
\]
out of $p_{r+1}\le \frac{\ell}{2^r}$ many surviving players.
In particular, $g_{r+1} \ge (1-\delta)p_{r+1} - 5p_{r+1}^{2/3}$.
So, in stage 2, we have $p_{r+1}$ many players remaining where the fraction of bad players is $\delta' = \delta + 5p_{r+1}^{-1/3}$.
Applying \cref{lemma: leader alon naor} with these parameters, we infer that probability of electing a good leader is at least 
\[
1 - \left(\delta + 5p_{r+1}^{-1/3} + 12\left(\delta + 5p_{r+1}^{-1/3}\right)^{3/2}\right) \ge 
1 - \delta - 12\delta^{3/2} - 6p_{r+1}^{-1/3}
\]
where the last inequality follows because $p_{r+1} \ge \omega(1)$.
Hence, our overall probability of electing a good leader is at least 
\[
1 - \delta - 12\delta^{3/2} - 6p_{r+1}^{-1/3} - \exp(-(1/10)\cdot (1-\delta)p_{r+1})
\ge 1 - \delta - 12\delta^{3/2} - \log(\ell)^{-1/3} = 1 - \eps
\]
where the last inequality follows because we let $p_{r+1} = C_0\log(\ell)$ for a large constant $C_0$.
We check that the number of rounds in the first stage is no more than $\log(\ell)$ and in stage 2, as guaranteed by \cref{lemma: leader alon naor}, the number of rounds is no more than $O(\log(\ell))$. These together give us our universal constant $C$ that we use in the claim.

\begin{proof}[Proof of \cref{claim: lower bound gi}]
Fix either of the two bins. We apply \cref{lemma: chernoff} with $\delta = \mu^{-1/3}$ to infer that with probability at least $1 - \exp(-(g_i/2)^{1/3}/2)$, it holds that the number of good players in that bin is $\ge g_i/2 - (g_i/2)^{2/3}$.
Applying this to both bins, we infer that with probability at least $1 - 2\exp(-(g_i/2)^{1/3}/2)$, it holds that $g_{i+1}\ge g_i/2 - (g_i/2)^{2/3}$.
By unravelling this recurrence and lower bounding, we see that
\[
g_{i+1} \ge \frac{g}{2^i} - \sum_{j=1}^i \frac{(g/2^j)^{2/3}}{2^{i-j}}
\]
Hence,
\begin{align*}
g_{i+1} 
& \ge \frac{g}{2^i} - g^{2/3}\sum_{j=1}^i 2^{j/3-i}\\
& = \frac{g}{2^i} - \left(\frac{g}{2^i}\right)^{2/3}\sum_{j=1}^i (2^{1/3})^{j-i}\\
& = \frac{g}{2^i} - \left(\frac{g}{2^i}\right)^{2/3}\sum_{j=0}^{i-1} (2^{-1/3})^{j}\\
& \ge \frac{g}{2^i} - \left(\frac{g}{2^i}\right)^{2/3}\frac{1}{1-2^{-1/3}}\\
& \ge \frac{g}{2^i} - 5\left(\frac{g}{2^i}\right)^{2/3}. \qedhere
\end{align*}
\end{proof}

By union bound, the overall probability that the claim holds is at least 
\begin{align*}
1 - \sum_{i=1}^{r+1}2\exp(-(g_i/2)^{1/3}/2)
&\ge 1 - \exp(-g_{r+1}/6)\\
&\ge 1 - \exp(-(1/10)\cdot (g/2^r)). \qedhere
\end{align*}
\end{proof}

\subsection{Multiple Bits per Round}\label{subsec:many bits}

If the players are allowed to send $O(\log \ell)$ bits per round, then the number of rounds can be significantly improved.
\begin{lemma}\label{lemma: leader log ell bits per round}
There exists a universal constant $C$ and an explicit protocol over $\ell$ players where each player sends $n = \log \ell$ bits per round, that lasts for $C\cdot \log^* \ell$ rounds such that for any constant $\delta>0$, if $\delta \ell$ players are bad, then a good leader is chosen with probability $1 - \eps$ where $\eps = \delta + 13\delta^{3/2}$.
\end{lemma}

\begin{proof}
Our protocol and proof is similar to \cref{lemma: leader 1 bit per round} with the key difference being that the larger value of $n$ allows us to increase the number of bins and simplify our analysis. Here, we end up being verbose and repeating ourselves for clarity.
Just like earlier, our protocol will have two stages, one using the lightest bin protocol from \cite{feige99} until the number of players is small enough and then resorting to the protocol from \cref{lemma: leader alon naor}.
Let $C_0, C_1$ be large constants that we set later.
Our final protocol will be:
\begin{enumerate}
\item 
Let $P_1 = [\ell]$.
\item
In round $i$ of stage $1$, all players in $P_i$ will present a number between $1$ and $b_i = \abs{P_i} / \log(\abs{P_i})^{C_0}$. Based on this value, they will be divided into sets$P_i^j$ where $j\in[b_i]$.
\item
Set $P_{i+1}$ equal to the smallest set amongst $P_i^1, \dots, P_i^{b_i}$ (breaking ties arbitrarily).
\item
Repeat this until the number of players becomes at most $\exp\left(\left(\log(1/\delta)\right)^{C_1}\right)$ (stop right before it goes below this value). 
Let this happens after $r$ rounds. This marks the end of the first stage.

\item
In the second stage, apply the protocol from \cref{lemma: leader alon naor} to $P_{r+1}$ and output the leader from that protocol.
\end{enumerate}

We now analyze this protocol. 
We argue that at the end of the first stage, with high probability, the fraction of good players in $P_{r+1}$ will be at least $(1-\delta) - o(1)$. For the second stage, the correctness of the protocol follows from \cref{lemma: leader alon naor}.

For $1\le i\le r+1$, let $g_i$ be the number of good players in $P_i$ and let $p_i = \abs{P_i}$.
As we always choose the lightest bin at each stage, $p_{i+1} \le p_i / b_i$.
Hence, we infer that $p_{r+1} \le \ell / \prod_{i=1}^r b_i$.
Let $g = g_1$. We first bound $g_i$:

\begin{claim}\label{claim: lower bound gi large buckets}
For any constant $C_1$, with probability at least $1 - \exp(-\log(p_{r+1})^{1/5})$, it holds that for all $1\le i\le r+1$, $g_i \ge \frac{g}{\prod_{j=1}^{i-1}b_j} - 2\left(\frac{g}{\prod_{j=1}^{i-1}b_j}\right)^{2/3}$.
\end{claim}

We prove this claim using concentration bounds later, and we remark that $C_0$ will be a growing function of $C_1$.
Using this claim, we see that in $P_{r+1}$, the number of good players will be at least
\[
\frac{(1-\delta)\ell}{\prod_{i=1}^{r} b_i} - 2\left(\frac{(1-\delta)\ell}{\prod_{j=1}^r b_i}\right)^{2/3}
\]
out of $p_{r+1}\le \frac{\ell}{\prod_{i=1}^{r} b_i}$ many surviving players.
In particular, $g_{r+1} \ge (1-\delta)p_{r+1} - 2p_{r+1}^{2/3}$.

So, in stage 2, we have $p_{r+1}$ many players remaining where the fraction of bad players is $\delta' = \delta + 2p_{r+1}^{-1/3}$.
Applying \cref{lemma: leader alon naor} with these parameters, we infer that probability of electing a good leader is at least 
\[
1 - \left(\delta + 2p_{r+1}^{-1/3} + 12\left(\delta + 2p_{r+1}^{-1/3}\right)^{3/2}\right) \ge 
1 - \delta - 12\delta^{3/2} - 3p_{r+1}^{-1/3}
\]
where the last inequality follows because $p_{r+1} \ge \omega(1)$.
Hence, our overall probability of electing a good leader is at least 
\[
1 - \delta - 12\delta^{3/2} - 3p_{r+1}^{-1/3} - \exp(-\log(p_{r+1})^{1/5})
\ge 1 - \delta - 12\delta^{3/2} - \exp(-\log(p_{r+1})^{1/6})
\ge 1 - \delta - 13\delta^{3/2}
= 1 - \eps
\]
where the first inequality follows because $C_1$ is a large enough universal constant, and $\delta < 1/4$.
We check that the number of rounds in the first stage is no more than $O(\log^*(\ell))$ and in stage 2, as guaranteed by \cref{lemma: leader alon naor}, the number of rounds is no more than $ c \log^*(\ell)$, where $c$ is a constant that just depends on $\delta$ and $C_1$ (and is independent of $\ell$). These together give us our universal constant $C$ of \Cref{lemma: leader log ell bits per round}.

\begin{proof}[Proof of \cref{claim: lower bound gi large buckets}]
Fix any of the $b_i$ bins in round $i$. We apply \cref{lemma: chernoff} with $\delta = \mu^{-1/3}$ to infer that with probability at least $1 - \exp(-(g_i/b_i)^{1/3}/2)$, it holds that the number of good players in that bin is $\ge g_i/b_i - (g_i/b_i)^{2/3}$.
Applying this to all $b_i$ bins, we infer that with probability at least $1 - b_i\exp(-(g_i/b_i)^{1/3}/2)$, it holds that $g_{i+1}\ge g_i/b_i - (g_i/b_i)^{2/3}$.
By unraveling this recurrence and lower bounding, we see that
\[
g_{i+1} \ge \frac{g}{\prod_{j=1}^i b_j} - \sum_{j=1}^i \frac{(g/\prod_{k=1}^j b_k)^{2/3}}{\prod_{k=j+1}^i b_k}
\]
For ease of notation, let $\alpha(u, v) = \prod_{j=u}^v b_j$.
Hence,
\begin{align*}
g_{i+1} 
& \ge \frac{g}{\alpha(1, i)} - g^{2/3}\sum_{j=1}^i \frac{(1/\alpha(1, j))^{2/3}}{\alpha(j+1, i)}\\
& = \frac{g}{\alpha(1, i)} - \left(\frac{g}{\alpha(1, i)}\right)^{2/3}\sum_{j=1}^i \frac{(\alpha(1, i)/\alpha(1, j))^{2/3}}{\alpha(j+1, i)}\\
& = \frac{g}{\alpha(1, i)} - \left(\frac{g}{\alpha(1, i)}\right)^{2/3}\sum_{j=1}^i \alpha(j+1, i)^{-1/3} \qedhere.
\end{align*}
\end{proof}

We observe that each term in the summand is exponentially decreasing. Hence, we can upper bound the the sum by $2\left(\frac{g}{\alpha(1, i)}\right)^{2/3}$.

This means 
\[
    g_{i+1}\ge \frac{g}{\alpha(1, i)} - 2\left(\frac{g}{\alpha(1, i)}\right)^{2/3}.
\]

By union bound, the overall probability that the claim holds is at least 
\begin{align*}
1 - \sum_{i=1}^{r+1}b_i \exp(-(g_i/b_i)^{1/3}/2)
& = 1 - \sum_{i=1}^{r+1}\exp(-(g_i/b_i)^{1/3}/2 + \log(b_i)).
\end{align*}
By our choice of parameters, in particular by letting $C_0$ to be a large enough constant, we can ensure that $g_i / b_i \ge \poly(b_i)$. Thus, we can ensure that the probability that the claim holds is at least 
\begin{align*}
 1 - \sum_{i=1}^{r+1}\exp(-(g_i/b_i)^{1/3}/2 + \log(b_i))
 & \ge 1 - \sum_{i=1}^{r+1}\exp(-\log(p_i)^{1/4})
\end{align*}
where we get the constant $1/4$ by appropriately increasing $C_0$ and we used the fact that $\delta < 1/4$.
As $p_i$ is exponentially decreasing, we infer that the overall probability that the desired conclusion holds is at least
\[
    1 - \exp(-\log(p_{r+1})^{1/5}). \qedhere
\]
\end{proof}

\dobib

\section{Online Influence and Extraction Lower Bounds}\label{sec:extracting oNOBF lower bound}


 Towards proving lower bounds on the possiblity of extraction from \oNOSFs, we introduce a new, natural notion of influence of Boolean functions, which we call \emph{online influence}. For simplicity, we first start by considering the class of \oNOBFs, which corresponds to  \unioNOSFs[g, \ell, n=1].  
 
We believe this is an interesting new measure and is worth studying in its own right, and we refer the reader to \cref{example: tight oI ineqs} for a couple of interesting examples. For monotone functions (and more generally, unate functions), it is not hard to see that online influence equals the usual notion of influence (see \cref{lem:monotone implies oI=I} for a proof). Thus, to find interesting properties of online influence (compared to standard influence, \cref{defn: influence}), one must look at non-monotone (in fact, non-unate) Boolean functions.

The following natural question arises towards our goal of proving extractor lower bounds: for a function $f$, what is the maximum online influence out of all $n$ bits? For the usual notion of influence, this question was resolved by the well-known theorem of \cite{kahn_influence_1988}, who showed there always exists a bit with influence at least $\Var(f) \cdot \Omega\left(\frac{\log \ell}{\ell}\right)$.  

We show that surprisingly, there exists a balanced function, namely the address function, where every bit has online influence at most $O\left(\frac{1}{\ell}\right)$ (see \cref{lem: address function tight oI} for a proof). This provides a separation between the usual notion of influence and online influence.

\paragraph{Organization} We formally define the notion for Boolean functions and discuss some  basic properties in \cref{sec:online influence basics}. We establish tight bounds on the online influence for general functions, including a Poincar\'e style inequality, in \cref{sec:Poincare style inequality}. We provide an example exhibiting a separation between maximum (standard) influence and online influence in \cref{sec:online influence tight examples}. Finally, in \cref{sec:oNOBF condensing and extraction lower bounds}, we extend the definition of online influence to  subsets of coordinates (and functions from $\Sigma^n \rightarrow \zo^m$, for arbitrary alphabet $\Sigma$). This allows us to prove the required lower bounds on extraction (and condensing) from \oNOSFs.

\paragraph{Notation} For convenience, we introduce some  notation that we use for the rest of this section. For any bit $b \in \zo$, let $e(b) = (-1)^{b}$. For any Boolean function $f: \bits^{\ell} \rightarrow \bits$, let $e(f)$ denote the function $e(f)(x) =(-1)^{f(x)}$.  

\subsection{Basic Properties}\label{sec:online influence basics}

In this section, for a function $f:\bits^{\ell} \to\bits$, we will freely use commas to indicate concatenation in its input. For example, for $x\in\bits^{i-1}$ and $y\in\bits^{{\ell}-i}$, we write $f(x,1,y)$ to indicate $f$ applied to the tuple $(x_1,\dots,x_{i-1},1,y_1,\dots,y_{{\ell}-i})$. 

When asking about the influence of a single bit, such as the $i$-th bit, previous work has specifically looked at whether the $i$-th bit still has the ability to change the output of some function $f:\bits^{\ell}\to\bits$ after all other ${\ell}-1$ bits have been set. In other words, if the $i$-th bit is a non-oblivious adversary (that is, it can look at the values of all the other bits before setting its own value), how much power does it have? This has led to a standard notion of influence defined below. 

\begin{definition}[Influence]\label{defn: influence}
    For a function $f:\bits^{\ell}\to\bits$, the \emph{influence} of the $i$-th bit is
    \begin{align*}
        \I_i[f]= \E_{\substack{x\sim\U_{i-1}\\y\sim\U_{n-i}}}\left[\abs{f(x,1,y)-f(x,0,y)}\right]
    \end{align*}
    and the \emph{total influence} is
    \begin{align*}
        \I[f]&=\sum_{i=1}^{\ell}\I_i[f].
    \end{align*}
\end{definition}

However, in our setting of \oNOSFs and \oNOBFs, an adversarial bit can only depend on the bits that come before it. This motivates our new definition of online influence, where we prevent the $i$-th bit from depending on bits that come after it by independently sampling subsequent bits.

\begin{definition}[Online influence]
    For a function $f:\bits^{\ell}\to\bits$, the \emph{online influence} of the $i$-th bit is
    \begin{align*}
        \oI_i[f]=   \E_{x\sim\U_{i-1}}\left[\abs{\E_{y\sim\U_{{\ell}-i}}[f(x,1,y)]-\E_{y\sim\U_{{\ell}-i}}[f(x,0,y)]}\right]
    \end{align*}
    and the \emph{total online influence} is
    \begin{align*}
        \oI[f]&=\sum_{i=1}^{\ell}\oI_i[f].
    \end{align*}
\end{definition}

\begin{remark}\label{rmk:last bit influences same}
    It is easy to see that for any $f:\bits^{\ell} \to \bits$, and any $i \in {\ell}$, we have $\oI_i(f) \le \I_i(f)$. Further, they are the same for the last bit: $\I_{\ell}[f]=\oI_{\ell}[f]$. 
\end{remark}

Many results for the influence of a function are based on working with monotone functions. In contrast, it turns out that monotone functions are not very interesting for online influence as the definition collapses to that of regular influence.

\begin{lemma}\label{lem:monotone implies oI=I}
    If $f:\bits^{\ell}\to\bits$ is monotone, then $\oI_i[f]=\I_i[f]$ for all $i\in[{\ell}]$.
\end{lemma}
\begin{proof}
Using the monotonicity of $f$, note that for any $x \in \bits^{i-1}$ and any $y \in \bits^{{\ell}-i}$, $f(x,1,y) \ge f(x,0,y)$. Thus,  $\oI_i[f]=  \E_{x\sim\U_{i-1},y\sim\U_{{\ell}-i}}[f(x,1,y)- f(x,0,y)] = \I_i(f)$.
\end{proof}

    

Thus, any difference between influence and online influence can only be demonstrated by non-monotone functions. 

    


\subsection{A Poincar\'e Inequality for Online Influence}\label{sec:Poincare style inequality}

Similar to regular influence, we prove  a Poincar\'e-style inequality holds for online influence, and also provide an upper bound on online influence. The following is the main result of this subsection.

\begin{theorem}\label{thm:oI ineqs}
    For any  $f:\bits^{\ell}\to\bits$, we have $\Var(e(f)) \leq \oI[f]\leq  \sqrt{{\ell} \Var(e(f))}$.
\end{theorem}

Before proving the above result, we observe that the MAJORITY and PARITY functions provide tight examples  for the upper and lower bound respectively for \cref{thm:oI ineqs}. 
\begin{example}\label{example: tight oI ineqs}
    The majority function on $\ell$ bits $\operatorname{Maj}_{\ell}:\bits^{\ell}\to\bits$, is monotone, and hence by \cref{lem:monotone implies oI=I},  has total online influence $\oI[\operatorname{Maj}_{\ell}]=\I[\operatorname{Maj}_{\ell}]=\sqrt{2{\ell}/\pi}+O(1/\sqrt{{\ell}})$, achieving the upper bound (up to constants).

    The PARITY function on ${\ell}$ bits $\bigoplus_{\ell}:\bits^{\ell}\to\bits$ for $i\in[{\ell}-1]$ has online influence $\oI_i[\bigoplus_{\ell}]=0$, while $\oI_{\ell}[\bigoplus_{\ell}]=1$. Thus, PARITY meets the lower bound of \cref{thm:oI ineqs}. We note that this is starkly different from regular influence where $\I_i[\bigoplus_\ell]=1$ for all $i$.
\end{example}

To prove \cref{thm:oI ineqs}, we will use Boolean Fourier analysis.  For any $f: \zo^n \to \zo$, $e(f)$ has a unique Fourier expansion given by: $e(f(x)) = \sum_{S \subseteq [{\ell}]} \widehat{f}(S) \chi_S(x)$, where $\chi_S(x) = (-1)^{\sum_{i\in S} x_i}$ and $\widehat{f}(S) = \E_{y \sim \U_{\ell}}[e(f)(y)\chi_S(y)]$.\footnote{For simplicity of notation, we use $\widehat{f}(S)$ for $\widehat{e(f)}(S)$.} Also recall that $\widehat{f}(\emptyset) = \E_{x \sim \U_n}[e(f)(x)]$, $\Var(e(f)) = \sum_{ S \subseteq [{\ell}], S \neq \emptyset}\widehat{f}(S)^2$, and for any $S \neq T$, $\E_{x \sim \U_{\ell}}[\chi_S(x) \chi_T(x)] = 0$. For more background, we refer the reader to the excellent book by O'Donnell \cite{o2014analysis}.

The following is our key lemma, from which \cref{thm:oI ineqs} is easy to derive. 
\begin{lemma}\label{lemma:Fourier bb}
For any $f: \bits^{\ell} \to \bits$ and $i \in [{\ell}]$, $\oI_i(f)^2 \le \sum_{\substack{S\subseteq[i]\\S\ni i}}\widehat{f}(S)^2 \le  \oI_i(f)$.
\end{lemma}

We first derive  \cref{thm:oI ineqs}   using \cref{lemma:Fourier bb}.
\begin{proof}[Proof of \cref{thm:oI ineqs}]
   We start with the lower bound. We have, 
      

    \begin{equation*}
    \oI[f] = \sum_{i=1}^{\ell}\oI_i[f] 
    \geq \sum_{i=1}^{\ell} \sum_{\substack{S\subseteq[i]\\S\ni i}}\widehat{f}(S)^2 
      = \sum_{\substack{S\subseteq[{\ell}]\\S\neq\varnothing}}\widehat{f}(S)^2 
    = \Var(e(f)),
\end{equation*}
where the inequality uses   \cref{lemma:Fourier bb}.

    The upper bound is easy to derive as well.   

    \begin{align*}
        \oI[f]&=\sum_{i=1}^{\ell}\oI_i[f]\\
        &\leq\sqrt{\ell\sum_{i=1}^{\ell}\left(\oI_i[f]\right)^2} && \text{(Cauchy-Schwarz inequality)}\\
        &\leq \sqrt{\ell\sum_{i=1}^{\ell} \sum_{\substack{S\subseteq[i]\\S\ni i}}\widehat{f}(S)^2} && \text{(\cref{lemma:Fourier bb})} \\
        &=\sqrt{{\ell}\Var(e(f))}.
    \end{align*}
    
    This completes the proof.
\end{proof}
We now focus on proving \cref{lemma:Fourier bb}. We need the following useful  characterization of~$\oI_i(f)$.
\begin{claim}\label{lem:Fourier oIif}
    For any $f:\bits^{\ell}\to\bits$, we can write the online influence of its $i$-th bit as
    \begin{align*}
        \oI_i[f]&= \E_{x\sim\U_{i-1}}\left[\abs{\sum_{\substack{T\subseteq[i]\\T\ni i}}\widehat{f}(T)\chi_{T\setminus\{i\}}(x)}\right].
    \end{align*}
\end{claim}

Assuming the above claim, let us prove \cref{lemma:Fourier bb}. We supply the proof of \cref{lem:Fourier oIif} below.
\begin{proof}[Proof of \cref{lemma:Fourier bb}]
We first prove the inequality $\oI_i(f) \ge   \sum_{\substack{S\subseteq[i]\\S\ni i}}\widehat{f}(S)^2$.  Since for any $x\in\bits^{i-1}$ we have $  \abs{\E_{y\sim\U_{{\ell}-i}}[e(f|_{x,1})(y)]-\E_{y\sim\U_{{\ell}-i}}[e(f|_{x,0})(y)]}= 2 \abs{\sum_{\substack{T\subseteq[i]\\T\ni i}}\widehat{f}(T)\chi_{T\setminus\{i\}}(x)}$ by \cref{lem:Fourier oIif}, and the fact that $\E_{y\sim\U_{{\ell}-i}}[e(f|_{x,b})(y)]$ is in $[-1,1]$ for all $x\in \zo^{i-1}, b\in \zo$, it follows that $\left|\sum_{\substack{T\subseteq[i]\\T\ni i}}\widehat{f}(T)\chi_{T\setminus\{i\}}(x)\right|$ is in $[0,1]$.

Thus, 

    \begin{align*}
        \oI_i[f]&=  \E_{x\sim\U_{i-1}}\left[\abs{\sum_{\substack{T\subseteq[i]\\T\ni i}}\widehat{f}(T)\chi_{T\setminus\{i\}}(x)}\right]\\
        &\geq \E_{x\sim\U_{i-1}}\left[\left(\sum_{\substack{T\subseteq[i]\\T\ni i}}\widehat{f}(T)\chi_{T\setminus\{i\}}(x)\right)^2\right]\\
        &= \sum_{\substack{T\subseteq[i]\\T\ni i}}\sum_{\substack{S\subseteq[i]\\S\ni i}}\widehat{f}(T)\widehat{f}(S) \cdot
          \E_{x\sim\U_{i-1}}[\chi_{T\setminus\{i\}}(x) \chi_{S\setminus\{i\}}(x)] \\
        &= \sum_{\substack{S\subseteq[i]\\S\ni i}}\widehat{f}(S)^2.
    \end{align*}

Next, we prove $\oI_i(f)^2 \le \sum_{\substack{S\subseteq[i]\\S\ni i}}\widehat{f}(S)^2$. We have, 
\begin{align*}
        \oI_i[f]^2&=  \left(\E_{x\sim\U_{i-1}}\left[\abs{\sum_{\substack{T\subseteq[i]\\T\ni i}}\widehat{f}(T)\chi_{T\setminus\{i\}}(x)}\right]\right)^2 && \text{(\cref{lem:Fourier oIif})}\\
        &\le   \E_{x\sim\U_{i-1}}\left[\left(\sum_{\substack{T\subseteq[i]\\T\ni i}}\widehat{f}(T)\chi_{T\setminus\{i\}}(x)\right)^2\right]\qquad  \\
        &= \sum_{\substack{S\subseteq[i]\\S\ni i}}\widehat{f}(S)^2 && \text{(derived above).} \qedhere
    \end{align*}
\end{proof}

Next, we show how to rewrite $\oI_i[f]$ in terms of the Fourier coefficients of $f$.

\begin{proof}[Proof of \cref{lem:Fourier oIif}]
    We begin by defining the restriction $f|_{x,b}(y)=f(x,b,y)$ for $x\in\bits^{i-1}$, $b\in\bits$, and $y\in\bits^{{\ell}-i}$. Thus, we can rewrite $\oI_i[f]$ as 
    \begin{align}
        \oI_i[f]&=\frac{1}{2} \cdot \E_{x\sim\U_{i-1}}\left[\abs{\E_{y\sim\U_{{\ell}-i}}[e(f|_{x,1})(y)]-\E_{y\sim\U_{{\ell}-i}}[e(f|_{x,0})(y)]}\right].\label{eq:oIf with restrictions}
    \end{align}
    We would like to put the above expression in terms of Fourier coefficients of $f$. This motivates us to find the Fourier coefficients of $f|_{x,b}(y)$ in terms of those of $f$, which we do via computation. We manipulate the Fourier expansion of $f(z)$ for $z=(x,b,y)\in\bits^{\ell}$ to get
    \begin{align*}
        e(f)(z)&=\sum_{S\subseteq[{\ell}]}\widehat{f}(S)\chi_S(z)\\
        &=\sum_{S\subseteq[{\ell}]}\widehat{f}(S)\chi_S(x,b,y)\\
        &=\sum_{S\subseteq[{\ell}]}\widehat{f}(S)\chi_{S\cap[i]}(x,b)\chi_{S\setminus[i]}(y)\\
        &=\sum_{S\subseteq\{i+1,\dots,{\ell}\}}\left(\sum_{T\subseteq[i]}\widehat{f}(S\cup T)\chi_T(x,b)\right)\chi_S(y).\numberthis\label{eq:coefficients f manipulated}
    \end{align*}
    We also have that
    \begin{align*}
        e(f)(z)&=e(f)(x,b,y)\\
        &=e(f|_{x,b})(y)\\
        &=\sum_{S\subseteq\{i+1,\dots,{\ell}\}}\widehat{f|_{x,b}}(S)\chi_S(y).\numberthis\label{eq:coefficients f restriction}
    \end{align*}
    Therefore, \cref{eq:coefficients f manipulated} and \cref{eq:coefficients f restriction} allow us to conclude that
    \begin{align*}
        \widehat{f|_{x,b}}(S)&=\sum_{T\subseteq[i]}\widehat{f}(S\cup T)\chi_T(x,b).
    \end{align*}

    Thus, we have
    \begin{align*}
        \E_{y\sim\U_{{\ell}-i}}\left[e(f|_{x,b})(y)\right]&=\widehat{f|_{x,b}}(\varnothing)\\
        &=\sum_{T\subseteq[i]}\widehat{f}(T)\chi_T(x,b)\\
        &=\sum_{\substack{T\subseteq[i]\\T\ni i}}\widehat{f}(T)\chi_{T\setminus\{i\}}(x)b+\sum_{T\subseteq[i-1]}\widehat{f}(T)\chi_T(x).
    \end{align*}

    We now plug this in to our definition of $\oI_i[f]$ in \cref{eq:oIf with restrictions} to get a simplified expression. Recalling the fact that for any $x \in \zo^n$, $f(x) = (1-e(f)(x))/2$, we have
    \begin{align*}
         \oI_i[f]&=\frac{1}{2}  \E_{x\sim\U_{i-1}}\left[\abs{\E_{y\sim\U_{{\ell}-i}}[e(f|_{x,1})(y)]-\E_{y\sim\U_{{\ell}-i}}[e(f|_{x,0})(y)]}\right]\\
        &=\frac{1}{2}   \E_{x\sim\U_{i-1}}\left[\abs{\left(-\sum_{\substack{T\subseteq[i]\\T\ni i}}\widehat{f}(T)\chi_{T\setminus\{i\}}(x)+\sum_{T\subseteq[i-1]}\widehat{f}(T)\chi_T(x)\right)-   \left(\sum_{\substack{T\subseteq[i]\\T\ni i}}\widehat{f}(T)\chi_{T\setminus\{b\}}(x)+\sum_{T\subseteq[i-1]}\widehat{f}(T)\chi_T(x)\right)}\right]\\
        &=  \E_{x\sim\U_{i-1}}\left[\abs{\sum_{\substack{T\subseteq[i]\\T\ni i}}\widehat{f}(T)\chi_{T\setminus\{i\}}(x)}\right].
    \end{align*}
\end{proof}

\subsection{A Tight Example for Maximum Online Influence}\label{sec:online influence tight examples}


The lower bound on total online influence from \cref{thm:oI ineqs} allows us to conclude that for balanced functions, there must be at least one bit with online influence $\Omega(1/{\ell})$. We can phrase this in terms of maximum influence.
\begin{definition}[Maximum influence]
    For a function $f:\bits^{\ell}\to\bits$, we define its \emph{maximum influence} as $\MaxI[f]=\max_{i\in[\ell]}\I_i[f]$ and its \emph{maximum online influence} as $\MaxoI[f]=\max_{i\in[{\ell}]}\oI_i[f]$.
\end{definition}
In terms of maximum online influence, we get the following corollary from \cref{thm:oI ineqs}.
\begin{corollary}\label{cor:single bit oIi lower bound}
    For a function $f:\bits^{\ell}\to\bits$, we have $\MaxoI[f] \ge \Var(e(f))/{\ell}$.
\end{corollary}
\begin{proof}
    By \cref{thm:oI ineqs} we have that $\oI[f]=\sum_{i=1}^{\ell}\oI_i[f]\geq \Var(e(f))$, and the conclusion follows via an averaging argument.
\end{proof}

We show that the bound in \cref{cor:single bit oIi lower bound} is in fact tight (up to constants), as witnessed by the address function.

\begin{definition}
    We define the \emph{address function} $\Addr_{\ell}:\bits^{\log({\ell})+{\ell}}\to\bits$ as follows: For $z\in\bits^{\log({\ell})+{\ell}}$, split $z$ up as $z=(x,y)$ with $x$ of length $\log({\ell})$ and $y$ of length ${\ell}$. Then interpret $x$ as a binary number which gives us an index $i(x)\in[{\ell}]$. The output of $\Addr_{\ell}$ is the $i(x)$-th bit of $y$, so $\Addr_{\ell}(x,y)=y_{i(x)}$.
\end{definition}

\begin{lemma}\label{lem: address function tight oI}
    Let $m={\ell}+ \log {\ell}$ and $\Addr_{\ell}$ be the function defined above. Then,
    \begin{itemize}
        \item for $1\le i \le \log {\ell}$, $\oI_i[\Addr_{\ell}]= 0$.
        \item for $\log {\ell} < i \le m$, $\oI_i[\Addr_{\ell}]= 1/{\ell}$.
    \end{itemize} 
    Thus, $\MaxoI(\Addr_{\ell}) = \Theta(1/m)$.
\end{lemma}
\begin{proof}
    For $i\in[\log {\ell}]$, no matter what the value of the $i$-th bit of $\Addr_{\ell}$ is set to, the output bit will be a uniform bit, so we immediately get that $\oI_i[f]=0$. 
    For $i\in\{\log {\ell}+1,\dots,m\}$, the $i$-th bit only has control if it's selected by the first $\log {\ell}$ address bits, meaning it has a $1/{\ell}$ chance of controlling the output (and otherwise the output is uniform). Hence, $\oI_i[f]=\frac{1}{{\ell}}$. 
\end{proof}

Compared with the result of \cite{kahn_influence_1988} that $\MaxI[f]\geq\Var(f) \cdot \Omega\left(\frac{\log {\ell}}{{\ell}}\right)$, this exhibits a separation between maximum (standard) influence and the online influence (of balanced functions).

Moreover, this analysis of the address function also shows us that it is an extractor for \unioNOSFs[\ell-1,\ell].
\begin{lemma}\label{lem:address extracts (l-1;l)-oNOSF}
For all $\ell, n$ where $\ell \ge 2$ and $n \ge \log(\ell-1)$, there exists an explicit extractor $\Ext: (\zo^{n})^{\ell}\to \zo^n$ such that for any \unioNOSF[\ell-1, \ell, n] $\X$, we have $\Ext(\X) \approx_{\eps} \U_n$ where $\eps = \frac{1}{\ell-1}$.
\end{lemma}

\begin{proof}
Let $\Ext$ be defined as follows: From the first block, use the first $\log(\ell-1)$ bits and interpret them as an index $j\in [\ell-1]$. Then, output the block with index $j+1$.
For a source $\X$ with first block controlled by an adversary, the output will be truly uniform and for a source $\X$ with adversary controlling one of the last $\ell-1$ blocks, that block will be outputted with probability $\frac{1}{\ell-1}$ while a uniform block will be outputted otherwise. This makes our total error at most $\frac{1}{\ell-1}$ as desired.
\end{proof}


\subsection{Online Influence of Sets and Extraction Lower Bounds}\label{sec:oNOBF condensing and extraction lower bounds}

For convenience we restate the definition of online influence of sets of coordinates.

\begin{definition}[Online influence, \cref{def: Online influence of coalitions} restated]
    For any function $f: \Sigma^{\ell} \to \bits$, and any $B \subset [{\ell}]$, where $B = \{ i_1 < i_2 < \ldots < i_k\}$, define $\oI_B(f)$ as follows: an online adversary $\mathcal{A}$ samples a distribution $\X$ in online manner. It starts by sampling the variables $x_{1},x_2,\ldots,x_{i_1-1}$ independently and uniformly from $\Sigma$, then picking the value of $x_{i_1}$ depending on $x_{<i_1}$. Next, the variables $x_{i_1+1},\ldots,x_{i_2-1}$ are sampled independently and uniformly from $\Sigma$, and $\mathcal{A}$ sets the value of $x_{i_2}$ based on all variables set so far, and so on. Define the advantage of $\mathcal{A}$ to be $\textrm{adv}_{f,B}(\mathcal{A}) = |\E[f(\X)] - \E[f(\U_{\ell})]|$. Then $\oI_B(f)$  is defined to be $\max_{\mathcal{A}} \{\textrm{adv}_{f,B}(\mathcal{A})\}$, where the maximum is taken over all online adversaries $\mathcal{A}$ that control the bits in~$B$.

    We say a function $f$ is $(b,\varepsilon)$-online-resilient if $\oI_B(f) \le \varepsilon$ for every $B$ of size at most~$b$. Moreover, if $B$ is an unsorted set, we sort it before considering it in this definition.
\end{definition}
In the special case where $\Sigma=\bits$ and we are considering the online influence of a single coordinate, the definition simplifies nicely.
\begin{definition}
    For a function $f:\bits^{\ell}\to\bits$, the \emph{online influence} of the $i$-th bit is
    \begin{align*}
        \oI_i[f]= \E_{x\sim\U_{i-1}}\left[\abs{\E_{y\sim\U_{{\ell}-i}}[f(x,1,y)]-\E_{y\sim\U_{{\ell}-i}}[f(x,0,y)]}\right]
    \end{align*}
    and the \emph{total online influence} is
    \begin{align*}
        \oI[f]&=\sum_{i=1}^{\ell}\oI_i[f].
    \end{align*}
\end{definition}

Online-resilient functions are equivalent to extractors (with $1$ output bit) for \oNOSFs.

\begin{lemma}[online-resilient functions yield extractors]\label{lem:online-resilient functions yield extractors}
	Let $f:\Sigma^{\ell}\to\zo$ be a $(b,\varepsilon_1)$-online-resilient function with the property that $\abs{f(\U_{\ell})-\U_1}\leq\varepsilon_2$. Then $f$ can extract from \oNOSFs[g=\ell-b, \ell] with error at most $\varepsilon_1+\varepsilon_2$.
\end{lemma}



\begin{proof}
Consider a \oNOSF[g=\ell-b, \ell] $\X$. Recall that $\X$ is created by choosing some set of bad indices $B$ of size $b$, letting the symbols in $\overline{B}$ be uniform, and finally setting the symbols in $B$ adversarially while only depending on uniform symbols to the left of them. Using the triangle inequality for total variation distance, we get that
\begin{align*}
	\abs{f(\X)-\U_1}&\leq \abs{f(\X)-f(\U_{\ell})}+\abs{f(\U_{\ell})-\U_1}\\
	&\leq\varepsilon_1+\varepsilon_2,
\end{align*}
as claimed.
\end{proof}
\begin{remark}
    We note that the other direction is immediate from definitions. If $\Ext: \Sigma^\ell \to \zo$ is an extractor with error $\eps$ for $(g=\ell-b,\ell)$-\oNOSFs, then $\Ext$ is a $(b,2\eps)$-online-resilient function.
\end{remark}    

\begin{remark} Our results below on \textrm{oNOBF} extraction impossibility can be interpreted as a limit on online-resilience of balanced Boolean functions.
\end{remark}
For $B\subset [\ell]$, we use the notation $f|_{\overline{B}}$ to indicate the function obtained from $f$ by letting an online adversary control the indices in $B$.
\begin{theorem}\label{thm:how many bad bits needed to get to beta}
    Let $f:\bits^{\ell}\to\bits$ be such that $\E_{x\sim\U_{\ell}}[f(x)=1] = \alpha$. Then for any $1\geq\beta>\alpha$, there exists  a coalition $B\subseteq[{\ell}]$ such that $\oI_B(f) \ge \beta -\alpha$, where $\abs{B}\leq\gamma \ell$ and $\gamma=\frac{\beta-\alpha}{4\alpha(1-\beta)}$.
\end{theorem}
\begin{proof}
    We greedily collect the bits with the most online influence and add them to $B$ until our goal of $\E_{x\sim\U_{\ell}|_{\overline{B}}}[f|_{\overline{B}}(x)=1]\geq\beta$ is achieved. Our first step is as follows: let $B_0=\varnothing$,  $f_0=f$, and $i_1=\argmax_{i\in[{\ell}]}\{\oI_i[f]\}$. \cref{cor:single bit oIi lower bound} tells us that $\oI_{i_1}[f_0]\geq\Var(e(f_0))/\ell$. Recall that if $\E_{x\sim\U_{\ell}}[f(x)=1]=p$ then $\Var(e(f))=4p(1-p)$. Because we have not yet achieved our goal of $\E_{x\sim\U_{\ell}|_{\overline{B}}}[f|_{\overline{B}}(x)=1]\geq\beta$, we have that $\Var(e(f_0))\geq 4\alpha(1-\beta)$. Thus, we collect $i_1$ as $B_1=\{i_1\}$, let $f_1=f_0|_{\overline{B_1}}$ and see that $\E_x[f_1(x)]\geq\E_x[f_0(x)]+\oI_{i_1}[f_0]\geq \alpha+\frac{4\alpha(1-\beta)}{{\ell}}$. 

    We now repeat this process $t$ times to get $B_t=\{i_1,\dots,i_t\}$ until our goal is achieved. For general $t$, let $f_t=f|_{B_t}$ where $B_t=B_{t-1}\cup \{i_t\}$ and $i_t=\argmax_{i\in[n]\setminus B_{t-1}}\{\oI_i[f_{t-1}]\}$. At the $(t-1)$-th step, since we have not stopped, it means that $\E_{x}[f_{t-1}(x)=1]<\beta$, but we of course have $\E_{x}[f_{t-1}(x)=1]\geq\alpha$ as well. Thus, by \cref{cor:single bit oIi lower bound}, collecting $i_t$ as a bad bit gives us that
    \begin{align*}
        \E_x[f_t(x)]&\geq \E_x[f_{t-1}(x)]+\oI_{i_t}[f_{t-1}]\\
        &\geq \alpha+\frac{4\alpha(1-\beta)}{{\ell}}(t-1)+\frac{4\alpha(1-\beta)}{{\ell}}\\
        &=\alpha+\frac{4\alpha(1-\beta)}{{\ell}}\cdot t.
    \end{align*}
    We repeat this process until $\Pr_x[f_t(x)=1]\geq\beta$. Therefore, the number of steps is the smallest $b$ such that $\alpha+\frac{4\alpha(1-\beta)}{{\ell}}\cdot b\geq\beta$, meaning that the number of steps is at most  $b\leq {\ell}\cdot\frac{\beta-\alpha}{4\alpha(1-\beta)}$. We let $B=B_b$ and get the desired coalition. 
\end{proof}
 We can also ask the dual question of how large we are able to make $\beta$ given some budget $b$ of bad bits.

\begin{corollary}\label{cor:min prob beta given bad bits b}
    Let $f:\bits^{\ell}\to\bits$ be such that $\Pr_{x\sim\U_{\ell}}[f(x)=1]\geq\alpha$. If we are able to control $b$ bits in an online adversarial manner, then there exists a set $B\subseteq[{\ell}]$ of indices of size $\abs{B}=b$ such that $\Pr_{x\sim\U_{\ell}|_{\overline{B}}}[f|_{\overline{B}}(x)=1]\geq\beta$ where $\beta\geq\frac{\alpha( {\ell}+4b)}{{\ell}+4\alpha b}$.
\end{corollary}
\begin{proof}
    For a fixed $\beta$, \cref{thm:how many bad bits needed to get to beta} tells us that $b\leq {\ell}\cdot\frac{\beta-\alpha}{4\alpha(1-\beta)}$. Solving for $\beta$ gives the desired~bound.
\end{proof}

We now immediately obtain our \textrm{oNOBF} extraction impossibility result.

\begin{corollary}\label{cor:impossible_nobf_ext}
    For any balanced function $f:\zo^\ell\to\zo$ and $0< \varepsilon < 1/3$, there exists a \oNOBF[g=\ell-b, \ell] $\X$ with $b\leq 3 \varepsilon \ell$ such that $\abs{f(\X)-\U_1}\geq\varepsilon$.
\end{corollary}
\begin{proof}
    It is enough to find a set $B$ of indices such that $ \oI_B(f) \ge \beta$. By \cref{thm:how many bad bits needed to get to beta}, there exists such a set $B$ of size $b=\abs{B}\leq {\ell}\cdot\frac{\varepsilon}{1-2\varepsilon}$. The bound on $|B|$ follows since $\varepsilon\leq\frac{1}{3}$.
\end{proof}

\begin{remark}
    By essentially following our Fourier analytic proof, one can similarly obtain a Poincar\'e inequality for functions $f:\Sigma^n \rightarrow \zo$, for arbitrary alphabet $\Sigma$. To obtain extraction impossibility for such \unioNOSFs with constant $\delta$ fraction of corrupt blocks, we do the following: Let $f$ be a candidate extractor for \unioNOSFs[(1-\delta)\ell, \ell, n]. Then, $f$ also extracts from \unioNOSF[\ceil{1/\delta}-1, \ceil{1/\delta}, \ell n / \ceil{1/\delta}]. Since there exists an influential coordinate with influence $O(\delta)$, we let the adversary control that coordinate and infer that there exists constant $\eps = O(\delta)$ for which it is impossible to extract with error less than $\eps$.
\end{remark}

\dobib

\section{Open Problems}\label{sec:open problems}

We list here some  interesting open problems left by our work:

\begin{itemize}
    \item 
    While we obtain explicit condensers for almost all parameter regimes, it remains open to construct them when the bock length is constant, matching the parameters of our existential results.
    As we show, one way of achieving this would be to  explicitly construct a seeded condenser with dependence on seed length being $1\cdot \log(1/\eps)$.

    \item 
    All our condensers have entropy gap much larger than a constant. It will be interesting to show there exist condensers with constant entropy gap (for any values of $n, \ell$) for \unioNOSFs. A slightly weaker but equally interesting question is to construct seeded extractors for \unioNOSFs with constant seed length.

    \item 
    Show that there exist non-trivial condensers for \oNOBFs or show no such condenser exists. We conjecture that no condenser exists with output entropy rate larger than the input entropy rate for such sources.

    \item 
    Construct $\eps$-collective sampling protocols with fewer rounds than the ones obtained using \unioNOSF condensers. It will also be interesting to explicitly construct such protocols when the number of players are very large compared to the number of bits each player has access to. Further, proving lower bounds for $\eps$-collective sampling protocols is a natural direction to explore.

    \item 
    Determine the exact threshold for extracting from \oNOBFs and \oNOSFs. Our lower bounds show extraction is impossible when $g\le 0.99\ell$ while our constructions using leader election protocols require $g\ge \ell - \Omega\left(\frac{\ell}{\log \ell}\right)$ for \oNOBFs and $g\ge \ell - \Omega\left(\frac{\ell}{\log^*(\ell)}\right)$ for \oNOSFs[g, \ell, n] where $n\ge \log(\ell)$. Using the connection between extractors and leader election protocols, lower bounds for extraction imply lower bounds for leader election protocols. In particular, matching lower bounds for extraction would imply all current leader election protocols are tight, a long standing open problem.
    
\end{itemize}

\dobib

\subsection*{Acknowledgements}
We thank Madhur Tulsiani for asking a question  that motivated us to consider the model of local oNOSF sources in \cref{sec:local oNOSFs}.
We thank Miranda Christ for helpful feedback on an earlier version of the manuscript.
We thank the organizers of the Dagstuhl Seminar on Algebraic and Analytic Methods in Computational Complexity and Schloss Dagstuhl for providing a stimulating research environment, where discussions between R.S. and E.C. contributed to this collaboration.

\dobib


\printbibliography
\appendix
\section{\texorpdfstring{Constructing $\Reduce'$}{Constructing Reduce'}}\label{subsubsec:RZ reduce base}

In this section we construct $\Reduce'$ which has the properties as guaranteed by \cref{lem:RZ reduce base}.
In \cite{russellzuckerman01}, the authors use hitting sets for combinatorial rectangles to reduce $\ell$-length \oNOSFs to shorter min-entropy sources. We provide a proof of their lemma for completeness here.

Let's first define combinatorial rectangles.
\begin{definition}[Combinatorial rectangle]
    Let $a,d\in\N$. We say that a set $R\subseteq[a]^d$ is a \emph{combinatorial rectangle} if $R=R_1\times R_2\times\cdots\times R_d$ for some sets $R_i\subseteq[a]$ for $i\in[d]$. The \emph{density} of $R$ is $\operatorname{Density}(R)=\frac{1}{a^d}\prod_{i=1}^d\abs{R_d}$.
\end{definition}

A hitting set for a family of combinatorial rectangles is a subset of $[a]^d$ such that it has an intersection with every combinatorial rectangle in the family. Formally:
\begin{definition}[Hitting sets for combinatorial rectangles]
    A set $\mathcal{H}\subseteq[a]^d$ is a \emph{$(a,d,\delta)$-hitting set for combinatorial rectangles} if for every combinatorial rectangle $R\subseteq[a]^d$ with $\operatorname{Density}(R)\geq\delta$ we have that $R\cap\mathcal{H}\neq\varnothing$. 
\end{definition}

Of course, taking $\mathcal{H}=[a]^d$ is a trivial hitting set for any combinatorial rectangle, so the difficulty lies in decreasing the cardinality of $\mathcal{H}$ while keeping the density requirement $\delta$ of the combinatorial rectangle low. In \cite{linial1997efficient}, the authors create a small enough hitting set for our use.
\begin{lemma}[\cite{linial1997efficient}]\label{lem:LLSZ hitting set}
    There exists a universal constant $C$ such that for any $\delta>0$ and $a,d\in\N$, there exists an explicit construction of an $(a,d,\delta)$-hitting set $\mathcal{H}\subseteq[a]^d$ such that $\abs{\mathcal{H}}\leq\left(\frac{a\log(d)}{\delta}\right)^C$.
\end{lemma}

Let's see how using all of these ingredients we can construct $\Reduce'$.
\begin{proof}[Proof of \cref{lem:RZ reduce base}]
    To construct $\Reduce'$, we begin by defining a family of functions $\mathcal{F}\subseteq\{f:[a]^d\to\zo\}$ and a hitting set $\mathcal{H}\subseteq[a]^d$ of size $\abs{\mathcal{H}}=2^t=T$. For every $x\in \supp(\X)$, we will select a $f_x\in\mathcal{F}$ and output the smallest $y\in\mathcal{H}$ such that $f_x(y)=1$, where we consider our output as an element of $[\abs{\mathcal{H}}]=\zo^t$. Then, for all $y\in\mathcal{H}$, we will show that $\Pr_{x\sim\X}[f_x(y)=1]\leq 2^{-k}$, meaning that $\Reduce'(\X)$ is a $(t,k)$-source.

    Formally, we let $\mathcal{F}$ be the following family of combinatorial rectangles on $[a]^d$. Given an $x\in[a]^d$, we define the combinatorial rectangle $\Rect_x=\{y\in[a]^d\mid\forall i\in[d], y_i\neq x_i\}$ and the associated function $f_x:[a]^d\to\zo$ for this rectangle as $f_x(y)=1_{y\in\Rect_x}$. Then, we let $\mathcal{F}=\{f_x\mid x\in[a]^d\}$.

    Note that the density of any particular rectangle $\Rect_x$ is $\delta=\operatorname{Density}(\Rect_x)=\frac{\Rect_x}{a^d}=\left(1-\frac{1}{a}\right)^d$. We can lower bound $\delta$ as $\delta\geq(\exp(-d/a))^{C_\delta}$ for some universal constant $C_\delta$. Rearranging then gives us that $\log(1/\delta)\leq C_\delta\cdot\frac{d}{a}$. With this in mind, we set up our hitting set $\mathcal{H}$ for $\mathcal{F}$. From \cite{linial1997efficient}, we know that there exists a universal constant $C_1$ and an explicit hitting set $\mathcal{H}$ such that $T=\abs{\mathcal{H}}\leq\left(\frac{a\log(d)}{\delta}\right)^{C_1}$. Simplifying this expression yields
    \begin{align*}
        T&\leq\left(\frac{a\log(d)}{\delta}\right)^{C_1}\\
        t&\leq C_1(\log(a)+\log\log(d)+\log(1/\delta))\\
        t&\leq C_1(\log(a)+\log\log(d)+C_\delta d/a)\\
        t&\leq C'(\log(a)+\log\log(d)+d/a),\numberthis\label{eq:t general UB}
    \end{align*}
    where $C'$ is a sufficiently large universal constant, depending only on $C_1, C_\delta$. To analyze the min-entropy of $\Reduce'(\X)$, we note that for all $y\in[a]^d$

    \begin{align*}
        \Pr_{x\sim\X}[f_x(y)=1]&\leq\left(1-\frac{1}{a}\right)^{\gamma d}\\
        &\leq\exp(-\gamma d/a) \le 2^{-\gamma d / a},
    \end{align*}
    which directly implies that the min-entropy $k$ of $\Reduce'(\X)$ is $\geq\gamma d/a$, as desired.
\end{proof}




\section{Extracting from Local \texorpdfstring{\oNOSFS}{online NOSF Sources}}\label{sec:local oNOSFs}

A natural variation on our definition of \oNOSFs is to consider the case where the adversary cannot remember the value of every good block in the past; rather, it can only remember the value of the most recent $s$ blocks. Arguably, this is a realistic assumption in the setting of many short blocks, where it could be difficult to introduce long range correlation. 
\begin{definition}[Local \oNOSFs]
    We call a \oNOSF[g,\ell,n,k] $\X=(\X_1,\dots,\X_\ell)$ an \emph{$s$-local \oNOSF[g,\ell,n,k]} if each bad block $\X_i$ can only depend on at most $s$ blocks $\X_{i-s},\dots,\X_{i-1}$ that come before it.
\end{definition}

Interestingly, weakening the adversary in this way  converts our \oNOSF into a small-space source. These sources were first studied by \cite{kamp_deterministic_2011} and we refer the reader to them for a definition and background. Since the adversarial blocks of an $s$-local \oNOSF[g,\ell,n,k] can only depend on the binary string of length at most $sn$ to its left, we easily see that an $s$-local 
\oNOSF[g,\ell,n,k] is samplable by a space-$sn$ source. 

Using  recent explicit extractors for low-space sources provided by \cite{chattopadhyay_extractors_2022, li2023two} and the fact that a $\oNOSF[g,\ell,n,k]$ has entropy at least $gk$, we get the following extraction result for these local online sources.
\begin{theorem}[Using the explicit extractor of \cite{chattopadhyay_extractors_2022}]
    There exists a universal constant $C$ such that for every $s$ and $k\geq\frac{2sn+\log^C(n\ell)}{g}$ there is an explicit extractor $\Ext:(\zo^n)^\ell\to\zo^m$ with error $\varepsilon=(n\ell)^{-\Omega(1)}$ and output length $m=(gk-2sn)^{\Omega(1)}$ for every $s$-local \oNOSF[g,\ell,n,k].
\end{theorem}
A similar result with slightly better entropy requirement, but constant error, can be obtained using the small-space extractor from \cite{li2023two}.

\end{document}